\documentclass[a4paper, 11pt]{article}
\usepackage{a4wide}
\usepackage[english]{babel}
\usepackage{tabularx}

\usepackage{float}
\restylefloat{figure}
\usepackage{amsfonts}
\usepackage{amsmath}
\usepackage{amssymb}
\usepackage{eurosym}
\usepackage{graphicx}
\usepackage[usenames,dvipsnames,svgnames,table]{xcolor}
\usepackage[colorlinks, bookmarks=true,linkcolor=MidnightBlue,citecolor=ForestGreen]{hyperref}
\usepackage{amsthm}
\usepackage{enumitem}
\usepackage[hypcap=false]{caption}
\usepackage{array, multirow, makecell}
\usepackage{csquotes}
\usepackage{mathtools}
\usepackage{dsfont}
\usepackage{tabularx,ragged2e}
\usepackage{longtable}
\usepackage{nth}
\usepackage{subcaption}
\usepackage{placeins}
\usepackage[backend=bibtex, citestyle=numeric, style=numeric, maxbibnames=99]{biblatex}
\usepackage{booktabs}
\usepackage{xargs}
\usepackage{xifthen}
\usepackage{tikz}

\setcellgapes{1pt}
\makegapedcells
\newtheorem{lemma}{Lemma}[section]

\newtheorem{theorem}{Theorem}[section]
\newtheorem{proposition}{Proposition}[section]
\newtheorem{remark}{Remark}[section]

\newtheorem{assumption}{Assumption}

\newtheorem{corollary}{Corollary}[section]

\DeclarePairedDelimiter{\ceil}{\lceil}{\rceil}

\newcommandx{\E}[2][1={}, 2={}]{\ifthenelse{\isempty{#1}}{\mathbb{E}}{\mathbb{E}^{#1}}\ifthenelse{\isempty{#2}}{}{\left[#2\right]}}
\newcommandx{\Proba}[2][1={}, 2={}]{\ifthenelse{\isempty{#1}}{\mathbb{P}}{\mathbb{P}^{#1}}\ifthenelse{\isempty{#2}}{}{\left[#2\right]}}
\newcommand{\diff}{\mathrm{d}}
\newcommand{\defeq}{\vcentcolon=}

\usepackage{ltcaption}

\title{From Glosten-Milgrom to the whole limit order book and\\
applications to financial regulation\footnote{This research benefited from the financial support of the ERC grant 679836 Staqamof and the chairs ``Deep finance \& Statistics'' and ``Analytics and Models for Regulation'' of \'Ecole polytechnique. We thank Alexandra Givry, Philippe Guillot, Charles-Albert Lehalle, Julien Leprun and Ioanid Ro{\c{s}}u for their valuable comments. We thank BMLL Technologies for providing the historical market data which made this study possible.}}

\author{Weibing Huang\thanks{LPSM, Sorbonne Universit\'e. \textbf{Email:} weibing.huang@upmc.fr}
\and Sergio Pulido\thanks{Universit\'e Paris-Saclay, CNRS, ENSIIE, Univ Evry, Laboratoire de Math\'ematiques et Mod\'elisation d'Evry (LaMME). \textbf{Email:} sergio.pulidonino@ensiie.fr} 
\and Mathieu Rosenbaum\thanks{Centre de Mathématiques Appliquées (CMAP), CNRS, École polytechnique, Institut Polytechnique de
Paris. \textbf{Email:} mathieu.rosenbaum@polytechnique.edu}
\and Pamela Saliba\thanks{Centre de Mathématiques Appliquées (CMAP), CNRS, École polytechnique, Institut Polytechnique de
Paris and Autorité des Marchés Financiers. \textbf{Email:} pamela.saliba@polytechnique.edu}
\and Emmanouil Sfendourakis\thanks{Centre de Mathématiques Appliquées (CMAP), CNRS, École polytechnique, Institut Polytechnique de
Paris. \textbf{Email:} emmanouil.sfendourakis@polytechnique.edu}}

\begin{document}

\maketitle

\begin{abstract}
\noindent We build an agent-based model for the order book with three types of market participants: an informed trader, a noise trader and competitive market makers. Using a Glosten-Milgrom like approach, we are able to deduce the whole limit order book (bid-ask spread and volume available at each price) from the interactions between the different agents. More precisely, we obtain a link between efficient price dynamic, proportion of trades due to the noise trader, traded volume, bid-ask spread and equilibrium limit order book state. With this model, we provide a relevant tool for regulators and market platforms. We show for example that it allows us to forecast consequences of a tick size change on the microstructure of an asset. It also enables us to value quantitatively the queue position of a limit order in the book. 
\end{abstract}


\section{Introduction}\label{intro}
Limit order book (LOB) modeling has become an important research topic in quantitative finance. This is because market participants and regulators need to use LOB models for many different tasks such as optimizing trading tactics, assessing the quality of the various algorithms operating on the markets, understanding the behaviors of market participants and their impact on the price formation process or designing new regulations at the microstructure level. In the literature, there are two main ways to model the LOB: statistical and equilibrium models. In statistical models, agents order flows follow suitable stochastic processes. In this type of approach, the goal is to reproduce important market stylized facts and to be useful in practice, enabling practitioners to compute relevant quantities such as trading costs, market impact or execution probabilities. Most statistical models are so-called zero-intelligence models because order flows are driven by independent Poisson processes, see for example \cite{abergel2013mathematical,cont2013price,cont2010stochastic,lachapelle2016efficiency,smith2003statistical}. This assumption is relaxed in \cite{bayer2017functional,huang2015simulating,huang2017ergodicity} where more realistic dynamics are obtained introducing dependencies between the state of the order book and the behavior of market participants. Market generators, see \cite{kuo2021improving,labiad2024generative}, aim to generate realistic LOB data using deep learning.

In equilibrium models, see for instance \cite{foucault1999order,glosten1985bid, parlour1998price,rocsu2009dynamic}, LOB dynamics arise from interactions between rational agents acting optimally: the agents choose their trading decisions as solutions of individual utility maximization problems. For example in \cite{parlour1998price}, the author investigates a simple model where traders choose the type of order to submit (market or limit order) according to market conditions, and taking into account the fact that their decisions can influence other traders. In this framework, it becomes possible to analyze accurately market equilibria. However, the spread is exogenous and there is no asymmetric information on the fundamental value of the asset so that no adverse selection effect is considered.  This is the case in the order-driven model of \cite{rocsu2009dynamic} too, where traders can also choose between market and limit orders. In this approach, all information is common knowledge and the waiting costs are the driving force. This model leads to several very relevant predictions about the links between trading flows, market impact and LOB shape.

In this paper, we introduce an equilibrium-type model. It is a simple agent-based model for the order book where we consider three types of market participants like in \cite{kyle1985continuous}: an informed trader, a noise trader
and market makers. The informed trader receives market information such as the jumps of the
efficient price, which is hidden to the noise trader. He then takes advantage of this information
to gain profit by sending market orders. Market makers also receive the same information but
with some delay and they place limit orders as long as the expected gain of these orders
is positive (they are assumed to be risk-neutral). The informed trader and market makers
represent the strategic part in the trading activity, while the random part consists in the noise trader who is assumed to send market orders according to a compound Poisson process.

Interestingly, the above simple framework allows us to deduce a link between efficient price dynamic, proportion of trades due to the noise trader, traded volume, bid-ask spread and equilibrium state for the LOB. It enables us to derive the whole order book shape (bid-ask spread and volume present at each price) from the interactions between the agents. The question of how the bid-ask spread emerges from the behavior of market participants
has been discussed in many works. It is generally accepted that the bid-ask spread is non-zero because of the existence of three types of costs: order processing costs, see \cite{huang1997components,rocsu2009dynamic}, inventory costs, see \cite {ho1981optimal,wyart2008relation}, and adverse selection costs, see \cite{glosten1985bid}. In \cite{rocsu2009dynamic}, the spread is a consequence of order processing costs: to compensate their waiting costs, traders place their limit orders on different price levels (for example, a sell limit order at a higher level gets a better expected price than one at a lower level but needs longer time to be executed. Thus the case where both orders lead to the same expected utility can be considered).

In contrast, our model is inspired by \cite{glosten1985bid}. Liquidity is offered by market makers only and they face an adverse selection issue since a participant agreeing to trade at the market maker's ask or bid price may be trading because he is informed. Order processing and inventory costs are neglected
and we consider the bid-ask spread as a purely informational phenomenon: limit orders are placed at different levels because liquidity providers must protect themselves from traders with superior information. In this framework, in a very similar way as in \cite{glosten1985bid}, the bid-ask spread emerges naturally from the fact that limit orders placed too close to the efficient price have negative expected returns when being executed:
the presence of the informed trader and the potential large jumps of the efficient price prevent
market makers from placing limit orders too close to the efficient price. We also find that the bid-ask spread turns out to be the sum of the tick value and of the intrinsic bid-ask spread, which corresponds to a latent value of the bid-ask spread under infinitesimal tick size.

Let us emphasize that several models study the LOB assuming the presence of our three types of market participants and imposing, as we will do, a zero-profit type condition stating that limit orders can only be placed in the LOB if their expected return relative to the efficient price is non-negative. For instance, the papers \cite{glosten1985bid} and \cite{baruch2017tail} share multiple similarities with ours. Compared with \cite{glosten1985bid}, there are two main differences. First, in \cite{glosten1985bid}, the zero-profit assumption applies only to the two best offer limits: the bid and ask prices at each trade are set to yield zero-profit to the market maker, and time priority plays no role. In our model, we propose a generalized version of the zero-profit condition under which fast market makers can still make profits because of time priority. Second, in \cite{glosten1985bid}, one assumes that only unit trades can occur, which is quite restrictive. In our model we relax this assumption, which allows us to retrieve the whole LOB shape and not only the bid-ask spread. In addition to this, we also treat the crucial case for practice where the tick size is non-zero, whereas it is assumed to be vanishing in \cite{glosten1985bid}.

In \cite{baruch2017tail}, the authors investigate the consequences of a zero-profit condition at the level of the whole liquidity supply curve provided by each market maker. This is an intricate situation where standard equilibria cannot be reached since a profitable deviation (from a Nash equilibrium) for any market maker is to offer the shares at a slightly higher price as explained in \cite{bernhardt1997splitting}. In this work, we rather assume that when a market maker computes his expected profit, he takes into consideration the orders submitted by other market makers. In particular, queue priority plays a key role in our analysis. This is done so that the zero-profit condition holds only for the last order of each queue in the LOB. It in particular means that a market maker can still make positive profit. This enables us to obtain a very operational and tractable framework, where we can deduce the whole LOB shape, compute various important quantities such as priority values of limit orders, and make predictions about consequences of regulatory changes, for example on the tick size.

Note that an important point in our model is that we also consider the case where the tick size is non-zero. This allows us to analyze its role in the LOB dynamic. For instance, we derive a new and very useful relationship between the tick size and the spread. We validate this relationship on market data and show how to use it for regulatory purposes, in particular to forecast new spread values after tick size changes.

To estimate the model, we use a maximum likelihood approach, inverting the characteristic function of efficient price jumps between regularly spaced time intervals. The distribution of the sizes of trades submitted by noise traders is calibrated so that the average LOB shape in our model matches exactly the one observed empirically.

The discreteness of available price levels also enables us to value in a quantitative way the queue position of limit orders. LOBs use a priority system for limit orders submitted at the same price. Several priority rules can be employed such as price-time priority or price-size priority, see \cite{gould2013limit}. We consider here the widely used price-time mechanism which gives priority to the limit orders in a first in first out way. Therefore, it encourages traders to submit limit orders early. Our model is one of the few approaches allowing to quantify with accuracy the advantage of being
at the top of the queue compared to being at its end. A notable exception is the paper \cite{moallemi2016model}. In this work, the authors value queue positions at the best levels for large tick assets in a queuing model taking into account price impact and some adverse selection. In our setting, we are able to compute the effects of the strategic interactions between market participants on queue position valuation. Furthermore, we are not restricted to the best levels of large tick assets. However, as will be seen in our empirical results, our findings are in line with those of \cite{moallemi2016model}. 

Imposing a discrete tick grid also allows the model to reproduce a well-known stylized fact: the predictive power of the volume imbalance for future price moves, highlighting a relationship between volume imbalance and efficient price as in \cite{pulido2023understanding,stoikov2018micro}.

The paper is organized as follows. In Section \ref{continu}, we introduce our agent-based LOB model
with zero tick value. Based on a greedy assumption for the informed trader's behavior, a link
is deduced between traded volume, efficient price jump distribution and LOB shape. We then add
the zero-profit condition for market makers, which enables us to compute
explicitly the bid-ask spread as well as the LOB shape. In Section \ref{discrete}, the case of
non-zero tick value is considered. We show that the bid-ask spread is in fact equal to the
sum of the intrinsic bid-ask spread (without the tick value constraint) and the tick value. 
The LOB shape under positive tick size is also deduced, and we give an
explicit formula for the value of the queue position of a limit order. In Section \ref{sec:estimation}, we describe the estimation procedure to retrieve the parameters of the model. In Section \ref{sec:applications}, we present four applications of our model: spread forecasting under a tick size change, probabilities of price moves with respect to the volume imbalance, waiting time until the next trade and computation of queue position values. Finally, the proofs are relegated to an appendix, as well as the tables with the estimated parameters.

\subsection{Notations}
In this work, $\mathbb{N}$ denotes the set of natural integers including 0, $\mathbb{N}^*=\mathbb{N}\setminus \{0\}$. For an integrable function $f$ defined on $\mathbb{R}$, $\hat{f}$ denotes its Fourier transform: $\hat{f}(z) = \int_{\mathbb{R}}e^{izx} \diff x$. For a measurable function $f$ defined on $\mathbb{R}$, $f^{\ast n}$ is $f$ convoluted $n \in \mathbb{N}^*$ times with itself, if it exists. Thus,  $f^{\ast 1}=f$ and  $f^{\ast 2} = f \ast f$. For $x \in \mathbb{R}$, $\ceil*{x}$ denotes the smallest integer $m$ such that $x \leqslant m$.

\section{Model and assumptions}\label{continu}

In our model, we assume the existence of an efficient price modeled by a compound Poisson process and the presence of three different types of market participants: an informed trader, a noise trader and several market makers. In our approach, market makers choose their bid-ask quotes by computing the expected gain of potential limit orders at various price levels. This is done in a context of asymmetric information between the informed and the noise trader regarding the efficient price (the efficient price is actually used as a tool to materialize asymmetry of information). This framework enables us to obtain explicit formulas for the spread and LOB shape. These quantities essentially depend on the law of the efficient price jumps, the distribution of the noise trader's orders size, and the number of price jumps compared to that of orders sent by the noise trader. Note that contrary to most LOB models which deal only with the dynamics at the best bid/ask limits, or assume that the spread is constant, see for example \cite{cont2013price}, our model allows for spread variations and applies to the whole LOB shape. We present in this section the case where the tick size is assumed to be equal to zero. The obtained results will help us understand those in Section \ref{discrete} where we consider a positive tick size.

\subsection{Modeling the efficient price}

We write $P(t)$ for the market underlying efficient price, whose dynamic is described as follows:
$$P(t)=P_0+Y(t),$$
where $Y(t)=\sum_{j=1}^{N_t}B_j$ is a compound Poisson process and $P_0>0$. Here $\{N_t: t\ge 0\}$ is a Poisson process with intensity $\lambda^i>0$, and the $\{B_j: j\ge1\}$ are independent, identically distributed, integrable and independent of $N$ random variables with non-negative symmetric density $f_B$ on $\mathbb{R}$ and cumulative distribution function $F_B$. Hence, we consider that new information arrives on the market at discrete times given by a Poisson process with intensity $\lambda^i$. We assume that at the $j^{th}$ information arrival time, the efficient price $P(t)$ is modified by a jump of random size $B_j$.

\noindent Furthermore, since $\mathbb{E}[B_j]=0$, we have that $P(t)$ is a martingale. Thus, $\mathbb{E}[P(t)]=P_0$. If $B_j$ is square integrable, $\text{Var}[P(t)]=\lambda^i t \mathbb{E}[B_j^2]$. In this case, we view $\lambda^i \mathbb{E}[B_j^2]$ as the macroscopic volatility of our asset. In the sequel, for sake of simplicity, we write $B$ for $B_j$ when no confusion is possible.  

\subsection{Market participants}

We assume that there are three types of market participants:
\begin{itemize}
\item One informed trader: by this term, we mean a trader who undergoes low latency and is able to access market data and assess efficient price jumps faster than other participants, creating asymmetric information in the market. For instance, he can analyze external information, use better technology, or use lead-lag relationships between assets or platforms to evaluate the efficient price (for details about lead-lag see \cite{hayashi2018wavelet,hoffmann2013estimation,huth2014high}). Therefore, we assume that the informed trader receives the value of the price jump size $B$ (and
the efficient price $P(t)$) just before it happens. He then sends his trades based
on this information to make profit. He does not send orders at other times than those of price jumps, and we write $Q^i$ for his order size that will be strategically chosen later. Note that he may not send orders at a price jump time if he considers such action would not be profitable.

\item One noise trader: he sends market orders in a zero-intelligence random fashion. We assume
that these trades follow a compound Poisson process with intensity $\lambda^u$. We denote by $\{Q^u_j: j\ge1\}$ the noise trader's order sizes which are independent and identically distributed integrable random variables. They are also supposed independent of the efficient price $P$. We denote by $f_{Q^u}$ the density of the $Q^u_j$ which is strictly positive and symmetric on $\mathbb{R}$ (a positive volume represents a buy order, while a negative volume
represents a sell order) and write $F_{Q^u}$ for their cumulative distribution function. Remark that $r = \frac{\lambda^i}{\lambda^i + \lambda^u}$ corresponds to the average proportion of price jumps compared to the total number of events happening on the market (efficient price jumps and trades by the noise trader). Recall that informed trades can occur only when there is a price jump. We will assume throughout the paper that $r>0$.
We denote by $Q$ the order size independently of the issuer of the order (noise or informed trader). 

\item Market makers: they receive the value of the price jump size $B$ (and the
efficient price $P(t)$) right after it happens. We assume that they are risk neutral. In practice, market makers are often high frequency traders and considered informed too. However, contrary to our notion of informed trader, their analyses typically rely on order flows (notably through spread and imbalance) to extract the efficient price rather than on external information. This is because directional trading is not at the core of market making algorithms. We consider like in \cite{glosten1985bid} that market makers know the proportion of price jumps compared to the total number of events happening on the market, that they compete with each other, and that they are free to modify their limit orders at any time after a price jump or a transaction. Market makers place their orders according to their potential profit and loss with respect to the efficient price (no inventory aspects are considered here). Thus, they only send sell orders at price levels above the efficient price and buy orders at price levels below it.

\end{itemize}

We assume here that there is no tick size (this assumption will be relaxed in Section \ref{discrete}). The LOB is made of
limit orders placed by market makers around the efficient price $P(t)$. We denote the cumulative available liquidity between $P(t)$ and $P(t)+x$ by $L(x)$\footnote{This quantity actually depends on time $t$ but for sake of simplicity, we just write $L(x)$.} for $x \in \mathbb{R}$. When $L(x)\ge0$ (resp. $L(x)\le0$), it represents the total volume of sell (resp. buy) limit orders with price smaller (resp. larger) than or equal to $P(t)+x$. This function $L$ is called cumulative LOB shape function.

\subsection{Assumptions}

We suppose the cumulative LOB shape function $L$ is right-continuous and non-decreasing on $[0, \infty)$, and, symmetrically, it is left-continuous and non-decreasing on $(-\infty, 0]$. We define its inverse $L^{-1}$ by:
$$L^{-1}(q) = {\mathrm{argmin}}\{x|L(x) \ge q\},\quad q \in [0, \infty).$$
Most of the analysis will be carried on $q \geqslant 0$ (ask side) and the results concerning $q \leqslant 0$ (bid side) follow by symmetry.

Given the function $L$, we now specify the behavior of the informed trader in the next
assumption. This assumption links the traded volume of the informed trader $Q^i$ to the LOB
cumulative shape $L$ and the size of the price jump $B$ received by the informed trader.
\begin{assumption} \label{informed assump}
Let $t$ be a jump time of the efficient price. Based on the received value $B$ and the cumulative LOB shape function $L$
provided by market makers, the informed trader sends his trades in a greedy way such that he wipes
out all the available liquidity in the LOB until level $P(t)+B$. Thus, his trade size $Q^i$ satisfies: 
$$Q^i=L(B-)\text{ if }B > 0,\ Q^i=L(B+)\text{ if }B < 0.$$
\end{assumption}

The informed trader computes his gain according to the future efficient price. If he knows that the price will increase (resp. decrease), which corresponds to a strictly positive (resp. strictly negative) jump $B$, he consumes all the sell (resp. buy) orders leading to positive ex-post profit. In both cases, his profit is equal to the absolute value of the difference between the future efficient price and the price per share at which he bought or sold, multiplied by the consumed quantity. Note that in the spirit of this work, the informed trader does not accumulate position intraday. What we have in mind is that he unwinds his position passively, or alternates between buy and sell orders. As an illustration, if at a given moment the efficient price is equal to 10 euros and the future price jump is equal to 0.05 euros, the informed trader consumes all the sell orders at prices between 10 and 10.05 euros. He then can potentially unwind his position by submitting passive sell orders at a price equal to or higher than the new efficient price. Knowing that their latent profit is computed with respect to the efficient price, he can afford to submit them close to the new efficient price, thereby making their execution very likely.

\begin{remark}\label{remarque2}
    For a given order of size $Q^i > 0$ initiated by the informed trader and for a given quantity $q$, the probability that the trade size $Q^i$ is less than $q$ satisfies: 
    \begin{align*}
        \mathbb{P}[Q^i<q] &= \mathbb{P}[L(B-) < q]\\
        &= \mathbb{P}[B < L^{-1}(q)] \\
        &=  F_B(L^{-1}(q)).
    \end{align*}
\end{remark}

In the following, our goal is to compute the spread and LOB shape. We proceed in two steps. First, we derive the expected gain of potential limit orders of the market makers. Second, we consider a zero-profit assumption for market makers (due to competition). Based on these two ingredients, we show how the spread and LOB shape emerge.

\subsection{Computation of the market makers expected gain}\label{mmgain}
This part is the first step of our approach. We focus here on the gain of passive sell orders. The gain of passive buy orders can be readily deduced the same way.

Let $L$ be the shape of the order book. Our goal is to compute the conditional average profit of a new infinitesimal order if submitted at price level $x$  knowing that $Q>L(x)$ and without any information about the trade's initiator. We write $G(x)$ for this quantity\footnote{Note that the gain depends on time $t$ but we keep the notation $G(x)$ when no confusion is possible.}.

We consider the profit of new orders with total volume $\varepsilon>0$, placed between $P(t)+x - \delta p$ and $P(t)+x$ for some $x>0$ and $\delta p > 0$, given the fact that these orders are totally executed. The new submitted orders are represented by an additional cumulative LOB shape function denoted by $\tilde L(x)$. Note that we work with orders submitted between $x - \delta p$ and $x$ to take into account two cases: $L(x)$ is continuous at $x$ and $L(x)$ has a mass at $x$. The function $\tilde L(x)$ is defined as follows: 
\begin{itemize}
\item For $s < x - \delta p, \tilde L(s) = 0$ and the liquidity available in the LOB up to $s$ is equal to $L(s)$. 

\item For $x - \delta p \le s \le x$, the available liquidity is $L(s) + \tilde L(s)$, where $\tilde L(x - \delta p) = 0 $ and $\tilde L(x ) = \varepsilon$. 

\item For $s \ge x$, the liquidity available in the LOB up to $s$ is equal to $L(s) + \varepsilon $. 
\end{itemize}

\noindent Furthermore, we assume that for any $s<x$, $\tilde L(s)<\varepsilon$. Let us write:
\begin{itemize}
    \item $\nu$ for a random variable that is equal to 1 if the trade is initiated by the informed trader and 0 if it is initiated by the noise trader. 
    \item $G^u(x -\delta p, x)$ for the gain of new orders with total volume $\varepsilon$ submitted between $x - \delta p $ and $x$ in case the trade is initiated by the noise trader knowing that $Q^u \ge L(x)+\tilde L(x)$.
    
    \item $G^i(x- \delta p, x)$ for the gain of new orders with total volume $\varepsilon$ submitted between $x - \delta p $ and $x$  in case the trade is initiated by the informed trader knowing that $Q^i \ge L(x) + \tilde L(x)$.
    \item $G(x-\delta p, x)$ for the expected conditional gain of new orders with total volume $\varepsilon$ submitted between $x - \delta p $ and $x$  knowing that $Q \ge L(x) + \tilde L(x)$ without any information about the trade's initiator.
\end{itemize}

The quantity $G(x - \delta p, x)$ is equal to: 
$$ G^i(x - \delta p,x) \mathbb{P}[\nu = 1|Q \ge L(x) + \tilde{L}(x)]+ G^u(x - \delta p, x) \mathbb{P}[\nu =0|Q \ge L(x) + \tilde{L}(x)].$$

Our aim being to compute the expected gain of a new infinitesimal order if submitted at price level $x$, we make $\delta p$ and $\varepsilon$ tend to 0. Thus, we define $$G(x) = \underset{\varepsilon \rightarrow 0}{\text{lim}} \left(\underset{\delta p \rightarrow 0}{\text{lim}}\frac{G(x-\delta p, x)}{\varepsilon}\right).$$ 
We have the following proposition proved in Appendix \ref{appendix1}. It is a natural tail condition, similar to that in \cite{baruch2017tail, glosten1994electronic,sandaas2001adverse} but with a different parametrization.

\begin{proposition}\label{gain} 
For $x\geq 0$, the average profit of a new infinitesimal order if submitted at price level $x$ satisfies: 
$$G(x) = x-\frac{r \mathbb{E}[B1_{B>x}]}{r\mathbb{P}[B > x] + (1 - r)\mathbb{P}[Q^u > L(x)]}
$$
and for  $x \le 0$
$$
G(x) = -x+\frac{r \mathbb{E}[B1_{B<x}]}{r\mathbb{P}[B < x] + (1 - r)\mathbb{P}[Q^u < L(x)]}.
$$
\end{proposition}
Remark that the average profit $G(x)$ above is well defined even when $L(x) = 0$. In fact, when $L(x)=0$, $G(x)$ represents the expected gain of an infinitesimal order submitted in an empty order book at $x$. Note that for a given $x$, when $L(x)$ goes large, the expected gain of the limit orders becomes negative.

We now describe the way the LOB is built via a zero-profit type condition. Let us take the ask side of the LOB. For any point $x$, market makers first consider whether or not there should be liquidity between 0 and $x$. To do so, they compute the value $\hat{L}(x)$ which is so that we obtain $G(x)=0$ in the expression in Proposition \ref{gain}. If $\hat{L}(x)$ is positive, then competition between market makers takes place and the cumulative order book adjusts so that $L(x) = \hat{L}(x)$ in order to obtain $G(x) = 0$. If $\hat{L}(x)=0$, then there is no liquidity between 0 and $x$. If $\hat{L}(x)$ is negative, we deduce that there is no liquidity between 0 and $x$ since this liquidity should be positive. This mechanism makes sense since, as we will see in what follows, $\hat{L}(x)$ is a non-decreasing function of $x$, which implies two things. First, it is impossible to come across a situation where $x_1<x_2$ and where market makers are supposed to add liquidity between $P(t)$ and $P(t)+x_1$ but not between $P(t)$ and $P(t) + x_2$. Second, the cumulative shape function for the LOB is indeed non-decreasing.

We have that $G(x)=0$ is equivalent to:
$$
x  = \left\{
    \begin{array}{ll}
        \frac{r \mathbb{E}[B\mathds{1}_{B>x}]}{r\mathbb{P}[B > x] + (1 - r)\mathbb{P}[Q^u > \hat{L}(x)]}  & \mbox{if $x \ge 0$} \\
       \frac{r \mathbb{E}[B\mathds{1}_{B<x}]}{r\mathbb{P}[B < x] + (1 - r)\mathbb{P}[Q^u < \hat{L}(x)]} & \mbox{if $x \le 0$.}
    \end{array}
\right.
$$
This implies:
$$
\hat{L}(x)  = \left\{
    \begin{array}{ll}
        F_{Q^u}^{-1}\left(\frac{1}{1-r}-\frac{r}{1-r}\mathbb{E}[\text{max}(\frac{B}{x},1)]\right)  & \mbox{if $x \ge 0$} \\
      F_{Q^u}^{-1}\left(\frac{-r}{1-r}+\frac{r}{1-r}\mathbb{E}[\text{max}(\frac{B}{x},1)]\right)  & \mbox{if $x \le 0$.}
    \end{array}
\right.
$$

The details of the computation of $\hat{L}(x)$ are given in Appendix \ref{proof}.

We formalize now the zero-profit assumption introduced above. It is the second step of our approach in order to eventually compute the spread and LOB shape. 

\begin{assumption}\label{zeroprofit}
For every $x>0$ (resp. $x<0$), market makers compute $\hat{L}(x)$. If $\hat{L}(x)\leq 0$ (resp. $\hat{L}(x)\geq 0$), market makers add no liquidity to the LOB: $L(x)=0$. If $\hat{L}(x) >0$ (resp. $\hat{L} (x) <0$), because of competition, the cumulative order book adjusts so that $G(x) = 0$. We then obtain $ L(x) = \hat{L}(x)$.
\end{assumption}

The above zero-profit assumption can be seen as a generalized version of the zero-profit condition proposed in \cite{glosten1985bid}, in which zero-profit is only considered
for the two best offer limits. It is also interesting to point out that, under this more realistic setting, those very fast market makers can still make profit as their orders are placed earlier in the
LOB. \\

In this case where the tick size is zero, it can seem difficult to imagine how competition between different
market makers takes place. One can think that every market maker specifies his own $L(x)$
(cumulative liquidity that he provides). Then Assumption \ref{zeroprofit} means that, when
there is still room for future profit at $x$ $(G(x) > 0)$, other market makers will come to the market
and increase the liquidity in the LOB until $G(x)$ becomes null. Note again that we consider here that market makers can insert infinitesimal quantities in the LOB. These ideas will be made clearer in Section \ref{discrete} where the tick size is no longer zero.

\subsection{The emergence of the bid-ask spread and LOB shape }
Based on the expected gain of the market makers, see Proposition \ref{gain}, and the zero-profit condition (Assumption \ref{zeroprofit}), we can derive the bid-ask spread and LOB shape. We have the following theorem proved in Appendix \ref{proof}.



 
\begin{theorem}\label{spread_theorem}
The cumulative LOB shape satisfies $L(x)=-L(-x)$ for any $x\in\mathbb{R}$, $L(x)=0$ for $x \in [-\mu,\mu]$ and $L$ is continuous strictly increasing for $x > \mu$, where $\mu$ is the unique solution of the following equation:
\begin{equation}\label{equation}
\frac{1+r}{2r}=\E[][\max\left(\frac{B}{\mu},1\right)].
\end{equation} 

For $x > \mu$, $L(x) > 0$ and
\begin{equation}\label{shape1}
 L(x)=F_{Q^u}^{-1}\left(\frac{1}{1-r}-\frac{r}{1-r}\E[][\max\left(\frac{B}{x},1\right)]\right). 
\end{equation}

For $x < -\mu$, $L(x) < 0$ and
\begin{equation}\label{shape2}
L(x)=F_{Q^u}^{-1}\left(\frac{-r}{1-r}+\frac{r}{1-r}\E[][\max\left(\frac{B}{x},1\right)]\right).
\end{equation}

In particular, the bid-ask spread is equal to $2\mu$.
\end{theorem}

Equation \eqref{equation} shows that the spread is an increasing function of $r$. This means that market makers are aware of the adverse selection they risk when the number of price jumps increases. As a consequence, they enlarge the spread in order to avoid this effect due to the trades issued by the informed trader just before the price jumps take place. In particular, if there is no noise trader in the market, then $r=1$ and the spread tends to infinity. On the contrary, when the number of trades from the noise trader increases, market makers reduce the spread because they are less subject to adverse selection. All these results are consistent with the findings in \cite{glosten1985bid}.

Equations \eqref{shape1} and \eqref{shape2} show that the liquidity submitted by the market makers is a decreasing function of $r$. Let us take $x > \mu$ and define $h(r)= \frac{1}{1-r}-\frac{r}{1-r}\mathbb{E}[\text{max}(\frac{B}{x},1)]$. We have 
$$\frac{\partial h}{\partial r}(r) = \frac{1 - \mathbb{E}[\text{max}(\frac{B}{x}, 1)]}{(1-r)^2}  \le 0. $$
This means that $h$ is a decreasing function of $r$. The function $F_{Q^u}^{-1}$ being increasing, we deduce that $L(x)$ is a decreasing function of $r$. When the number of price jumps increases, market makers reduce the quantity of submitted passive orders. In contrast, when the number of trades from the noise trader is large, the market becomes very liquid. This is in line with the empirical results in \cite{megarbane2017behavior} where it is shown that just before certain announcements, in order to avoid adverse selection, market makers reduce their depth and increase their spread.

Finally, we recall that in our setting, we do not \textit{a priori} impose any condition on $L(x)$. Equations \eqref{equation}, \eqref{shape1} and \eqref{shape2} show that the cumulative LOB we obtain is continuous and strictly increasing beyond the spread. Remark also that $L(x)$ tends to infinity as $x$ goes to infinity. This implies that the noise trader can always find liquidity in the LOB, whatever the size of his market order. If the price jumps due to information are bounded, there is infinite liquidity in the order book at prices above that bound: these orders guarantee profit for the market-makers.

Proposition \ref{prop:any_shape_attainable}, proved in Appendix \ref{subsec:proof_any_shape_attainable}, shows that any (up to not very restrictive regularity conditions) LOB shape is attainable in our model, due to the freedom given by the very general shape the density $f_{Q^u}$ is allowed to take. Restricting $f_{Q^u}$ to a low-dimensional parametric family may make the model fail to capture the shape of the order book far from the mid-price, as we see in section \ref{subsubsec:gaussian_noise} (see also \cite{sandaas2001adverse}).

\begin{proposition}
    \label{prop:any_shape_attainable}
    Assume $f_B$ is strictly positive on $\mathbb{R}$. Let $\mu$ be the associated half-spread given by Equation \eqref{equation}. Let $\Lambda:[\mu, \infty) \to [0, \infty)$ be a bijective continuous function such that $\Lambda^{-1}$ is differentiable. Then, there exists a strictly positive symmetric density $f_{Q^u}$ on $\mathbb{R}$ such that for all $x \geqslant \mu$,
    \begin{equation*}
        \Lambda(x) = F_{Q^u}^{-1}\left(\frac{1}{1-r}-\frac{r}{1-r}\E[][\max\left(\frac{B}{x},1\right)]\right),
    \end{equation*}
    where $F_{Q^u}$ denotes the cumulative distribution function associated to $f_{Q^u}$.
\end{proposition}

Proposition \ref{prop:shape_tail} below, proved in Appendix \ref{sec:proof_shape_tail}, computes the shape of the LOB far from the best quotes in the case $1- F_B$ has power-law decay, which is the case of Pareto distributions and Lévy stables laws, see \cite[Theorem 1.2]{nolan2020univariate} which are the ones used in our empirical study in Sections \ref{sec:estimation} and \ref{sec:applications}.
\begin{proposition}
    \label{prop:shape_tail}
    Suppose that the tail probability of $B$ verifies $1-F_B(x) \underset{x\to \infty}{\sim} c x^{-a}$ for some $c > 0$, $a > 1$.\\
    (i) If $1-F_{Q^u}(x) \underset{x\to \infty}{\sim} c'x^{-b}$ for some $c', b > 0$, then
    \begin{equation*}
        L(x)  \underset{x\to \infty}{\sim} \left(\frac{1-r}{r}\frac{c'(a-1)}{c}\right)^{\frac{1}{b}}x^{\frac{a}{b}}.
    \end{equation*}
    (ii) If $Q^u$ follows a centered Gaussian law with variance $\sigma^2$, then  $1-F_{Q^u}(x) \underset{x\to \infty}{\sim} \frac{1}{x \sqrt{2\pi}\sigma}e^{-\frac{x^2}{2 \sigma^2}}$ and
    \begin{equation*}
        L(x) \underset{x\to \infty}{\sim} \sigma \sqrt{2 a \ln(x)}.
    \end{equation*}
\end{proposition}
When the sizes of the trades sent by noise traders follow a power-law distribution, the number of orders present at prices $[x,x+\diff x]$ is proportional to $x^{\frac{a}{b}-1}\diff x$. If $Q^u$'s law has a fatter tail than $B$'s, the liquidity present far for the best quotes tends to increase since the market makers bet on big trades sent by noise traders. On the contrary, if $Q^u$'s law has a thinner tail than $B$'s, fewer orders are posted far from the best quotes since the adverse selection risk dominates--recall that the market makers know the distribution of $Q^u$ and quote accordingly. If the trades sent by noise traders have sizes that follow a normal distribution, liquidity decreases rapidly, in $\frac{1}{x \sqrt{\ln(x)}}$, as $x$ increases because big trades are very rare.

\section{The case of non-zero tick size}\label{discrete}

In this section, we study the effect of introducing a tick size, denoted by $\alpha$, that constraints
the price levels in the LOB. The same efficient price dynamic as that described in the previous section still
applies, but the cumulative LOB shape becomes now a piecewise constant function. Due to
price discreteness, the discontinuity points of $L(x)$ will depend on the position
of the efficient price $P(t)$ with respect to the tick grid. 

\subsection{Notations and assumptions}

\paragraph{Notations}

To deal with the discontinuity points of $L(x)$, the following notations will
be used in the sequel. Let us denote by $\tilde{P}(t)$ the smallest admissible price level that is greater than or equal
to the current efficient price $P(t)$, and their distance by $d(t) := \tilde{P}(t)-P(t)$, where $d(t) \in  [0,\alpha)$. We drop the dependence in $t$ and simply write $d$ when no confusion is possible. The cumulative LOB shape function $L(x)$ is now defined by $L^d(i)$:

\begin{equation}\label{eq_def_descrete}
L^d(i) = \left\{
    \begin{array}{ll}
        L(d+(i-1)\alpha) & \mbox{for $i>0$} \\
        L(d+i\alpha) & \mbox{for $i<0.$}
    \end{array}
\right.
\end{equation}

The index $i=1$ (resp. $i=-1$) corresponds to the closest price level that is larger (resp. smaller) than or equal to $P(t)$. When $L^d(i) > 0$ (resp. $L^d(i) <0$), it represents the total volume of sell (resp. buy) passive orders with prices smaller (resp. larger) than or equal to the $i^{th}$ limit. \\

\noindent We write $l^d(i)$ for the quantity placed at the $i^{th}$ limit:

$$
l^d(i) = \left\{
    \begin{array}{ll}
       L^d(i)-L^d(i-1) & \mbox{for $i>0$} \\
        L^d(i)-L^d(i+1) & \mbox{for $i<0$.}
    \end{array}
\right.
$$

When $l^d(i) > 0$ (resp. $l^d(i) <0$), it represents the volume of sell (resp. buy) limit orders placed at the $i^{th}$ limit. Recall that $l^d(i) \geq 0$ (resp. $l^d(i) \leq 0$) for $i>0$ (resp. $i<0$).

\paragraph{Assumptions} We adapt Assumption \ref{informed assump} to our tick size setting. We again assume that when he receives new information, the informed trader sends his trades in a greedy way such that he wipes out all the available liquidity at limits where the price is smaller than the new efficient price. This can be translated as follows.

\begin{assumption}\label{informed assump tick}
When the informed trader sends a market order, then $Q^i$ is equal to $L^d(i)$ for some $i\in\mathbb{Z}^*$. We have
$Q^i=L^d(i)$ if and only if $B \in [d+(i-1)\alpha,d+i\alpha)$.
\end{assumption}
\begin{remark}
In practice, it is rare that a trade consumes more than one limit in the LOB. Such trade in our model should be interpreted in practice as a sequence of transactions, each of them consuming one limit.
\end{remark}


Proposition \ref{prop:invariant_fractional_part}, proved in Appendix \ref{subsec:proof_invariant_frac_part}, will be used to compute various quantities of interest, putting a uniform prior on $d$.

\begin{proposition}
    \label{prop:invariant_fractional_part}
    The process $(d(t))_{t \geqslant 0}$ is a time-homogenous Markov process. Its unique invariant probability is the uniform distribution on $(0,\alpha)$.
\end{proposition}

\subsection{Computation of the market makers expected gain }\label{section gain}

As in the previous section, let us compute the conditional average profit of a new infinitesimal passive order submitted at the $i^{th}$ limit, knowing that $Q > L^d(i)$, and without any information about the trade's initiator. This quantity is denoted by $G^d(i)$ and defined in a similar fashion as $G(x)$ in Section \ref{mmgain}. The computation of $G^d(i)$ is comparable to that of $G(x)$, and actually even easier since we now have that the volume at the $i^{th}$ limit cannot be infinitesimal. This means that different orders can be submitted at the same price with disparities in their gain according to their position in the queue.  For instance, the order placed on top of the queue has the highest expected gain, while we will impose later that the gain of a new order submitted at the rear of the queue is null. We have the following proposition proved in Appendix \ref{proofproptick}.

\begin{proposition}\label{gain tick}
Under Assumption \ref{informed assump tick}, for $i \in \mathbb{Z}^*$, the expected gain of a new infinitesimal passive order placed at the $i^{th}$ level, given that it is executed, satisfies:
$$G^d(i)  = G(d+(i-1)\alpha)=d+(i-1)\alpha - \frac{r\mathbb{E}[B1_{B > d+(i-1)\alpha}]}{r\mathbb{P}[B > d+(i-1)\alpha] + (1 -r) \mathbb{P}[Q^u > L^d(i)]}$$
for  $i > 0$ and
$$
G^d(i)  =      G(d+i\alpha)=d+i\alpha - \frac{r\mathbb{E}[B1_{B < d+i\alpha}]}{r\mathbb{P}[B < d+i\alpha] + (1 -r) \mathbb{P}[Q^u <L^d(i)]}$$
for $i < 0$.
\end{proposition}

The quantity $G^d (i)$ can be understood as the expected gain of a newly inserted infinitesimal limit order at
the $i^{th}$ limit, under the condition that it is executed against some market order. For this situation with non-zero tick size, we follow the same reasoning as in the case with zero tick size. Indeed, for all $i \in \mathbb{Z}^*$, market makers compute $\hat{L}^d(i)$ so that $G^d(i)=0$ in Proposition \ref{gain tick}. The equality $G^d(i)=0$ is equivalent to:
$$d+(i-1)\alpha = \frac{r\mathbb{E}[B1_{B > d+(i-1)\alpha}]}{r\mathbb{P}[B > d+(i-1)\alpha] + (1 -r) \mathbb{P}[Q^u > \hat{L}^d(i)]}$$
if $i > 0$ and
$$d+i\alpha = \frac{r\mathbb{E}[B1_{B < d+i\alpha}]}{r\mathbb{P}[B < d+i\alpha] + (1 -r) \mathbb{P}[Q^u < \hat{L}^d(i)]}$$
if $i < 0$.

This is equivalent to: 
$$
\hat{L}^d(i)  = \left\{
    \begin{array}{ll}
        F_{Q^u}^{-1}\left(\frac{1}{1-r}-\frac{r}{1-r}\mathbb{E}[\text{max}(\frac{B}{d + (i-1)\alpha},1)]\right)  & \mbox{if $i > 0$} \\
      F_{Q^u}^{-1}\left(\frac{-r}{1-r}+\frac{r}{1-r}\mathbb{E}[\text{max}(\frac{B}{d + i\alpha},1)]\right)  & \mbox{if $i \le 0$.}
    \end{array}
\right.
$$
As in the case without tick size, this leads to the following zero-profit assumption.

\begin{assumption} \label{zeroprofit tick}
For every $i \in \mathbb{Z}^+$ (resp. $i \in \mathbb{Z}^-$), market makers compute $\hat{L}^d(i)$. If $\hat{L}^d(i)\leq 0$ (resp. $\hat{L}^d(i)\geq 0$), market makers add no liquidity to the LOB: $L^d(i) = 0$. If $\hat{L}^d(i)>0$ (resp. $\hat{L}^d(i)<0$), because of competition, the cumulative order book adjusts so that $G^d(i)=0$. We then obtain then $L^d(i) = \hat{L}^d(i)$.  
\end{assumption}

The zero-profit condition applies only to a new order submitted at the bottom of the queue. The expected profit of the other orders is non-zero, maximum gain being obtained for the one on top of the queue.

\subsection{Bid-ask spread and LOB formation}

Based on the expected gain of the market makers, see Proposition \ref{gain tick}, and the zero-profit condition (Assumption \ref{zeroprofit tick}), as previously, we deduce the bid-ask spread and LOB shape. We have the following theorem proved in Appendix \ref{slt}.

\begin{theorem}\label{spread_lob_tick}

The LOB shape function satisfies $l^d(i)=0$ for all $-k_l^d<i<k_r^d$, where $k_l^d$ and $k_r^d$ are two positive integers
determined by the following equations:
$$k_r^d=1+ \ceil*{\frac{\mu-d}{\alpha}},~~k_l^d=\ceil*{\frac{\mu+d}{\alpha}},$$
with $\mu$ defined by \eqref{equation}, and where $ \lceil x \rceil $ denotes the smallest integer that is larger than $x$ (which can be equal to 0). Furthermore, for $i \ge k_r^d:$
$$L^d(i) = F_{Q^u}^{-1}\left(\frac{1}{1-r}-\frac{r}{1-r}\E[][\max\left(\frac{B}{d+(i-1)\alpha},1\right)]\right)$$
and for $i \le -k_l^d:$
$$L^d(i) = F_{Q^u}^{-1}\left(\frac{-r}{1-r}+\frac{r}{1-r}\E[][\max\left(\frac{B}{d + i\alpha},1\right)]\right).$$

For given $d$, the bid-ask spread $\phi_{\alpha}^d$ satisfies:
$$\phi_{\alpha}^d=\alpha\left(\ceil*{\frac{\mu - d}{\alpha}}+ \ceil*{\frac{\mu + d}{\alpha}}\right).$$
\end{theorem}

Let us consider the approximation that $d$ follows its stationary distribution and is uniformly distributed on $[0,\alpha]$  (Proposition \ref{prop:invariant_fractional_part}). In this case, we obtain the following corollary proved in Appendix \ref{pcs}.

\begin{corollary} \label{cor_spread}
If $d$ follows a uniform distribution on $[0, \alpha)$, the average spread $\phi_{\alpha}$ satisfies:
\begin{equation}\label{spread}
  \phi_{\alpha}=2\mu+\alpha.  
\end{equation}
\end{corollary}

When the tick size is vanishing, we have seen in Theorem \ref{spread_theorem} that the spread is equal to $2\mu$. When it is not, the spread cannot necessarily be equal to $2\mu$ because of the tick size constraint. What is particularly interesting is that even if $\alpha\leq 2\mu$, the equilibrium spread is not $2\mu$. There is always a tick size processing cost leading to a spread value of $2\mu+\alpha$.

\section{Parameters estimation}
\label{sec:estimation}

This section is dedicated to the calibration of the model on market data. The estimation procedure is carried over 103 stocks traded on the New-York Stock Exchange (NYSE) and European markets operated by Euronext. The stocks are listed in Appendix \ref{appendix:stocks}. The prices are not normalized, and expressed in their respective currencies.

For each stock, the estimation is done on two periods: on trading days between October \nth{1}, 2022 and March \nth{31} 2023 and between April \nth{1} 2023 and September \nth{30} 2023, with LOB data between 10 am and 3 pm in local time.

Our estimation procedure requires a constant tick size on the whole estimation period. It is the case for stocks traded on NYSE where the tick is always of $\$0.01$. For stocks traded on European exchanges, days with non-constant tick size (at the best quotes) are discarded. We keep the days having the most frequent tick size in the trading period.

The size of the sample $(N = 36000$ for each stock with 120 trading days observed) yield an accurate and stable estimation.

\subsection{Estimation of the efficient price dynamics}
\label{subsec:estimation_efficient_price}

The parameter $\mu$ can be estimated in a straightforward way using Corollary \ref{cor_spread}: $\hat{\mu}= \frac{\bar{\phi}_{\alpha} - \alpha}{2}$ where $\bar{\phi}_{\alpha}$ is the average spread over the observation period. Having an estimate of the law of the price jumps $B$, one can estimate $r$ using Equation \eqref{equation}:
\begin{equation*}
    \hat{r} = \frac{1}{2\E[][\max\left(\frac{B}{\hat{\mu}},1\right)-1]},
\end{equation*}
where the expectation is computed using the estimate of the law of $B$.

To estimate $f_B$, we observe the reference price $\tilde{P}(t)$ every $\Delta t = 1 \text{ min}$. We do not observe the efficient price directly, just its rounded version. We do not use a continuous observation of $\tilde{P}(t)$ to avoid short-term effects not captured by our model: after a trade of a noise trader, there is a non-zero delay for the LOB to return to the theoretical state. Also, the new price information is not processed immediately.

The quantity $\tilde{P}(t)$ itself is not immediately observable, but it can be recovered using the formulas in Theorem \ref{spread_lob_tick}. Precisely, write $\mu = \alpha m + p$ where $m \in \mathbb{N}$ and $p \in [0, \alpha)$. Denote by $V^a$ and $V^b$ the pending volumes at the best ask and bid respectively. Let $P^{mid}$ be the mid-price in the conventional sense: the average between the best ask price $P^a$ and the best bid price $P^b$. $P^{mid}$ is directly observable. According to Tables \ref{tab:P_tilde_1} and \ref{tab:P_tilde_2}, if the spread is odd, $\tilde{P}(t) = P^{mid} + \frac{\alpha}{2}$. If it is even and $V^a > V^b$, then $\tilde{P}(t) = P^{mid}$, otherwise, $\tilde{P}(t) = P^{mid}+\alpha$. Figure \ref{fig:LOB_illustration} provides an illustration of this observation.

\begin{table}
    \centering
\begin{tabular}{c|c|c|c}
    $d$ & $(0, p)$ & $(p, \alpha-p)$ & $(\alpha - p, \alpha)$ \\
    \hline \hline
    $P^a$ & $\tilde{P}+(m+1)\alpha$ & $\tilde{P}+m\alpha$ & $\tilde{P}+m\alpha$ \\
    \hline
    $P^b$ & $\tilde{P}-(m+1)\alpha$ & $\tilde{P}-(m+1)\alpha$  & $\tilde{P}-(m+2)\alpha$\\
    \hline
    Spread     & $2(m+1)\alpha$  & $(2m+1)\alpha$ & $2(m+1)\alpha$ \\
    \hline
    $\tilde{P}$ & $P^{mid}$ & $P^{mid}+\frac{\alpha}{2}$  & $P^{mid}+\alpha$ \\
    \hline
    Volumes  &  $V^a > V^b$  &     X        & $V^b > V^a$
    \end{tabular}
    \caption{Quantities of interest at the best prices when $p \leqslant \frac{\alpha}{2}$ ($\mu = \alpha m + p$, $m \in \mathbb{N}$, $p \in [0, \alpha)$).}\label{tab:P_tilde_1}
\end{table}
\begin{table}
    \centering
    \begin{tabular}{c|c|c|c}
        $d$ & $(0, \alpha - p)$ & $(\alpha-p, p)$ & $(p, \alpha)$ \\
        \hline \hline
        $P^a$ & $\tilde{P}+(m+1)\alpha$ & $\tilde{P}+(m+1)\alpha$ & $\tilde{P}+m\alpha$ \\
        \hline
        $P^b$ & $\tilde{P}-(m+1)\alpha$ & $\tilde{P}-(m+2)\alpha$  & $\tilde{P}-(m+2)\alpha$\\
        \hline
        Spread     & $2(m+1)\alpha$  & $(2m+3)\alpha$ & $2(m+1)\alpha$ \\
        \hline
        $\tilde{P}$ & $P^{mid}$ & $P^{mid}+\frac{\alpha}{2}$  & $P^{mid}+\alpha$ \\
        \hline
        Volumes  &  $V^a > V^b$  &     X        & $V^b > V^a$
        \end{tabular}
    \caption{Quantities of interest at the best prices when $p \geqslant \frac{\alpha}{2}$ ($\mu = \alpha m + p$, $m \in \mathbb{N}$, $p \in [0, \alpha)$).}\label{tab:P_tilde_2}
\end{table}

\begin{figure}
    \centering
    \begin{tikzpicture}
        \def\P{7}
        \def\p{0.3*4}
        \pgfmathsetmacro\tildeP{4*ceil(\P / 4)}
        \pgfmathsetmacro\d{\tildeP - \P}
    
        \draw[fill=red!50] (\tildeP+4-0.25,0) rectangle (\tildeP+4+0.25,2);
        \draw[fill=blue!50] (\tildeP-4-0.25,0) rectangle (\tildeP-4+0.25,1);
    
        \draw[->] (0,0) -- (13,0) node[right] {\(P\)};
    
        \draw[thick, blue] (\P,-0.25) -- (\P,0.25);
        \draw[thick, red] (\tildeP,-0.25) -- (\tildeP,0.25);
        \draw[thick, red] (\tildeP-4,-0.25) -- (\tildeP-4,0.25);
        \draw[thick, red] (\tildeP-8,-0.25) -- (\tildeP-8,0.25);
        \draw[thick, red] (\tildeP+4,-0.25) -- (\tildeP+4,0.25);
        \draw[thick, green] (\tildeP-\p,-0.15) -- (\tildeP-\p,0.15);
        \draw[thick, green] (\tildeP-4+\p,-0.15) -- (\tildeP-4+\p,0.15);
    
        \draw[<-, thick] (\P, 1.0) -- (\tildeP, 1.0) node[midway, above] {\(d\)};
        \draw[<-, thick] (\tildeP-\p, 0.5) -- (\tildeP, 0.5) node[midway, above] {\(p\)};
        \draw[->, thick] (\tildeP-4, 0.5) -- (\tildeP-4+\p, 0.5) node[midway, above] {\(p\)};
    
        \node[blue] at (\P, -0.5) {\(P\)};
        \node[red] at (\tildeP, -0.5) {\(\tilde{P}\)};
        \node[red] at (\tildeP-4, -0.5) {\(\tilde{P}-\alpha\)};
        \node[red] at (\tildeP-8, -0.5) {\(\tilde{P}-2\alpha\)};
        \node[red] at (\tildeP+4, -0.5) {\(\tilde{P}+\alpha\)};
    
    \end{tikzpicture}
    
    \begin{tikzpicture}
        \def\P{6}
        \def\p{0.3*4}
        \pgfmathsetmacro\tildeP{4*ceil(\P / 4)}
        \pgfmathsetmacro\d{\tildeP - \P}
    
        \draw[fill=red!50] (\tildeP-0.25,0) rectangle (\tildeP+0.25,2);
        \draw[fill=blue!50] (\tildeP-4-0.25,0) rectangle (\tildeP-4+0.25,2);
    
        \draw[->] (0,0) -- (13,0) node[right] {\(P\)};
    
        \foreach \x in {0,4,8,12} {
            \draw (\x,0.1) -- (\x,-0.1);
        }
    
        \draw[thick, blue] (\P,-0.25) -- (\P,0.25);
        \draw[thick, red] (\tildeP,-0.25) -- (\tildeP,0.25);
        \draw[thick, red] (\tildeP-4,-0.25) -- (\tildeP-4,0.25);
        \draw[thick, red] (\tildeP-8,-0.25) -- (\tildeP-8,0.25);
        \draw[thick, red] (\tildeP+4,-0.25) -- (\tildeP+4,0.25);
        \draw[thick, green] (\tildeP-\p,-0.15) -- (\tildeP-\p,0.15);
        \draw[thick, green] (\tildeP-4+\p,-0.15) -- (\tildeP-4+\p,0.15);
    
        \draw[<-, thick] (\P, 1.0) -- (\tildeP, 1.0) node[midway, above] {\(d\)};
        \draw[<-, thick] (\tildeP-\p, 0.5) -- (\tildeP, 0.5) node[midway, above] {\(p\)};
        \draw[->, thick] (\tildeP-4, 0.5) -- (\tildeP-4+\p, 0.5) node[midway, above] {\(p\)};
    
        \node[blue] at (\P, -0.5) {\(P\)};
        \node[red] at (\tildeP, -0.5) {\(\tilde{P}\)};
        \node[red] at (\tildeP-4, -0.5) {\(\tilde{P}-\alpha\)};
        \node[red] at (\tildeP-8, -0.5) {\(\tilde{P}-2\alpha\)};
        \node[red] at (\tildeP+4, -0.5) {\(\tilde{P}+\alpha\)};
    
    \end{tikzpicture}

    \begin{tikzpicture}
        \def\P{5}
        \def\p{0.3*4}
        \pgfmathsetmacro\tildeP{4*ceil(\P / 4)}
        \pgfmathsetmacro\d{\tildeP - \P}
    
        \draw[fill=red!50] (\tildeP-0.25,0) rectangle (\tildeP+0.25,1);
        \draw[fill=blue!50] (\tildeP-8-0.25,0) rectangle (\tildeP-8+0.25,2);
    
        \draw[->] (0,0) -- (13,0) node[right] {\(P\)};
    
        \foreach \x in {0,4,8,12} {
            \draw (\x,0.1) -- (\x,-0.1);
        }
    
        \draw[thick, blue] (\P,-0.25) -- (\P,0.25);
        \draw[thick, red] (\tildeP,-0.25) -- (\tildeP,0.25);
        \draw[thick, red] (\tildeP-4,-0.25) -- (\tildeP-4,0.25);
        \draw[thick, red] (\tildeP-8,-0.25) -- (\tildeP-8,0.25);
        \draw[thick, red] (\tildeP+4,-0.25) -- (\tildeP+4,0.25);
        \draw[thick, green] (\tildeP-\p,-0.15) -- (\tildeP-\p,0.15);
        \draw[thick, green] (\tildeP-4+\p,-0.15) -- (\tildeP-4+\p,0.15);
    
        \draw[<-, thick] (\P, 1.0) -- (\tildeP, 1.0) node[midway, above] {\(d\)};
        \draw[<-, thick] (\tildeP-\p, 0.5) -- (\tildeP, 0.5) node[midway, above] {\(p\)};
        \draw[->, thick] (\tildeP-4, 0.5) -- (\tildeP-4+\p, 0.5) node[midway, above] {\(p\)};
    
        \node[blue] at (\P, -0.5) {\(P\)};
        \node[red] at (\tildeP, -0.5) {\(\tilde{P}\)};
        \node[red] at (\tildeP-4, -0.5) {\(\tilde{P}-\alpha\)};
        \node[red] at (\tildeP-8, -0.5) {\(\tilde{P}-2\alpha\)};
        \node[red] at (\tildeP+4, -0.5) {\(\tilde{P}+\alpha\)};
    
    \end{tikzpicture}
    \caption{Illustration of the LOB, in the case $m=0$, $p \leqslant \frac{\alpha}{2}$ ($\mu = \alpha m + p$). In blue the first filled bid pile, in red the first filled ask pile. Upper: $d \in (0,p)$. Middle: $d \in (p - \alpha - p)$. Lower: $d \in (\alpha - p, \alpha)$.}
    \label{fig:LOB_illustration}
\end{figure}

$\tilde{P}(t)$ takes values in $\alpha \mathbb{Z}$. The probability of observing a variation of $k\alpha$ on a time interval on length $\Delta t$ is given by Proposition \ref{prop:likelihood}, proved in Appendix \ref{sec:proof_formula_likelihood}. It can be computed using the characteristic function of $B$, taking advantage of the compound Poisson model used for the efficient price jumps.

\begin{proposition}
    \label{prop:likelihood}
    Suppose $d(t)$ follows a uniform law on $[0, \alpha)$. Let $t \in [0, \infty)$ and $k \in \mathbb{Z}$. Then,
    \begin{equation*}
        \Proba[][\tilde{P}(t + \Delta t) - \tilde{P}(t)= k\alpha]
        = \frac{4}{\pi \alpha}\int_{0}^{\infty} e^{\lambda^i \Delta t \left(\hat{f}_{B}(z) -1\right)}\frac{1}{z^2}\sin\left(\frac{z \alpha}{2}\right)^2 \cos\left(k \alpha z\right) \diff z.
    \end{equation*}
\end{proposition}

Having a parametric model for $f_B$, we use a maximum likelihood approach to recover our parameters. Specifically, observing $k_1,\dots k_n$ for $\tilde{P}(t + \Delta t) - \tilde{P}(t)$, we minimize
\begin{equation*}
    \sum_{j = 1}^n \log \int_{0}^{\infty} e^{\lambda^i \Delta t \left(\hat{f}_{B}(z) -1\right)}\frac{1}{z^2}\sin\left(\frac{z \alpha}{2}\right)^2 \cos\left(k_j \alpha z\right) \diff z,
\end{equation*}
to recover $\lambda^i$ and the parameters of $f_B$.

We choose to look for $f_B$ in the family of symmetric Lévy-stable distributions, parametrized by $a \in (1, 2)$ and $\sigma > 0$. They have the following characteristic function:
\begin{equation*}
    \xi \mapsto \exp\left(-\xi^a \sigma^a \right).
\end{equation*}
They give two degrees of freedom: the scale $\sigma$ and the decay of the tail, in power law with exponent $a$ (\cite[Theorem 1.2]{nolan2020univariate}). The explicit form of its characteristic function allows to compute easily the integral in Proposition \ref{prop:likelihood}, using a quadrature method. By Proposition \ref{prop:identifiability} below, proved in Appendix \ref{sec:proof_identifiability}, it also satisfies an identifiability property: $(\lambda^i, a, \sigma)$ are uniquely determined by the law of $\tilde{P}(t + \Delta t) - \tilde{P}(t)$.

\begin{proposition}
    \label{prop:identifiability}
    Let $t \in [0, \infty)$. Let $\lambda_1, \lambda_2, \sigma_1, \sigma_2 > 0$ and $a_1, a_2 \in (0,2)$ .Suppose that for all $k \in \mathbb{N}$,
    \begin{equation*}
        \int_{0}^{\infty} e^{\lambda_1 \left(e^{-\sigma_1^{a_1} z^{a_1}} -1\right)}\frac{\sin\left(\frac{z \alpha}{2}\right)^2}{z^2} \cos\left(k \alpha z\right) \diff z
        = \int_{0}^{\infty} e^{\lambda_2 \left(e^{-\sigma_2^{a_2} z^{a_2}} -1\right)}\frac{\sin\left(\frac{z \alpha}{2}\right)^2}{z^2} \cos\left(k \alpha z\right) \diff z.
    \end{equation*}
    Then, $\lambda_1 = \lambda_2$, $\sigma_1 = \sigma_2$ and $a_1 = a_2$.
\end{proposition}

\begin{remark}
    Proposition \ref{prop:identifiability} is stated for $a \in (0,2)$ but the $a \leqslant 1$ are discarded in our model since they break the integrability assumption.
\end{remark}

Estimated values $(\hat{\lambda}, \hat{a}, \hat{\sigma})$ are recovered optimizing the log-likelihood using the CMA-ES algorithm \cite{hansen2016cma,nomura2024cmaes} and reported in Appendix \ref{appendix:estimation_results}. The majority of the stocks have an estimated $\hat{a}$ around 1.8.

\subsection{Estimation of $F_{Q^u}$}

\subsubsection{The case of a Gaussian distribution}
\label{subsubsec:gaussian_noise}

We first consider a Gaussian distribution $\mathcal{N}(0,\sigma_{noise}^2)$ for the trade sizes from the noise traders. For $y \geqslant \mu$, we introduce the notation
\begin{equation}
    \label{eq:definition_g}
    g(y) \defeq \frac{1}{1-r}-\frac{r}{1-r}\E[][\max\left(\frac{B}{y}, 1\right)].
\end{equation}
The function $g$ is non-decreasing.

When $d$ follows its stationary uniform distribution, the expected volume pending on the best ask pile is
\begin{equation*}
    V^1=\frac{1}{\alpha}  \int_{\mu}^{\mu + \alpha} F^{-1}_{Q^u}\left(g(y)\right) \diff y = \frac{\sigma_{noise}}{\alpha}  \int_{\mu}^{\mu + \alpha} F^{-1}_{N}\left(g(y)\right) \diff y
\end{equation*}
where $F_N$ is the cumulative distribution function of $\mathcal{N}(0,1)$. Taking the estimator
\begin{equation*}
    \hat{\sigma}_{noise} \defeq \frac{\alpha\hat{V}^1}{\int_{\mu}^{\mu + \alpha} F^{-1}_{N}\left(g(y)\right)\diff y}
\end{equation*}
where $\mu$ and $g$ are estimated with the procedure from Section \ref{subsec:estimation_efficient_price}, and $\hat{V}^1$ is the average volume pending on the best pile, we obtain an estimated model that matches exactly the order book at the best piles. The estimated $\hat{\sigma}_{noise}$ for each stock are reported in Appendix \ref{appendix:estimation_results}.

\begin{figure}
    \centering
    \begin{subfigure}[b]{0.45\textwidth}
        \centering
        \includegraphics[width=\textwidth,page=2]{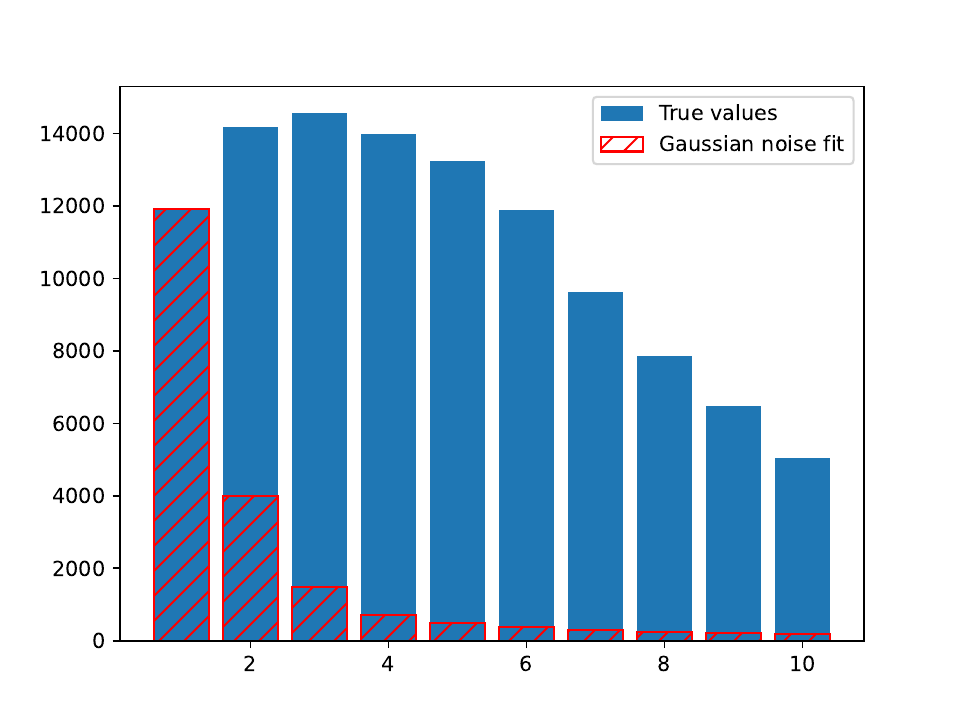}
        \caption{Carnival Corporation}
    \end{subfigure}
    \hfill
    \begin{subfigure}[b]{0.45\textwidth}
        \centering
        \includegraphics[width=\textwidth,page=2]{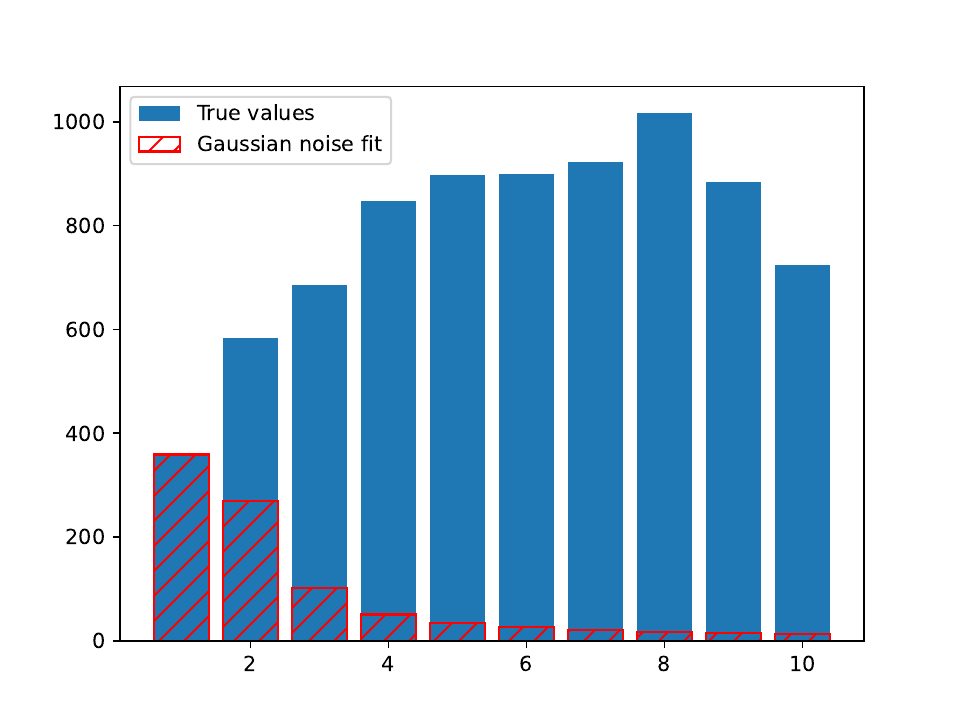}
        \caption{ESSILORLUXOTTICA}
    \end{subfigure}

    \vspace{0.5cm}
    \begin{subfigure}[b]{0.45\textwidth}
        \centering
        \includegraphics[width=\textwidth,page=2]{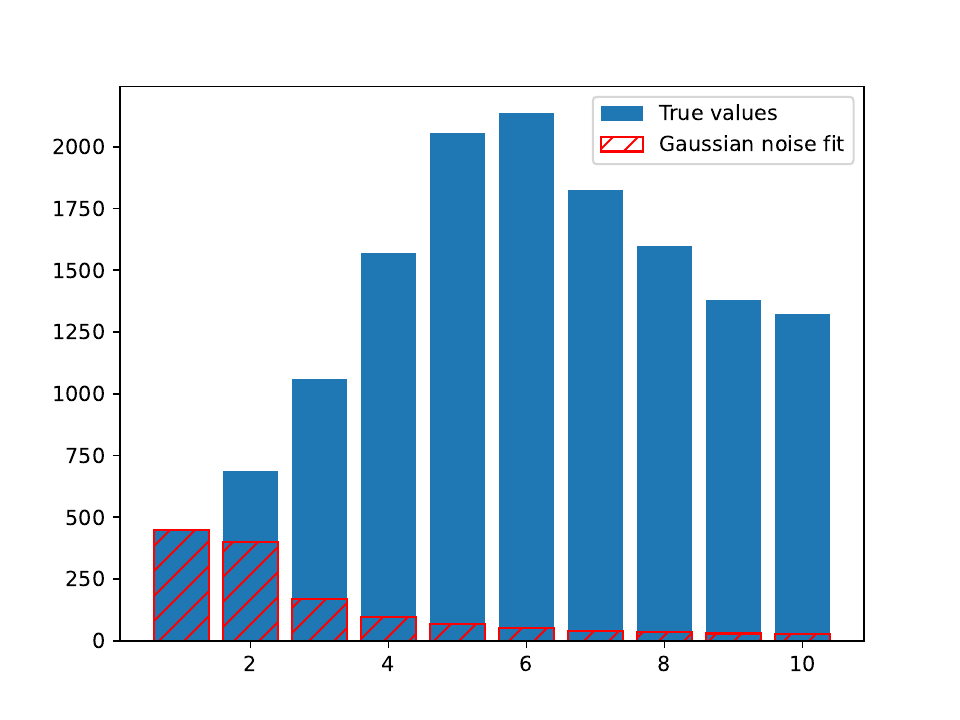}
        \caption{OCI}
    \end{subfigure}
    \hfill
    \begin{subfigure}[b]{0.45\textwidth}
        \centering
        \includegraphics[width=\textwidth,page=2]{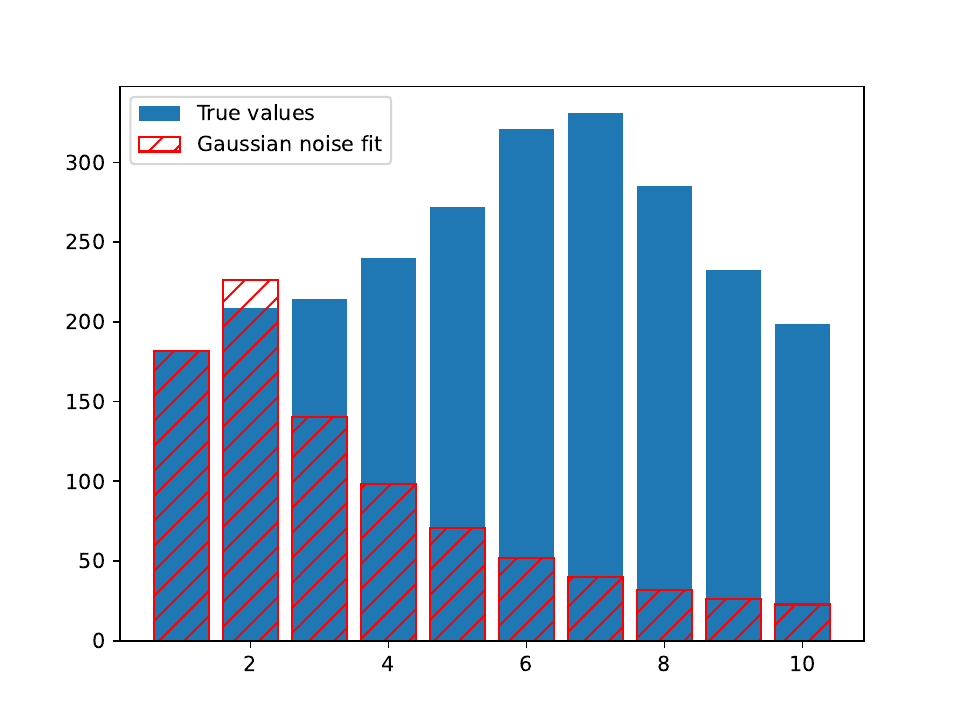}
        \caption{Royal Caribbean Group}
    \end{subfigure}
    
    \vspace{0.5cm}
    \begin{subfigure}[b]{0.45\textwidth}
        \centering
        \includegraphics[width=\textwidth,page=2]{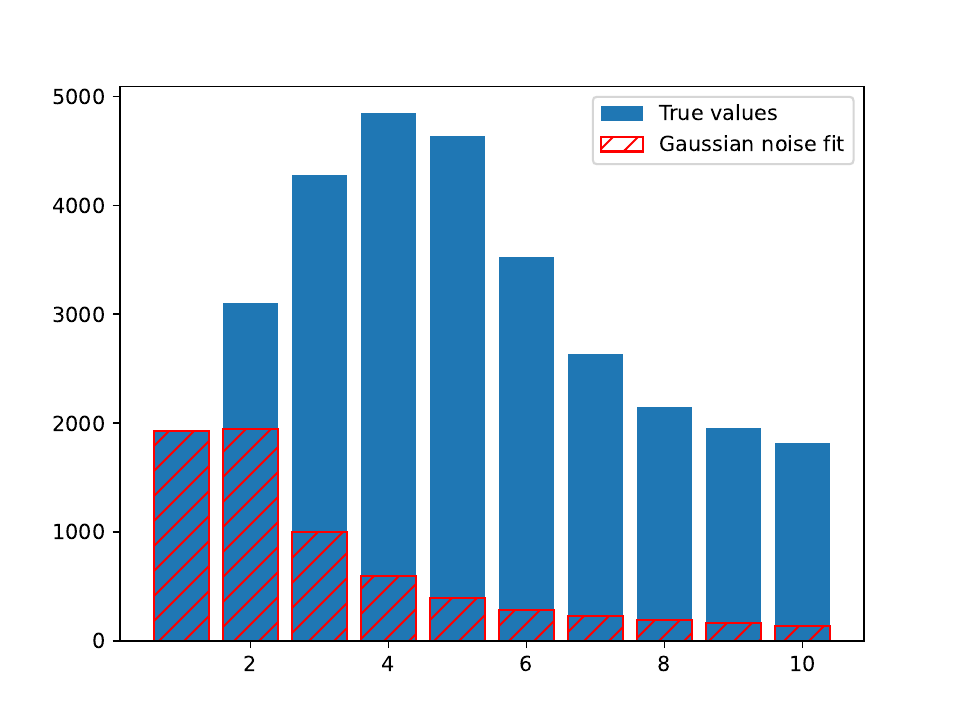}
        \caption{STELLANTIS NV}
    \end{subfigure}
    \hfill
    \begin{subfigure}[b]{0.45\textwidth}
        \centering
        \includegraphics[width=\textwidth,page=2]{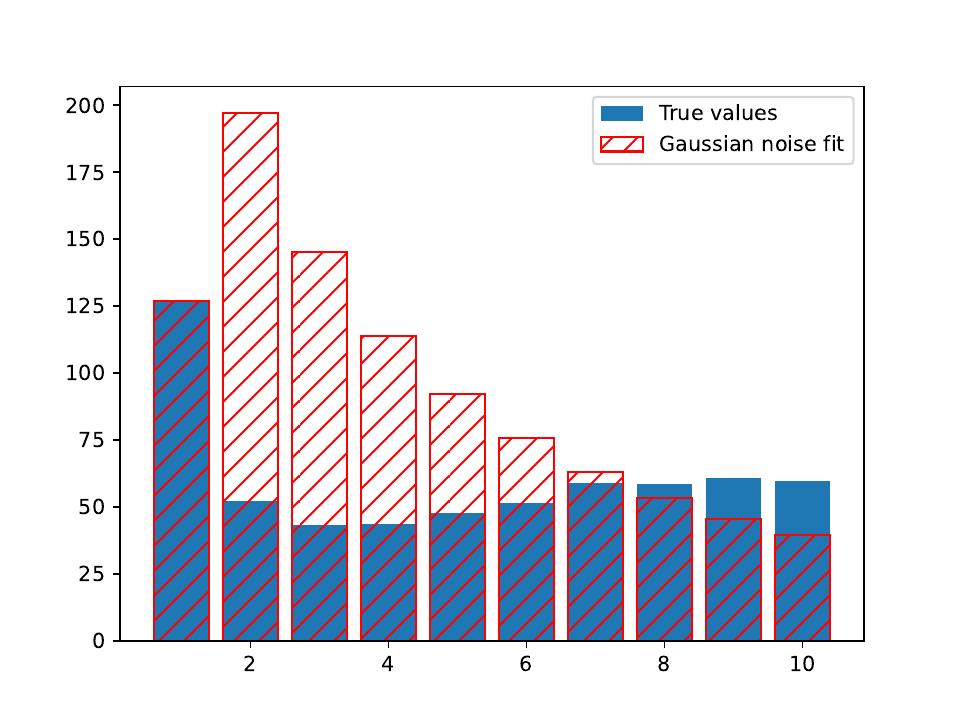}
        \caption{The Boeing Company}
    \end{subfigure}
    
    \caption{Mean volume pending on each pile up to 10 ticks from the best (in blue) and the mean volumes given by the model when $Q^u$ follows a normal distribution calibrated on the volume pending at the best price. Estimated on data from 2022-10-01 to 2023-03-31.}
    \label{fig:gaussian_noise_fit}
\end{figure}

As shown in Figure \ref{fig:gaussian_noise_fit}, the Gaussian model for $Q^u$ is tailored to capture the size of the best pile, but not the ones beyond.

\subsubsection{Calibrating multiple piles}

According to Proposition \ref{prop:any_shape_attainable}, at least in the model with zero tick size, when $B$ satisfies mild assumptions (that Lévy-stable laws satisfy here), every shape of the LOB should be reproduced by some distribution $F_{Q^u}$ of the sizes of the trades submitted by noise traders\footnote{A similar property should hold for the version with non-zero tick size: for any LOB shape $L$, the existence of $F_{Q^u}$ such that the average LOB shape matches $L$.}.

Here, we present a method to calibrate a piecewise constant density $f_{Q^u}$ such that the average LOB shape is exactly reproduced by our model. Of course, in the averaged case, such a general density is not unique. There is no guarantee that our method succeeds in finding such a density, but we keep it for its simplicity: there is only a limited number of integrals to compute numerically, and the parameters are computed sequentially.

For $i \in \mathbb{N}$, we define $\mu_i \defeq \mu + i \alpha$.

We track $M=10$ piles. The average volume (supposing $d(t)$ has uniform distribution) pending on the $i$-th queue, $2 \leqslant i \leqslant M$ is
\begin{equation*}
    V^{i} = \begin{aligned}[t]
        \frac{1}{\alpha}  \int_{\mu_{i-1}}^{\mu_i} &F^{-1}_{Q^u}\left(g(y)\right) \diff y 
        -\frac{1}{\alpha}\int_{\mu_{i-2}}^{\mu_{i-1}} F^{-1}_{Q^u}\left(g(y)\right) \diff y.
    \end{aligned}
\end{equation*}

We focus on modelling $F^{-1}_{Q^u}$ on $\left[\frac{1}{2}, g(\mu_M)\right]$ since what is beyond $g(\mu_M)$ has no involvement on our observation and $F^{-1}_{Q^u}$ on $\left[-g(\mu_M), \frac{1}{2}\right]$ can be deduced by symmetry.

We suppose that $F^{-1}_{Q^u}$ is continuous on $\left[-g(\mu_M), g(\mu_M)\right]$ and affine on every subinterval $\left[g(\mu_{i}), g(\mu_{i+1})\right]$. More specifically, we suppose there exists $(b_i)_{0 \leqslant i \leqslant M-1} \in (0, \infty)^M$ such that for each $i$,
\begin{equation*}
    F^{-1}_{Q^u}(p) = b_i (p - g(\mu_i)) + \sum_{k=0}^{i-1} b_k (g(\mu_{k+1}) - g(\mu_{k})),\quad p \in \left[g(\mu_i), g(\mu_{i+1})\right].
\end{equation*}
With $I_i \defeq \frac{1}{\alpha}\int_{\mu_i}^{\mu_{i+1}}\left(g(y) - g(\mu_i) \right) \diff y$, we have $V^1 = b_0 I_0$ and, for $2 \leqslant i \leqslant M$,
\begin{equation*}
    V^i = b_{i-2}\left(g(\mu_{i-1}) - g(\mu_{i-2})\right) + b_{i-1}I_{i-1} - b_{i-2}I_{i-2}.
\end{equation*}
Observing an average volume $\hat{V}^i$ on the $i$-th pile, and computing the $I_i$ with the parameters estimated in Section \ref{subsec:estimation_efficient_price}, we can estimate the $b_i$ sequentially: $\hat{b}_0 = \frac{\hat{V}^1}{I_0}$ and for $i \geqslant 1$,
\begin{equation}
    \label{eq:bi_computation}
    \hat{b}_i = \frac{\hat{V}^{i+1} + \hat{b}_{i-1}I_{i-1} - \hat{b}_{i-1}\left(g(\mu_{i}) - g(\mu_{i-1})\right)}{I_i}.
\end{equation}
Then, the density $f_{Q^u}$, is constant and equal to $\frac{1}{b_i}$ on every interval $[F^{-1}_{Q^u}(g(\mu_i)), F^{-1}_{Q^u}(g(\mu_{i+1}))]$. By construction, the average shape of the LOB (up to $M$ piles) in the calibrated model matches exactly the empirical one.

The fit failed for some stocks, in the sense that some the computed $\hat{b}_i$ are negative. These stocks are listed in Tables \ref{table:failed_histogram_period_1} and \ref{table:failed_histogram_period_2} in Appendix \ref{appendix:failed_histogram}. They have in common that they are very small tick stocks: their spread is bigger than 4 ticks. That makes the tick increment quite negligible to the market makers, and the liquidity is spread out, leaving many empty piles. Consequently, the average volume posted at non-best prices is low compared to the volume posted at the best price, see Figure \ref{fig:failed_histogram_LOB_shape}. Consequently, in Equation \eqref{eq:bi_computation}, $\hat{V}^2$ does not make up for the large $\hat{b}^0$ and $\hat{b}^1$ is negative.

In practice, when the fit fails, the density and the concerned histogram bin can be put to 0, and the procedure \eqref{eq:bi_computation} can be resumed. The fit will not be perfect, but the average shape of the LOB will be closer to reality.

\begin{figure}
    \centering
    \begin{subfigure}[b]{0.45\textwidth}
        \centering
        \includegraphics[width=\textwidth,page=4]{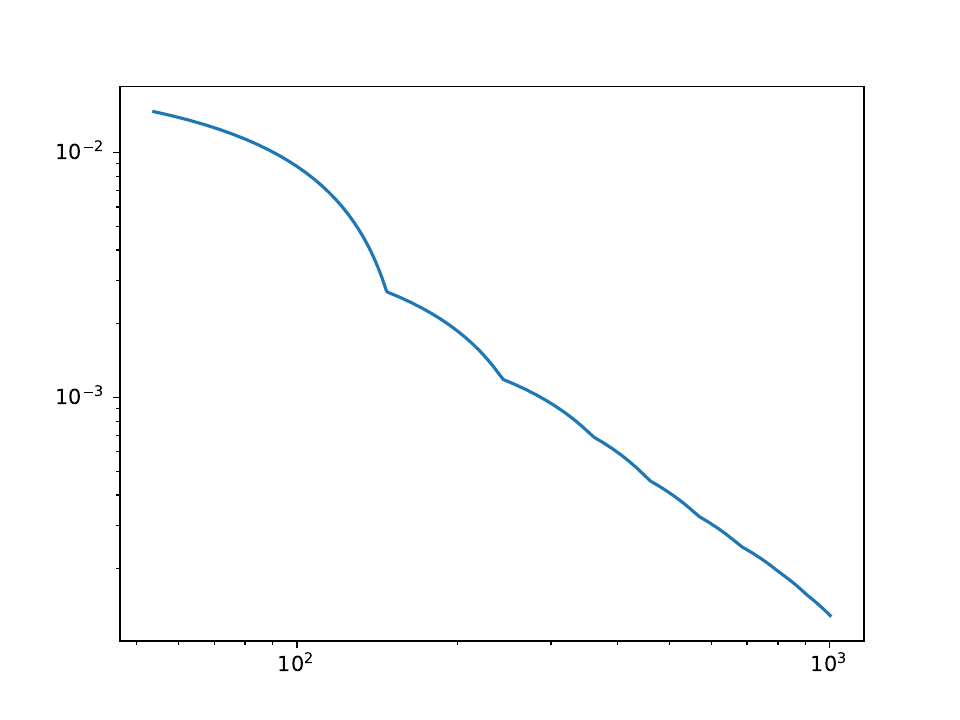}
        \caption{HERMES INTL}
    \end{subfigure}
    \hfill
    \begin{subfigure}[b]{0.45\textwidth}
        \centering
        \includegraphics[width=\textwidth,page=4]{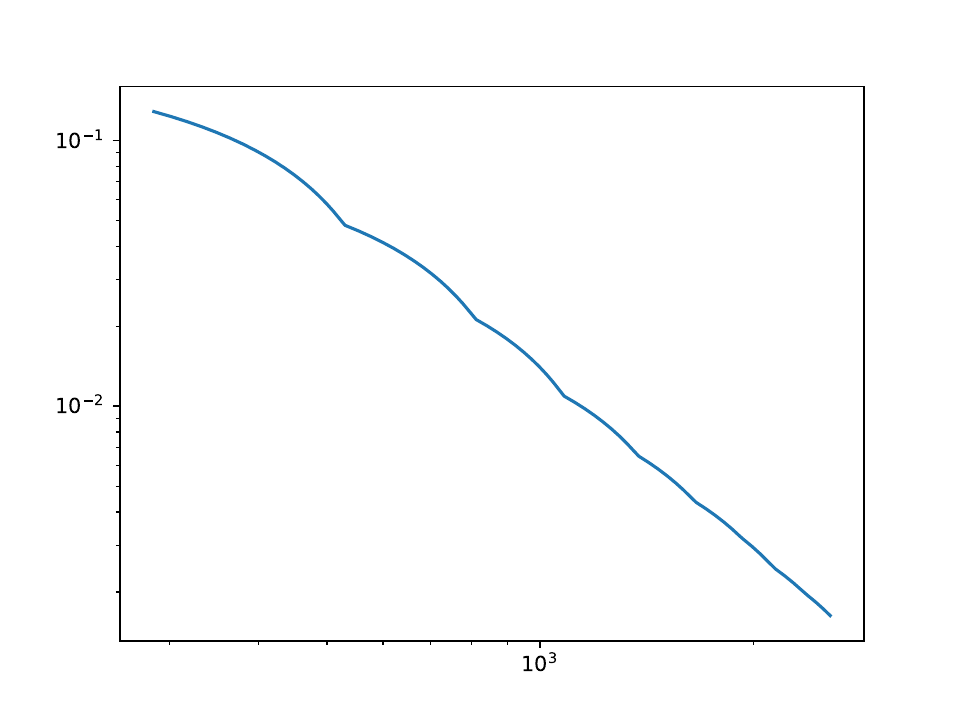}
        \caption{Johnson \& Johnson}
    \end{subfigure}

    \vspace{0.5cm}
    \begin{subfigure}[b]{0.45\textwidth}
        \centering
        \includegraphics[width=\textwidth,page=4]{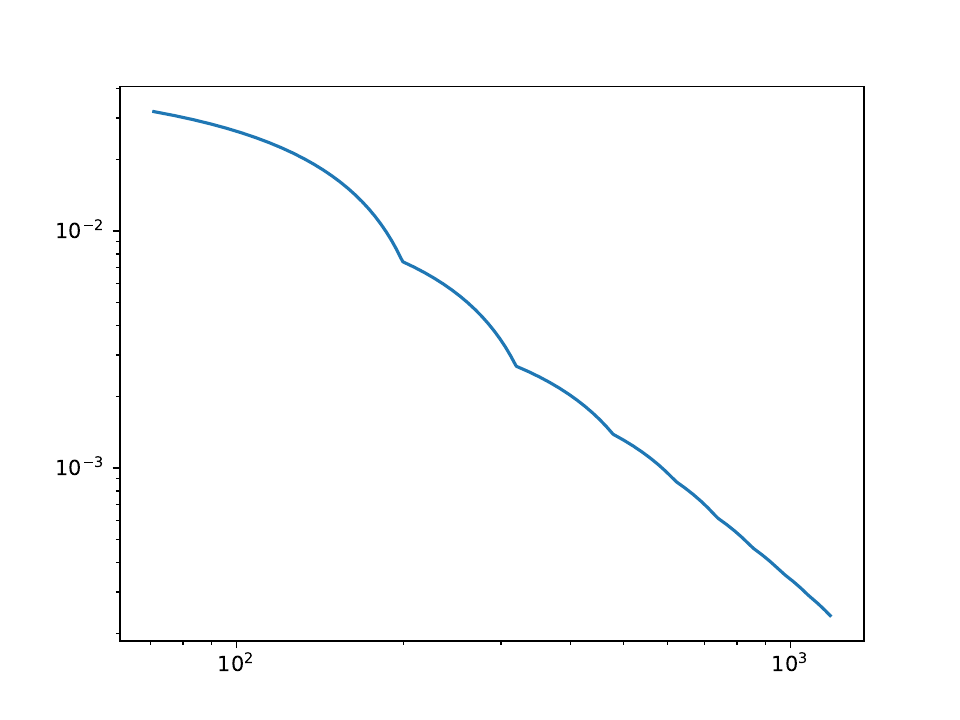}
        \caption{LVMH}
    \end{subfigure}
    \hfill
    \begin{subfigure}[b]{0.45\textwidth}
        \centering
        \includegraphics[width=\textwidth,page=4]{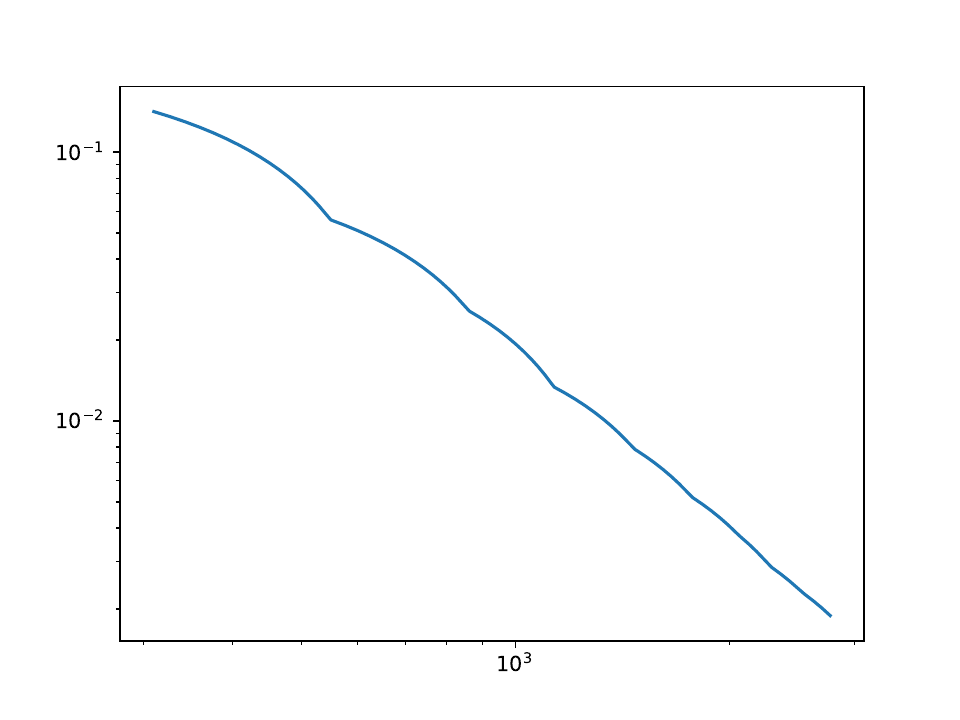}
        \caption{NIKE, Inc.}
    \end{subfigure}
    
    \vspace{0.5cm}
    \begin{subfigure}[b]{0.45\textwidth}
        \centering
        \includegraphics[width=\textwidth,page=4]{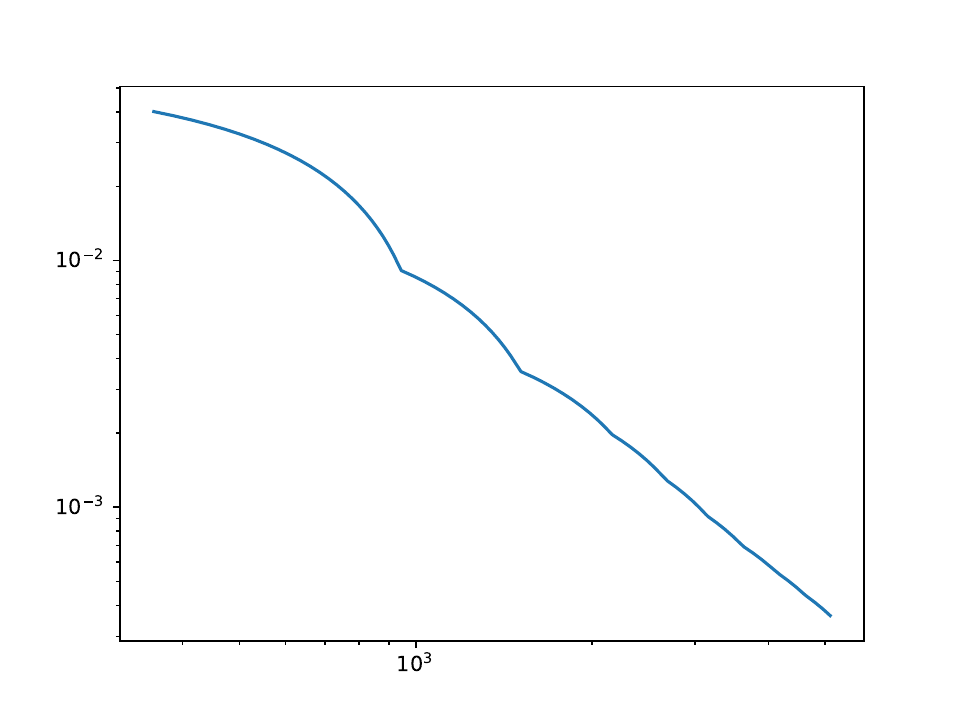}
        \caption{SAFRAN}
    \end{subfigure}
    \hfill
    \begin{subfigure}[b]{0.45\textwidth}
        \centering
        \includegraphics[width=\textwidth,page=4]{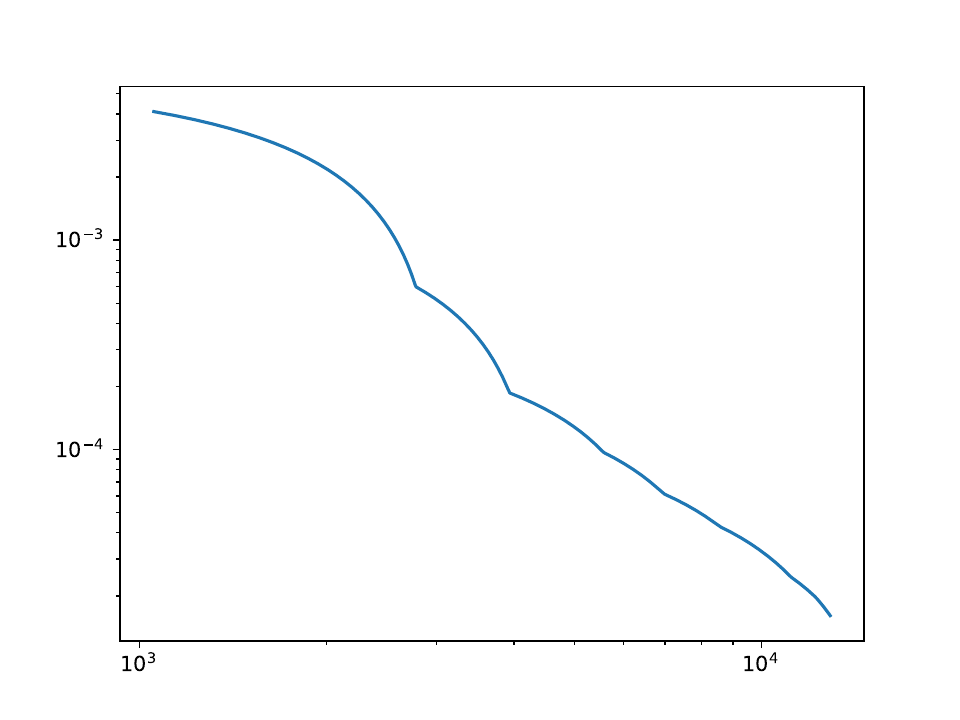}
        \caption{The Coca-Cola Company}
    \end{subfigure}
    
    \caption{$1-F_{Q^u}$ in log-log scale for $f_{Q^u}$ fitted piecewise constant. Estimated on data from 2022-10-01 to 2023-03-31.}
    \label{fig:histogram_fit_cdf}
\end{figure}

\begin{table}
    \centering
    \begin{tabular}{|l|c|c|}
        \hline
         & Period 1 & Period 2\\
        \hline
        Estimation succeeded & 80 stocks & 82 stocks\\
        \hline
        Estimation failed & 23 stocks & 21 stocks\\
        \hline
    \end{tabular}
    \caption{Outcome of the calibration of $M=10$ piles of the LOB.}
    \label{table:whole_lob_estimation_summary}
\end{table}

For the majority of the stocks, the estimation succeeds, see Table \ref{table:whole_lob_estimation_summary}. The cumulative distribution function of the resulting fitted distribution, plotted in Figure \ref{fig:histogram_fit_cdf}, suggests that the law of $Q^u$ has fat tails, that decay in power law, much slower than the Gaussian distribution. Note that this may be highly dependent on the Lévy-stable law model chosen for the efficient price jumps.

\subsection{Summary of the estimation procedure}

Observing the reference price $\tilde{P}$ at discrete times $t_j = j \Delta t$, we are able to infer the parameters of the efficient price dynamics--both their jump intensity $\lambda^i$ and the law of the jump sizes $B$--using a maximum likelihood approach. We consider Lévy-stable distributions for $B$ which are both flexible enough and computationally convenient.

Once they are estimated, we then estimate the noise trade size density $f_B$ in order to match the average LOB shape in the best way possible. Two models are considered: a Gaussian one, which allows reproducing perfectly the size of the best bid and ask piles, and a piecewise constant one allowing to match an arbitrary number of queues but is not guaranteed to be compatible in its simplest version.

\section{Applications}
\label{sec:applications}

\subsection{Spread forecasting}

Our model (in particular Equation \eqref{spread}) allows us to forecast the new value of the spread if the tick size is modified. In the following, we predict the spread changes due to the new tick size regime under the recent European regulation MiFID II, and compare our results to the effective spread values. We expect our model to be relevant for rather liquid assets since it is based on the presence of competitive market makers. We therefore restrict ourselves to this class. Note that there are other models in the literature enabling practitioners to forecast spreads, see notably \cite{dayri2015large} where the authors propose an approach designed for large tick assets. This methodology is applied for example in \cite{huang2016predict} on Japanese data and in \cite{roslaruelle} where spread values before and after MiFID II are compared. The advantage of our procedure is that it can be applied to both small and large tick assets.

\subsubsection{The tick size issue and MiFID II regulation}

In the recent years, trading platforms have raced to reduce their tick sizes in order to offer better prices and gain market share. This broad trend has had adverse effects on the overall market quality: a too small tick leads to unstable LOBs and a degradation of the price formation mechanism. However, a tick that is too large prevents the price from moving freely according to the views of market participants. Therefore, finding suitable tick values is crucial for the fluidity of financial markets. To solve this issue, some regulators tried to use pilot programs, as was the case in Japan and in the United States, see for example \cite{huang2016predict}. This is a costly practice which does not really rely on theoretical foundations. We believe that using quantitative results such as those presented in this work could lead to a much more efficient methodology.

In Europe, MiFID II (Markets in Financial Instruments Directive II) regulation introduced a harmonized tick size regime (Article 49) which is based on a two-entries table: price and liquidity (expressed in terms of number of transactions per day)\footnote{The number of transactions per day is computed on a yearly basis. The liquidity bin in which an asset belongs, and thus the tick size rule it has to follow is updated every year on April \nth{1}.}. Note that one of the targets for regulators was to obtain for liquid assets spreads between 1.5 and 2 ticks, see \cite{amf}.

\subsubsection{Empirical study}

In our dataset, 10 stocks changed liquidity bin and thus of tick size on April \nth{1}, 2024. We compute the average spread on the period 2022-10-01 to 2023-03-31. Then, we predict average spread for the period 2023-04-01 to 2023-09-30 in two ways: with our model (Equation \eqref{spread}), and supposing it stays constant in nominal value.

The results are reported in Table \ref{table:spread_forecast}. Our model outperforms a constant spread prediction on 9 out of 11 stocks and gives globally a more accurate prediction (relative error of $14\%$ versus $39\%$ on average). LHYFE is a notable outlier. It is the only stock for which the tick size and the average spread go in opposite directions after the tick size modification: the average spread (in euros) decreased after the tick size went up.

\begin{table}
    \begin{scriptsize}
        \begin{tabular}{|c|p{1cm}|p{1cm}|p{1cm}|p{1cm}|p{1.5cm}|p{2cm}|p{2cm}|}
            \hline
            \textbf{Company}& \textbf{tick size period 1} & \textbf{spread period 1} & \textbf{tick size period 2} & \textbf{spread period 2} & \textbf{predicted spread} & \textbf{Relative prediction error, model} & \textbf{Relative prediction error, constant spread} \\ \hline
            ADP & 0.05 & 0.131 & 0.1 & 0.151 & 0.181 & 19.60\% & 13.45\% \\ \hline
            AMG & 0.02 & 0.039 & 0.01 & 0.031 & 0.029 & 5.11\% & 27.18\% \\ \hline
            BORR DRILLING & 0.01 & 0.092 & 0.05 & 0.148 & 0.132 & 10.41\% & 37.48\% \\ \hline
            ESSILORLUXOTTICA & 0.05 & 0.067 & 0.02 & 0.042 & 0.037 & 10.46\% & 61.63\% \\ \hline
            FDJ & 0.01 & 0.029 & 0.02 & 0.041 & 0.039 & 3.13\% & 27.72\% \\ \hline
            HERMES INTL & 0.5 & 0.674 & 0.2 & 0.461 & 0.374 & 18.85\% & 46.21\% \\ \hline
            LHYFE & 0.001 & 0.047 & 0.01 & 0.043 & 0.056 & 29.54\% & 8.62\% \\ \hline
            MICHELIN & 0.005 & 0.009 & 0.01 & 0.014 & 0.014 & 1.66\% & 36.53\% \\ \hline
            OCI & 0.02 & 0.039 & 0.01 & 0.028 & 0.029 & 1.40\% & 36.90\% \\ \hline
            OKEA & 0.05 & 0.131 & 0.02 & 0.073 & 0.101 & 38.17\% & 79.39\% \\ \hline
            REMY COINTREAU & 0.1 & 0.16 & 0.05 & 0.111 & 0.11 & 1.13\% & 43.95\% \\ \hline
            \end{tabular}
    \end{scriptsize}
    \caption{Forecasting assets spreads on Period 2 (2023-04-01 to 2023-09-30) using data from Period 1 (2022-10-01 to 2023-03-31).}
    \label{table:spread_forecast}
\end{table}

\subsection{Model validation: predictive power of imbalance}

It is a known stylized fact, see \cite{huang2015simulating,lehalle2021optimal,muni2017modelling,pulido2023understanding,sfendourakis_lob_2023}, that the volume imbalance, defined as $I=\frac{V^b -V^a}{V^b + V^a}$ where $V^b$ is the volume posted at the best bid and $V^a$ the volume posted at the best ask, is a good predictor of upcoming price moves at high frequency, especially for large-tick stocks.

Consider a stock that has an average spread below 1.5 ticks, that is a large-tick stock in the sense of \cite{huang2016predict}. In our model, that corresponds to $\mu \leqslant \frac{\alpha}{4}$. For $d \in (\mu, \alpha - \mu)$,
\begin{equation*}
    I(d) = \frac{F^{-1}_{Q^u}\left(\frac{1}{1-r}-\frac{r}{1-r}\E[][\max\left(\frac{B}{\alpha-d}, 1\right)]\right) - F^{-1}_{Q^u}\left(\frac{1}{1-r}-\frac{r}{1-r}\E[][\max\left(\frac{B}{d}, 1\right)]\right)}{F^{-1}_{Q^u}\left(\frac{1}{1-r}-\frac{r}{1-r}\E[][\max\left(\frac{B}{\alpha-d}, 1\right)]\right) + F^{-1}_{Q^u}\left(\frac{1}{1-r}-\frac{r}{1-r}\E[][\max\left(\frac{B}{d}, 1\right)]\right)}.
\end{equation*}
In the case where $f_B$ has unbounded support and $f_{Q^u}$ is strictly positive on $(0, g(\mu+a))$ ($g$ defined by \eqref{eq:definition_g}), which is the case of every example we have seen, $I:(\mu, \alpha - \mu) \to (-1,1)$ is strictly increasing and bijective. In this case, to each value $y$ of the imbalance corresponds one and only one value of $I^{-1}(i)$ of $d$. The probability of an upward jump of the reference price $\tilde{P}$ at horizon $\Delta t$ can be computed, recalling that $\tilde{P}(t) = P(t) + d(t)$:
\begin{align*}
    \Proba[][\tilde{P}(t+\Delta t)- \tilde{P}(t) > 0 | I(d(t)) = y]
    &= \Proba[][\tilde{P}(t+\Delta t) - \tilde{P}(t) > 0| d(t) = I^{-1}(y)]\\
    &= \Proba[][P(t+\Delta t) - P(t) > I^{-1}(y)].
\end{align*}
By symmetry, $\Proba[][\tilde{P}(t+\Delta t) - \tilde{P}(t) < 0 | I(d(t)) = y]
= \Proba[][P(t+\Delta t) - P(t) < -I^{-1}(y)]$.

We compare the empirical price move probabilities at horizon $\Delta t = 5 \text{s}$ to the ones given by our model. For this subsection only, the model parameters used were estimated on a sampling of 5 seconds instead of 1 minute: the latter gave smaller values of $\lambda^i$, failing to capture many short-term efficient price jumps.

\begin{figure}
    \centering
    \begin{subfigure}[b]{0.45\textwidth}
        \centering
        \includegraphics[width=\textwidth,page=1]{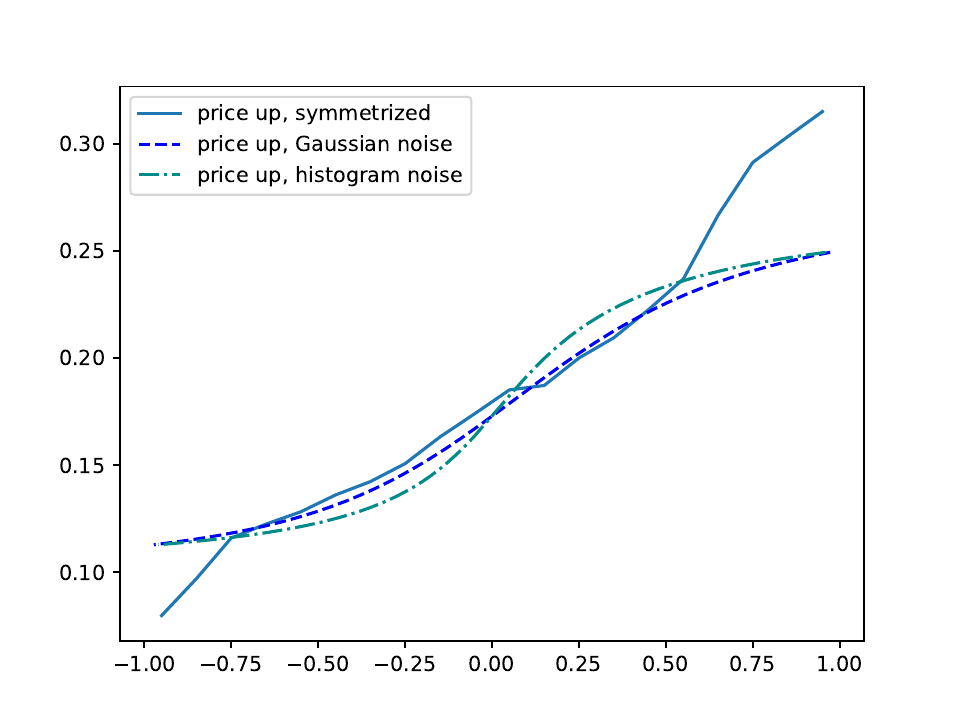}
        \caption{AIRBUS}
    \end{subfigure}
    \hfill
    \begin{subfigure}[b]{0.45\textwidth}
        \centering
        \includegraphics[width=\textwidth,page=1]{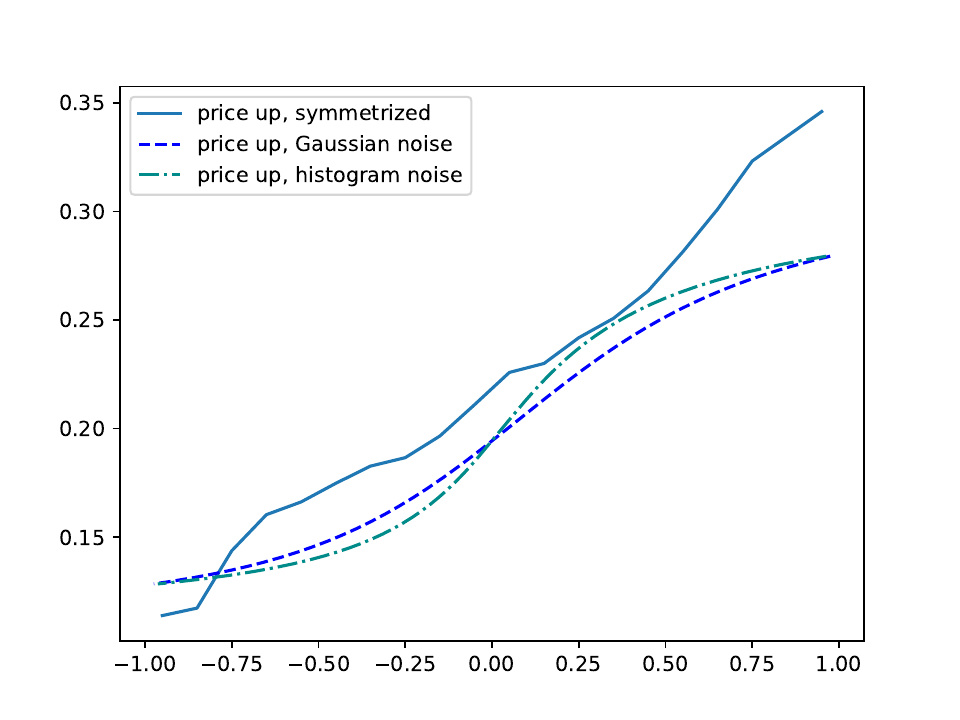}
        \caption{BNP PARIBAS ACT.A}
    \end{subfigure}

    \vspace{0.5cm}
    \begin{subfigure}[b]{0.45\textwidth}
        \centering
        \includegraphics[width=\textwidth,page=1]{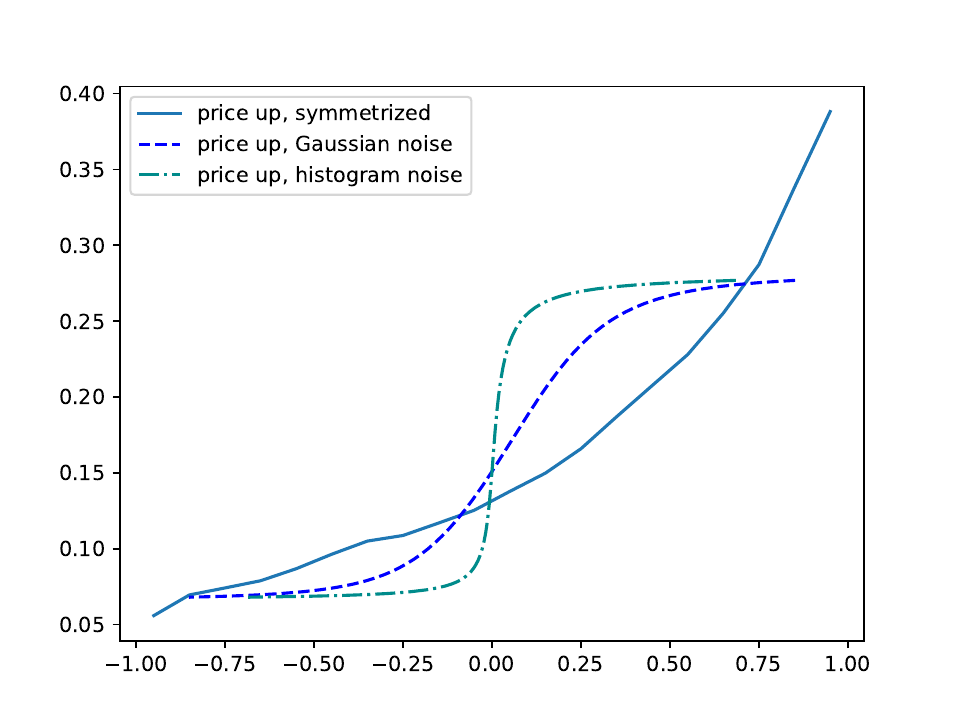}
        \caption{Citigroup Inc.}
    \end{subfigure}
    \hfill
    \begin{subfigure}[b]{0.45\textwidth}
        \centering
        \includegraphics[width=\textwidth,page=1]{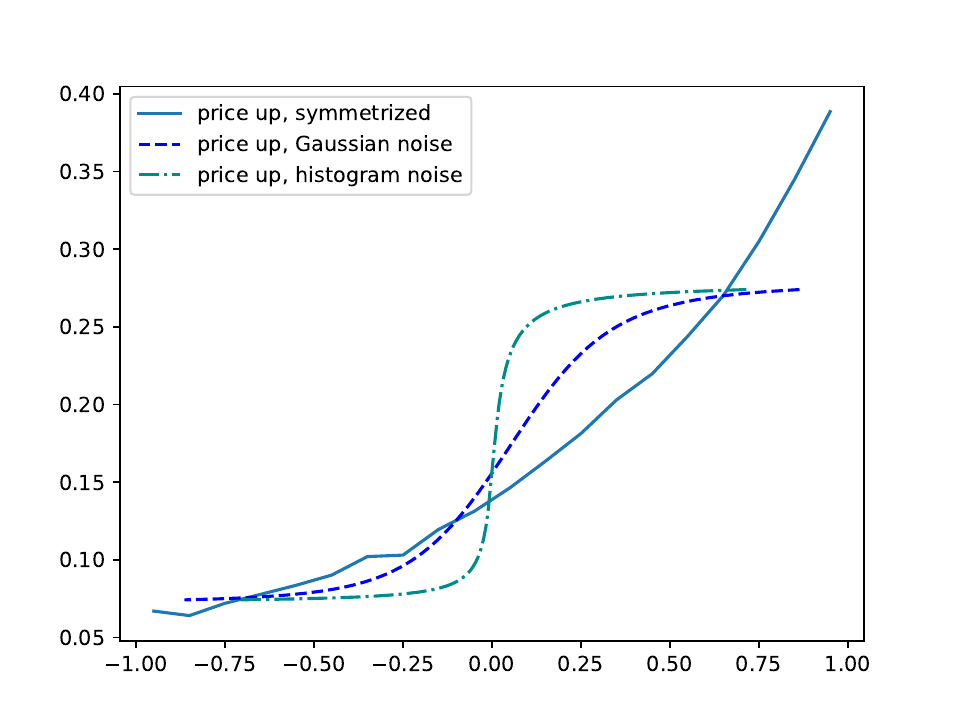}
        \caption{Delta Air Lines, Inc.}
    \end{subfigure}
    
    \vspace{0.5cm}
    \begin{subfigure}[b]{0.45\textwidth}
        \centering
        \includegraphics[width=\textwidth,page=1]{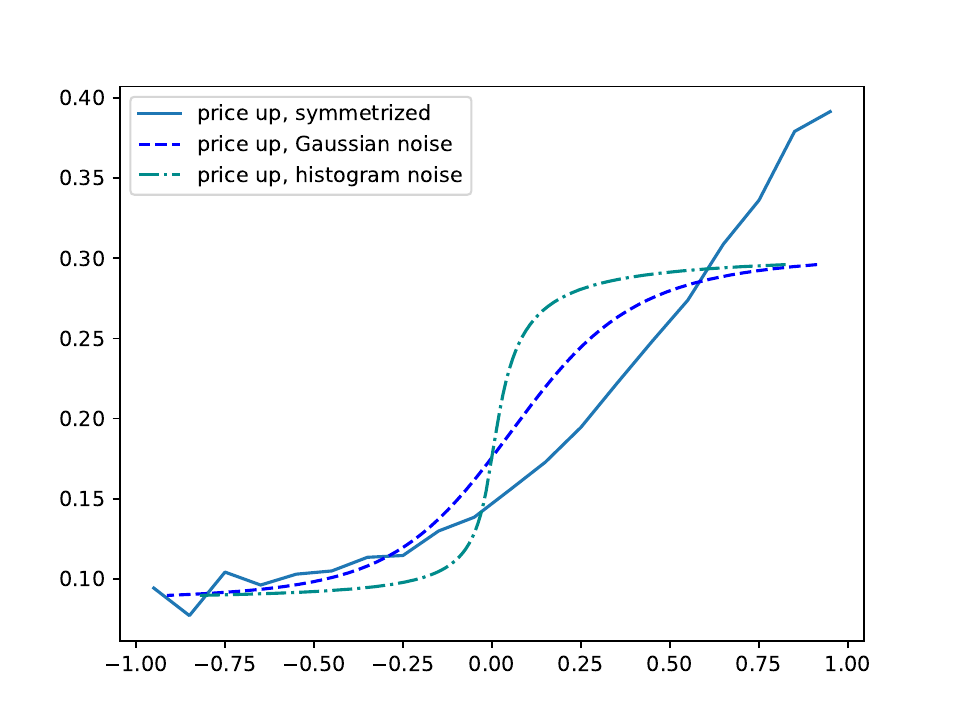}
        \caption{Halliburton Company}
    \end{subfigure}
    \hfill
    \begin{subfigure}[b]{0.45\textwidth}
        \centering
        \includegraphics[width=\textwidth,page=1]{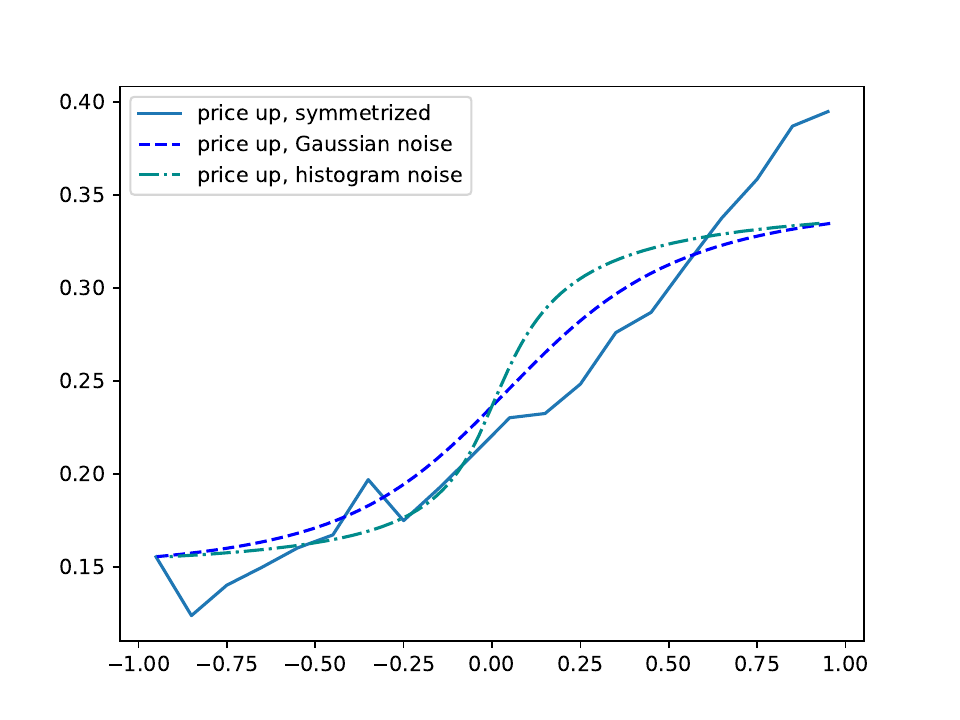}
        \caption{Schlumberger Limited}
    \end{subfigure}
    
    \caption{Probability of an upward price move in the next 5s with respect to the imbalance. In solid blue, the empirical probabilities, in dashed the probabilities given by the model when the noise is supposed Gaussian, in dot-dashed the probabilities given by the model when the noise has a piecewise constant density.}
    \label{fig:imbalance}
\end{figure}

In Figure \ref{fig:imbalance}, we compare the empirical probabilities of an upwards price with the ones given by the model. The empirical probabilities are symmetrized: to avoid any asymmetry between the bid and the ask, we compute the empirical version of 
\begin{equation*}
    \frac{1}{2}\left(\Proba[][\tilde{P}(t+\Delta t) - \tilde{P}(t) > 0 | I(d(t)) = y] + \Proba[][\tilde{P}(t+\Delta t) - \tilde{P}(t) < 0 | I(d(t)) = -y]\right).
\end{equation*}
To compute them, the values of the imbalance are grouped in intervals of length 0.1.
They are reproduced quite well by the model, and the Gaussian choice for the law of $Q^u$ seems to perform better than the version with piecewise constant density. The observed probability of an upward price for an imbalance close to 1 goes beyond the theoretical bound of $e^{-\lambda^i \Delta t}$ given by the model (there cannot be a jump of $\tilde{P}$ if $P$ does not jump).

Figure \ref{fig:imbalance} (c) and (d) present more lukewarm results: the strongly convex dependence of the price jump probability with respect to the imbalance is difficult to capture by our estimation procedure. The histogram model chosen for $f_{Q^u}$ prevents flexibility around 0 which is important for this exercise.

\subsection{Queue position valuation}

Introducing a tick size in our modeling enables us to study the value of the position of the limit orders in the queues. We can quantify the advantage of an order placed on top of a queue compared to another one placed at the bottom. The difference in the values of the positions in a queue is a crucial parameter for trading algorithms. It has actually led to a technological arms race among high-frequency traders and other algorithmic market participants to establish early (and hence advantageous) positions in the queues, see \cite{abergel2014understanding,moallemi2016model}. Placing limit orders at the front of a queue is very valuable for different reasons. It guarantees early execution and less waiting time. In addition, it reduces adverse selection risk. In fact, as explained in \cite{moallemi2016model}, when a limit order is placed at the end of a queue, it is likely that it will be executed against a large trade. In contrast, a limit order placed at the front of the (best) queue will be executed against the next trade independently of the trade size. Large trades are in general sent by informed traders aiming at consuming all limit orders which will generate profit for them. In this way, a limit order submitted at the front of the queue is less likely to undergo adverse selection.

In light of this, to optimize their execution, practitioners need to place limit orders in a relevant way. This requires an estimate of the value of a limit order according to its position in the queue. This very problem is studied in \cite{moallemi2016model} for the queues at the best limits for large tick assets. We complement here this nice work providing formulas valid for any queue of a large or small tick asset and taking into account strategic interactions between market participants.

Assumption \ref{zeroprofit tick} tells us that the expected profit of a new infinitesimal limit order placed at the bottom of a non-empty queue is
equal to zero. However, under our zero-profit condition, market makers may still make profit if their orders are placed before.
The value of queue position at the $i^{th}$ level, denoted by $\tilde{G}^d(i)$, can be formulated in this model as the difference between the expected profit of the order placed on top and that of a new one that would be placed at the bottom of the $i^{th}$ queue (which makes on average zero profit at equilibrium). Computing this quantity is very similar to deriving the equations in Proposition \ref{gain tick}. The difference is that now we no longer consider a traded volume totally depleting the limit but a traded volume consuming all the limits before the $i^{th}$ one. This leads to the following theorem.
  
\begin{theorem}\label{pv}
For $i\ge k_r^d$, we have
   \begin{equation*}
    \tilde{G}^d(i)=d+(i-1)\alpha-\frac{r\mathbb{E}[B\mathds{1}_{B > d+(i-1)\alpha}]}{1-rF_B(d+(i-1)\alpha)-(1-r)F_{Q^u}\left(L^d(i-1)\right)}.
\end{equation*} 

\end{theorem}

The formula for $i\leq -k_l^d$ is obviously deduced. An integral computation, similar to the ones in the proofs of Proposition \ref{prop:invariant_fractional_part} and Lemma \ref{lem:fractional_intergal}, shows Corollary \ref{corol:best_valuation}, which gives the average value of the highest priority order in the best queue. It is the object of study of \cite{moallemi2016model} and of our subsequent analysis on the data. Remark that since every trade sent by a noise trader consumes the top of the pile, the distribution $F_{Q^u}$ is not involved in this quantity.

\begin{corollary}
    \label{corol:best_valuation}
    If $d(t)$ follows its stationary distribution, the value of the best queue is on average
    \begin{equation*}
        \tilde{G}(best) = \frac{1}{\alpha}\int_{\mu}^{\mu+\alpha}
        \left(z - \frac{r \E[][B\mathds{1}_{B > z}]}{1 - r F_B(z)}\right) \diff z.
    \end{equation*}
\end{corollary}

A few examples are given in Table \ref{table:qpv_example}, while the complete results are reported in Appendix \ref{appendix:qpv_estimation}. Like \cite{moallemi2016model} we obtain values of the same order of magnitude as the spread.

\begin{table}
    \centering
    \begin{tabular}{|c|c|c|c|}
        \hline
        \textbf{Company} & \textbf{spread} & \textbf{QPV}& \textbf{tick size}\\
        \hline
        AIR LIQUIDE & 0.0301 & 0.0126 & 0.02 \\ \hline
        ENGIE & 0.0033 & 0.0013 & 0.002\\ \hline
        Freeport-McMoRan Inc. & 0.011 & 0.0053 & 0.01 \\ \hline
        Johnson \& Johnson & 0.0249 & 0.0083 & 0.01 \\ \hline
        OCI & 0.0386 & 0.0146 & 0.02 \\ \hline
        Roblox Corporation & 0.0197 & 0.0073 & 0.01  \\ \hline
    \end{tabular}
    \caption{Queue position valuation for some stocks. Period: from 2022-10-01 to 2023-03-31.}
    \label{table:qpv_example}
\end{table}








\section*{Conclusion}

In this article, we introduce an agent-based model for the LOB. Inspired by the seminal work by Glosten and Milgrom \cite{glosten1985bid} and its extension to the whole limit order book in \cite{glosten1994electronic}, we use a zero-profit condition for the market
makers which enables us to derive a link between proportion of events due to the noise trader, bid-ask spread, dynamic of the efficient price and equilibrium LOB state. The effect of introducing a tick size
is then discussed. We in particular show that the constrained bid-ask spread is equal to the sum
of the tick value and the intrinsic bid-ask spread that corresponds to the case of a vanishing tick size. This model allows us to do spread forecasting when one modifies the tick size.
We develop an estimation procedure allowing us to recover the parameters of the model, while only observing discrete price changes, using the characteristic function of the compound Poisson model chosen for the efficient price.

Price discreteness also enables us to value queue positions in the LOB. In the large-tick stock framework, the model describes accurately the probabilities of price moves conditionally to the volume imbalance. It also allows an exact fit of the average shape of the LOB.

In our approach, market makers only are allowed to insert limit orders.
In practice, the roles of informed trader and market makers are often mixed,
and the informed trader also has the possibility to place passive limit orders. By doing so, he may get
better prices but also leak some information to other market participants. Extending our model by taking into account accurately these intricate features is left for future work.

\printbibliography

\newpage
\begin{center}
{\Large \bf {\centering Appendix}}
\end{center}
\appendix
\label{appendix}

\section{Supplementary mathematical development: time until next trade}

A trader looking to sell an asset can either place a market order, incurring the bid-ask spread cost, or submit a limit order and wait for execution, risking a potential price decline. Accurately estimating the expected waiting time for execution is therefore crucial.

We first derive the formula for the expected duration until the next trade in a model with zero tick size, followed by an approximation for stocks with small tick sizes.

Let $\tau$ denote the time until the first trade in the model with continuous transaction prices. The expected value of $\tau$ is given in Proposition \ref{prop:next_trade_continuous_prices}, with proof in Appendix \ref{sec:proof_next_trade_continuous_prices}.

\begin{proposition}
    \label{prop:next_trade_continuous_prices}
    The average waiting time for a trade in the model with continuous transaction prices is
    \begin{equation*}
        \E[][\tau] = \frac{\mu}{\lambda^i \E[][|B| \mathds{1}_{\{|B| > \mu\}}]}.
    \end{equation*}
\end{proposition}

Introducing a discrete tick size in the model results in longer waiting times for trades. This occurs because larger quoted spreads reduce the number of efficient price movements that lead to executions. We define $\tau^{d}_{\alpha}$ as the time until the first trade when the efficient price is at a distance $d$ from $\tilde{P}(t)$. Our objective is to analyze the expected waiting time, given that $d$ follows its stationary uniform distribution
\begin{equation*}
    u(\alpha) \defeq \frac{1}{\alpha}\int_0^{\alpha} \E[][\tau^{z}_{\alpha}] \diff z.
\end{equation*}

Although we are not able to derive explicit expressions for $u(\alpha)$, we can characterize its asymptotic behavior in the cases where the tick size is either very small ($\alpha \to 0$) or very large ($\alpha \to \infty$).

Note that due to the presence of noise traders,
\begin{equation}
    \label{eq:noisy_trade_bound}
    \E[][\tau^d_{\alpha}] \leqslant \frac{1}{\lambda^u},\quad d \in [0,\alpha).
\end{equation}

The asymptotic formulas for $u(\alpha)$ are based on the following equation for $\E[][\tau^{d}_{\alpha}]$ proved in Appendix \ref{sec:fredholm}.
\begin{theorem}
    \label{thm:equation_trade_time}
    Let $d \in [0, \alpha)$. Then,
    \begin{equation*}
        \E[][\tau^d_{\alpha}]
        = \frac{1}{\lambda^u + \lambda^i} + r \int_{d - \alpha\ceil*{\frac{\mu + d}{\alpha}}}^{d + \alpha\ceil*{\frac{\mu - d}{\alpha}}} \E[][\tau_{\alpha}^{d - z + \alpha \ceil*{\frac{z - d}{\alpha}}}] f_B(z) \diff z.
    \end{equation*}
\end{theorem}

In the small-tick stock limit $(\alpha \to 0)$, the expected waiting time for a trade is close to the one of the model with no tick. Proposition \ref{prop:small_tick_trade_time}, proved in Appendix \ref{sec:proof_small_tick_trade_time} gives its first order approximation.

\begin{proposition}
    \label{prop:small_tick_trade_time}
    As $\alpha \to 0$,
    \begin{equation*}
        \begin{split}
            u(\alpha) = \frac{\mu}{\lambda^i\E[][|B| \mathds{1}_{\{|B| > \mu\}}]}
        &+ \frac{2\mu^2}{\lambda^i \alpha \E[][|B| \mathds{1}_{\{|B| > \mu\}}]^2}\int_{0}^{\alpha}(\alpha - z)f_B(\mu+z)\diff z\\
        &+ o_{\alpha\to 0}\left(\frac{1}{\alpha}\int_{0}^{\alpha}(\alpha - z)f_B(\mu+z)\diff z\right).
        \end{split}
    \end{equation*}
\end{proposition}
The following corollary follows directly from Proposition \ref{prop:small_tick_trade_time}. 
\begin{corollary}
    If $f_B$ is right-continuous at $\mu$, then,
    \begin{equation*}
        \begin{split}
            u(\alpha) = \frac{\mu}{\lambda^i\E[][|B| \mathds{1}_{\{|B| > \mu\}}]}
        &+ \frac{\mu^2 f_B(\mu)\alpha}{\lambda^i \E[][|B| \mathds{1}_{\{|B| > \mu\}}]^2} + o_{\alpha\to 0}\left(\alpha\right).
        \end{split}
    \end{equation*}
\end{corollary}

In the large-tick stock limit ($\alpha \to \infty$), Proposition \ref{prop:large_tick_trade_time}, proved in Appendix \ref{sec:proof_large_tick_trade_time}, establishes that the average waiting time for a trade matches that of a market driven solely by noise traders, as efficient price jumps rarely lead to executions.

\begin{proposition}
    \label{prop:large_tick_trade_time}
    As $\alpha \to \infty$,
    \begin{equation*}
        \lim_{\alpha \to \infty} u(\alpha) = \frac{1}{\lambda^u}.
    \end{equation*}
\end{proposition}

\section{Proofs}\label{proofs}

\subsection{Proof of Proposition \ref{gain}}\label{appendix1}

We focus on the gain from passive sell orders, as the corresponding results for buy orders can be derived analogously.



 

First, we compute $G^i(x- \delta p, x)$. We have
\begin{align*}
   G^i(x - \delta p, x) & = \int _{x - \delta p} ^{x} (P(t)+s) \mathrm{d}\tilde L(s) - \int _{x - \delta p} ^{x} (P(t)+\mathbb{E}[B|B > x]) \mathrm{d}\tilde L(s) \\
   & =   \int _{x - \delta p} ^{x} s  \mathrm{d}\tilde L(s) - \tilde L(x)\mathbb{E}[B|B>x ].
\end{align*}

For $G^u(x - \delta p, x)$ we obtain
$$
G^u(x - \delta p, x)=  \int _{x - \delta p} ^{x} (P(t)+s) \mathrm{d}\tilde L(s) - \int _{x - \delta p} ^{x}P(t) \mathrm{d}\tilde L(s) = \int _{x - \delta p} ^{x} s \mathrm{d} \tilde L(s).
$$
We deduce that
\begin{align*}
G(x - \delta p, x) & =   G^i(x -\delta p, x) \mathbb{P}[\nu = 1|Q \ge L(x) + \tilde L(x)]+ G^u(x - \delta p,x)\mathbb{P}[\nu =0|Q \ge L(x) + \tilde L(x)]\\
& =\int _{x - \delta p} ^{x} s \mathrm{d}\tilde L(s) -\mathbb{P}[\nu=1|Q \ge L(x) + \tilde L(x)] \tilde L(x) \mathbb{E}[B|B>x ]\\
& =\int _{x - \delta p} ^{x} s\mathrm{d}\tilde L(s) - \tilde L(x)  \mathbb{E}[B|B>x ] \frac{r\mathbb{P}[B>x]}{\mathbb{P}[Q \ge L(x) + \tilde L(x)]}\\
&= \int_{x - \delta p}^x s \mathrm{d}\tilde L(s) - \tilde L(x) \frac{r \mathbb{E}[B \mathds{1}_{B > x}]}{\mathbb{P}[Q \ge L(x) + \tilde L(x)]}\\
&= \int_{x - \delta p}^x s \mathrm{d}\tilde L(s) - \tilde L(x) \frac{r \mathbb{E}[B \mathds{1}_{B > x}]}{r\mathbb{P}[B > x] + (1-r)\mathbb{P}[Q^u > L(x) +\tilde{L}(x)]}.
\end{align*}
Integrating by parts we obtain
$$\int_{x - \delta p} ^x s  \mathrm{d}\tilde L(s) = \tilde L(x ) x - \int_{x - \delta p}^x \tilde L(s) \mathrm{d}s =\varepsilon x- \int_{x - \delta p}^x \tilde L(s) \mathrm{d}s.$$
When $\delta p$ tends to 0, the last expression tends to $\varepsilon x$. Consequently, we conclude that
$$\underset{  \delta p \rightarrow 0} {\text{lim}}G(x- \delta p, x) = \varepsilon \left(x - \frac{r \mathbb{E}[B\mathds{1}_{B > x}]}{r\mathbb{P}[B > x] + (1-r)\mathbb{P}[Q^u > L(x) + \tilde{L}(x)]}\right), $$
and
$$G(x) = \underset{\varepsilon \rightarrow 0}{\text{lim}} \left(\underset{\delta p \rightarrow 0}{\text{lim}}\frac{G(x-\delta p, x)}{\varepsilon}\right)\\
 = x - \frac{r\mathbb{E}[B\mathds{1}_{B > x}]}{r\mathbb{P}[B  > x] + (1-r)\mathbb{P}[Q^u  > L(x)] }.$$

\subsection{Proof of Theorem \ref{spread_theorem}}\label{proof}

We consider the passive sell orders ($x>0$). We first compute $\hat{L}(x)$. This is the theoretical liquidity that market makers should add in the LOB in order to obtain $G(x)=0$. Under Proposition \ref{gain}, $G(x)=0$ is equivalent to
\begin{align*}
\mathbb{P}[Q^u > L(x)] &= \frac{r}{1-r}\left( \E[][\frac{B}{x} \mathds{1}_{B>x}] - \mathbb{P}[B > x]\right)\\
&= \frac{r}{1-r}\left( \E[][\frac{B}{x} \mathds{1}_{B>x} ]- 1 + \mathbb{P}[B < x]\right)\\
&= \frac{r}{1-r}\left( -1 + \E[][\text{max}\left(\frac{B}{x}, 1\right)]\right).
\end{align*}

We deduce that
$$\hat{L}(x)=F_{Q^u}^{-1}\left(\frac{1}{1-r}-\frac{r}{1-r}\E[][\max\left(\frac{B}{x},1\right)]\right). $$

We now prove that the spread is positive and finite and deduce the shape of the whole LOB.

Recall that $\hat{L}(x)$, as computed above, is a theoretical value, and that market makers will add liquidity only when $\hat{L}(x)>0$.

We have $\hat{L}(x)>0$ when ($\frac{1}{1-r}-\frac{r}{1-r}\E[][\text{max}(\frac{B}{x},1)] )> \frac{1}{2}$. This holds for all $x$ such that $\mathbb{E}[\text{max}(\frac{B}{x},1)] < \frac{1+r}{2r}$. Equivalently, the inequality is satisfied for any $x$ such that $x>\mu$, where $\mu$ is unique solution of the following equation 
$$\E[][\max\left(\frac{B}{\mu},1\right)] = \frac{1+r}{2r}.$$

By Assumption \ref{zeroprofit}, we deduce that for any $x \le \mu$,
$L(x)=0$. Moreover, for any $x>\mu$,
$$L(x)=F_{Q^u}^{-1}\left(\frac{1}{1-r}-\frac{r}{1-r}\E[][\max\left(\frac{B}{x},1\right)]\right). $$

Therefore, $\mu$ is the half spread.




The cumulative LOB we obtain is unique, continuous and strictly increasing beyond the spread (since the laws of $B$ and $Q^u$ have positive densities on $\mathbb{R}$). 

\subsection{Proof of Proposition \ref{prop:any_shape_attainable}}
\label{subsec:proof_any_shape_attainable}
Define $g:x\in[\mu, \infty)\mapsto \frac{1}{1-r}-\frac{r}{1-r}\E[][\max\left(\frac{B}{x},1\right)]$. Suppose that $g$ is differentiable and define
\begin{equation*}
    f_{Q^u}(y) = (\Lambda^{-1})'(y)g'\left(\Lambda^{-1}(y)\right),\quad y \in [0, \infty).
\end{equation*}
For $y \in (-\infty, 0)$, set $f_{Q^u}(y) = f_{Q^u}(-y)$.
Since $\Lambda$ and $g$ are strictly increasing, $f_{Q^u} > 0$ almost everywhere. In addition,
\begin{equation*}
    \int_{\mathbb{R}}f_{Q^u}(y)\diff y =
    2 \int_0^{\infty}(\Lambda^{-1})'(y)g'\left(\Lambda^{-1}(y)\right) \diff y
    =2 \int_{\mu}^{\infty}g'(x) \diff x = 2 \lim_{x\to\infty} g(x) - 2 g(\mu) = 1.
\end{equation*}
Thus, $f_{Q^u}$ is a strictly positive and symmetric density.
For $x \in [0, \infty)$,
\begin{equation*}
    F_{Q^u}(x) =\frac{1}{2} + \int_0^{x}(\Lambda^{-1})'(y)g'\left(\Lambda^{-1}(y)\right) \diff y = g\left(\Lambda^{-1}(x)\right),
\end{equation*}
hence, for $x \in [\mu, \infty)$, $\Lambda(x) = F_{Q^u}^{-1}(g(x))$, as required. The differentiability of $g$ is established in Lemma \ref{lemma:g_differentiable}.

\begin{lemma}
    \label{lemma:g_differentiable}
    For all $x \in [\mu,  \infty)$, $g(x) = \frac{r}{1-r}\int_{\mu}^{x}\frac{1}{z^2}\int_z^{\infty} y f_B(y) \diff y \diff z$. Consequently, $g$ is continuously differentiable.
\end{lemma}

\begin{proof}
    Let $x \in [\mu,  \infty)$ and observe that
    \begin{align*}
        \int_{\mu}^{x}\frac{1}{z^2}\int_z^{\infty} y f_B(y) \diff y \diff z
        &= \int_{\mu}^{\infty}\int_{\mu}^{\min(x,y)}\frac{y}{z^2}  f_B(y) \diff z \diff y\\
        &=\int_{\mu}^{\infty}y  f_B(y) \left(\frac{1}{\mu} - \frac{1}{\min(x,y)}\right) \diff y \\
        &= \frac{1}{\mu}\int_{\mu}^{\infty}y f_B(y)\diff y  - \int_{\mu}^x f_B(y) \diff y - \int_{x}^{\infty}\frac{y}{x}  f_B(y)\diff y 
    \end{align*}
    By Equation \eqref{equation},
    \begin{equation*}
        \frac{r}{1-r}\int_{\mu}^{\infty}\frac{y}{\mu} f_B(y)\diff y = \frac{1}{1-r} - \frac{r}{1-r}\int_{-\infty}^{\mu}f_B(y)\diff y.
    \end{equation*}
    Combining these observations, we conclude that
    \begin{equation*}
        \frac{r}{1-r}\int_{\mu}^{x}\frac{1}{z^2}\int_z^{\infty} y f_B(y) \diff y \diff z
        =
        \frac{1}{1-r} - \frac{r}{1-r}\int_{-\infty}^{x}f_B(y)\diff y - \frac{r}{1-r} \int_{x}^{\infty}\frac{y}{x}  f_B(y)\diff y = g(x).
    \end{equation*}
\end{proof}

\begin{remark}
    Lemma \ref{lemma:g_differentiable} holds even if $f_B$ is not supposed to be strictly positive almost everywhere.
\end{remark}

\subsection{Proof of Proposition \ref{prop:shape_tail}}
\label{sec:proof_shape_tail}

Define $g$ as in the beginning of Section \ref{subsec:proof_any_shape_attainable}. We start by computing an asymptotic equivalent for $g$.

\begin{lemma}
    \label{lem:equivalent_g}
    Suppose that $1-F_B(x) \underset{x\to \infty}{\sim} c x^{-a}$ for some constants $c> 0$ and $a > 1$. Then
    \begin{equation*}
        1- g(x) \underset{x\to \infty}{\sim} \frac{r}{1-r}\frac{c}{a-1} x^{-a}.
    \end{equation*}
\end{lemma}

\begin{proof}
    Let $x \in \mathbb{R}$. We have, using Fubini's theorem,
\begin{equation*}
    \int_x^{\infty}(1-F_B(y)) \diff y
    = \int_x^{\infty}\int_{y}^{\infty}f_B(z) \diff z \diff y = \int_x^{\infty}(z-x)f_B(z)\diff z = \int_x^{\infty}zf_B(z)\diff z - x(1-F_B(x)).
\end{equation*}
By the definition of $g$,
\begin{equation*}
    g(x) = \frac{1}{1-r} - \frac{r}{1-r}F_B(x) - \frac{r}{1-r}\frac{1}{x}\int_x^{\infty}yf_B(y) \diff y,
\end{equation*}
therefore
\begin{equation*}
    g(x) = 1 - \frac{r}{1-r}\frac{1}{x}\int_x^{\infty}(1-F_B(y)) \diff y.
\end{equation*}
Integrating the equivalents, we have $\int_x^{\infty}(1-F_B(y)) \diff y \underset{x\to \infty}{\sim} \frac{c}{a-1}x^{1-a}$. The desired result follows immediately.
\end{proof}

We first prove case (i) where $1-F_{Q^u}(x) \underset{x\to \infty}{\sim} c'x^{-b}$ for some $c', b > 0$.
Let $x \in \mathbb{R}$. Since $g(x) = F_{Q^u}\left(F_{Q^u}^{-1}(g(x))\right)$ and $\lim_{x \to \infty}F_{Q^u}^{-1}(g(x)) = \infty$,
\begin{equation*}
    g(x) = 1 - c'F_{Q^u}^{-1}(g(x))^{-b} + o_{x\to \infty}\left(F_{Q^u}^{-1}(g(x))^{-b}\right).
\end{equation*}
Thus, by Lemma \ref{lem:equivalent_g}
\begin{equation*}
    c'F_{Q^u}^{-1}(g(x))^{-b} \underset{x\to \infty}{\sim} 1 - g(x) \underset{x\to \infty}{\sim} \frac{r}{1-r}\frac{c}{a-1}x^{-a}
\end{equation*}
and the desired result follows (recall that $L(x) = F_{Q^u}^{-1}(g(x))$).

For case (ii), suppose now that $Q^u$  has a normal distribution with variance $\sigma^2$. By \cite[Equation (3.41)]{nolan2020univariate},  $1-F_{Q^u}(x) \underset{x\to \infty}{\sim} \frac{\sigma}{x \sqrt{2\pi}}e^{-\frac{x^2}{2 \sigma^2}}$.

Let $x \in \mathbb{R}$. Since $g(x) = F_{Q^u}\left(F_{Q^u}^{-1}(g(x))\right)$ and $\lim_{x \to \infty}F_{Q^u}^{-1}(g(x)) = \infty$,
\begin{equation*}
    g(x) = 1 - \frac{\sigma}{F_{Q^u}^{-1}(g(x)) \sqrt{2\pi}}e^{-\frac{F_{Q^u}^{-1}(g(x))^2}{2 \sigma^2}}\left(1  + o_{x\to \infty}\left(1\right)\right).
\end{equation*}
Thus, by Lemma \ref{lem:equivalent_g},
\begin{equation*}
    \frac{\sigma}{F_{Q^u}^{-1}(g(x)) \sqrt{2\pi}}e^{-\frac{F_{Q^u}^{-1}(g(x))^2}{2 \sigma^2}}\left(1  + o_{x\to \infty}\left(1\right)\right)
     = \frac{r}{1-r}\frac{c}{a-1} x^{-a}\left(1  + o_{x\to \infty}\left(1\right)\right).
\end{equation*}
Taking the logarithm on both sides of the last equation, we obtain
\begin{equation*}
    -\frac{F_{Q^u}^{-1}(g(x))^2}{2 \sigma^2}
    +\ln(\sigma)- \ln\left(F_{Q^u}^{-1}(g(x)) \sqrt{2\pi}\right) + o_{x \to \infty}(1)
    = \ln\left(\frac{r}{1-r} \frac{c}{a-1}\right)
    -a \ln(x) + o_{x\to \infty}(1).
\end{equation*}
Since $\ln(\sigma)-\ln\left(F_{Q^u}^{-1}(g(x)) \sqrt{2\pi}\right) = o_{x \to \infty} \left(F_{Q^u}^{-1}(g(x))^2\right)$ and $\ln\left(\frac{r}{1-r} \frac{c}{a-1}\right) = o_{x \to \infty}(\ln(x))$,
\begin{equation*}
    \frac{F_{Q^u}^{-1}(g(x))^2}{2 \sigma^2}
    \underset{x\to \infty}{\sim}
    a \ln(x)
\end{equation*}
and the desired result follows.

\subsection{Proof of Proposition \ref{prop:invariant_fractional_part}}
\label{subsec:proof_invariant_frac_part}

Let $t \geqslant 0$. We have
\begin{equation*}
    d(t) = d(0) - \left(P(t) - P(0)\right) + \alpha \ceil*{\frac{-d(0)+P(t) - P(0)}{\alpha}}.
\end{equation*}
Thus, for $s \in [0, t]$,
\begin{align*}
    d(t) &= d(0) - \left(P(s) - P(0)\right) - \left(P(t) - P(s)\right)+ \alpha \ceil*{\frac{P(s) - P(0)}{\alpha} + \frac{P(t) - P(s)}{\alpha}}.\\
    &= d(s) - \left(P(t) - P(s)\right)+ \alpha \ceil*{\frac{-d(0)+P(s) - P(0)}{\alpha} + \frac{P(t) - P(s)}{\alpha}} - \alpha \ceil*{\frac{-d(0)+P(s) - P(0)}{\alpha}}\\
    &= d(s) - \left(P(t) - P(s)\right)+ \alpha \ceil*{-\frac{d(s)}{\alpha} + \frac{P(t) - P(s)}{\alpha}}.
\end{align*}
For $z \in [0, \alpha)$, $s \geqslant 0$, and $A \subset (0, \alpha)$ a measurable set, we define 
\begin{equation*}
    Q_s(z, A) \defeq \Proba[][z - (P(s)-P(0)) + \alpha\ceil*{-\frac{z}{\alpha} + \frac{P(s)-P(0)}{\alpha}} \in A].
\end{equation*}
For such $A$ and $s \leqslant t$, since $P(t) - P(s)$ is independent of $\mathcal{F}_s$ and has the same law as $P(s) - P(0)$,
\begin{equation*}
    \Proba[][d(t) \in A| \mathcal{F}_s]
    = Q_{t-s}(d(s), A).
\end{equation*}
Thus, $(d(t))_{t\geqslant 0}$ is a time-homogenous Markov process with $Q$ as transition kernel.

The claim about the invariant measure of $(d(t))_{t\geqslant 0}$ follows from Lemma \ref{lem:equivalent_invariant}, which provides conditions for a probability measure to be invariant. This reduces the problem to studying discrete-time Markov chain invariance.

\begin{lemma}
    \label{lem:equivalent_invariant}
    A probability measure is invariant for $(d(t))_{t\geqslant 0}$ if and only if for all $n \in \mathbb{N}^*$, it is invariant for the (discrete-time) Markov kernel $Q^{(n)}$ defined by
    \begin{equation*}
        Q^{(n)}(z, A) \defeq \int_{\mathbb{R}} \mathds{1}_A \left(z - y + \alpha\ceil*{-\frac{z}{\alpha} + \frac{y}{\alpha}}\right) f_B^{\ast n}(y)\diff y,\quad  z \in [0,\alpha),\ A \subset [0,\alpha) \text{ measurable}.
    \end{equation*}
\end{lemma}

\begin{proof}
    Let $z \in [0, \alpha)$, $t \geqslant 0$ and $A \subset (0, \alpha)$ be a measurable set. Using explicitly the law of $P(t)-P(0)$ we obtain
\begin{equation*}
    Q_t(z, A) = e^{-\lambda^i t} + e^{-\lambda^i t}\sum_{n=1}^{\infty}\frac{(\lambda^i t)^n}{n!}\int_{\mathbb{R}}\mathds{1}_A \left(z - y + \alpha\ceil*{-\frac{z}{\alpha} + \frac{y}{\alpha}}\right) f_B^{\ast n}(y)\diff y.
\end{equation*}
Let $\nu$ be a probability measure on $[0,\alpha)$. Using the previous equality yields
\begin{equation*}
    \int_{[0,\alpha)} Q_t(z, A) \nu(\diff z)
    = e^{-\lambda^i t}\nu(A) + e^{-\lambda^i t}\sum_{n=1}^{\infty}\frac{(\lambda^i t)^n}{n!} \int_{[0,\alpha)} Q^{(n)}(z, A) \nu(\diff z),
\end{equation*}
which implies
\begin{equation*}
    \int_{[0,\alpha)} Q_t(z, A) \nu(\diff z)\sum_{n=0}^{\infty}\frac{(\lambda^i t)^n}{n!} = \nu(A) + \sum_{n=1}^{\infty}\frac{(\lambda^i t)^n}{n!} \int_{[0,\alpha)} Q^{(n)}(z, A) \nu(\diff z).
\end{equation*}
Identifying the terms in the power series, we obtain the desired result.
\end{proof}

We first show that the uniform distribution is invariant for $(d(t))_{t\geqslant 0}$. Let $n \in \mathbb{N}^*$ and let $A \subset (0, \alpha)$ be a measurable set. Fix $y \in \mathbb{R}$. We write $y = m\alpha +p$ with $m \in \mathbb{Z}$, $p \in [0, \alpha)$. Then,
\begin{align*}
    \int_0^{\alpha} \mathds{1}_A \left(z - y + \alpha\ceil*{-\frac{z}{\alpha} + \frac{y}{\alpha}}\right) \diff z
    &= \int_0^p \mathds{1}_A \left(z - y + m\alpha + \alpha\right) \diff z + \int_p^{\alpha}\mathds{1}_A \left(z - y + m\alpha\right) \diff z\\
    &= \int_{m \alpha + \alpha}^{y+\alpha} \mathds{1}_A \left(z - y \right) \diff z + \int_y^{m\alpha+\alpha}\mathds{1}_A \left(z - y \right) \diff z\\
    &=\int_0^{\alpha}\mathds{1}_A \left(z\right) \diff z.
\end{align*}
Thus,
\begin{equation*}
    \frac{1}{\alpha}\int_0^{\alpha}Q^{(n)}(z, A) \diff z
    = \frac{1}{\alpha}\int_0^{\alpha}\mathds{1}_A \left(z\right) \diff z \int_{\mathbb{R}}f_B^{\ast n}(y)\diff y
    = \frac{1}{\alpha}\int_0^{\alpha}\mathds{1}_A \left(z\right) \diff z.
\end{equation*}
Thanks to Lemma \ref{lem:equivalent_invariant}, we conclude that the uniform distribution on $[0,\alpha)$ is invariant for $(d(t))_{t\geqslant 0}$.\\

To show uniqueness, it is sufficient to show that for some $n \in \mathbb{N}^*$, a Markov chain with transition kernel $Q^{(n)}$ is Harris recurrent. Indeed, if this is the case, then, by \cite[Theorem 10.0.01]{meyn2009markov}, $Q^{(n)}$ admits a unique invariant measure, and consequently, by Lemma \ref{lem:equivalent_invariant}, $(d(t))_{t\geqslant 0}$ admits at most one invariant measure.

The proof of the Harris recurrence property is based on Lemma \ref{lem:sum_variables_smoothing}.
\begin{lemma}
    \label{lem:sum_variables_smoothing}
    There exists $n \in \mathbb{N}^*$, $\delta > 0$ and $\beta \in \mathbb{R}$ such that $f_B^{\ast n} \geqslant \delta$ almost everywhere on $[\beta, \beta+\alpha]$.
\end{lemma}

\begin{proof}
    Since for all $n \in \mathbb{N}^*$, $f_B^{\ast n} \geqslant \min(f_B, 1)^{\ast n}$, we assume from now on (without loss of generality), that $f_B \leqslant 1$. We do not use the fact that $f_B$ integrates to 1, just that it is strictly positive.

Since $f_B$ is square integrable (thanks to our assumption), $f_B^{\ast 2}$ is almost everywhere equal to a continuous function, which is the version we use from now on. $f_B^{\ast 2}$ is non-negative and not identically zero because $\int f_B^{\ast 2} = \left(\int f_B\right)^2 > 0$. Hence, there exists an interval $[a, b]$ with $a < b$ and $\varepsilon > 0$ such that $f_B^{\ast 2} \geqslant \varepsilon$ on $[a,b]$. Thus,
\begin{equation}
    \label{eq:convolution_minoration}
    f_B^{\ast 2n} \geqslant \varepsilon^{n}\mathds{1}_{[a,b]}^{\ast n},\quad n \in \mathbb{N}^*.
\end{equation}

We will show that for all $n \in \mathbb{N}^*$, $\mathds{1}_{[a,b]}^{\ast n} > 0$ in $(na, nb)$. Then, for $n \in \mathbb{N}^*$ big enough, by the continuity of $\mathds{1}_{[a,b]}^{\ast n}$, there exists $c(n) > 0$ and an interval of length bigger than $\alpha$ such that $\mathds{1}_{[a,b]}^{\ast n} > c(n)$ on that interval, which implies the claim  we wanted to prove by injecting the last inequality in \eqref{eq:convolution_minoration}.

The claim is true for $n=1$. Let $n \geqslant 2$ and $x \in (na, nb)$. Let $\eta > 0$ be such that $[x-(n-1)\eta, x+(n-1)\eta] \subset (na, nb)$. In particular, $\left[\frac{x}{n}-\frac{\eta}{n}, \frac{x}{n} + \frac{\eta}{n}\right]^{n-1} \subset \left[\frac{x}{n}-\frac{(n-1)\eta}{n}, \frac{x}{n} + \frac{(n-1)\eta}{n}\right]^{n-1} \subset (a,b)^{n-1}$. We have
\begin{align*}
    \mathds{1}_{[a,b]}^{\ast n}(x)
    &= \int_{\mathbb{R}^{n-1}} \mathds{1}_{[a,b]}(x-y_1-\dots-y_{n-1})
    \mathds{1}_{[a,b]}(y_1)\dots \mathds{1}_{[a,b]}(y_{n-1})\diff y_1 \dots  \diff y_{n-1}\\
    &= \int_{[a,b]^{n-1}} \mathds{1}_{[a,b]}(x-y_1-\dots-y_{n-1})
    \diff y_1 \dots  \diff y_{n-1}\\
    &\geqslant \int_{\left[\frac{x}{n}-\frac{\eta}{n}, \frac{x}{n} + \frac{\eta}{n}\right]^{n-1}} \mathds{1}_{[a,b]}(x-y_1-\dots-y_{n-1})
    \diff y_1 \dots  \diff y_{n-1}.
\end{align*}

Let $(y_1,\dots,y_{n-1})\in \left[\frac{x}{n}-\frac{\eta}{n}, \frac{x}{n} + \frac{\eta}{n}\right]^{n-1}$. Then,
\begin{align*}
    a < \frac{x}{n} -\frac{n-1}{n}\eta \leqslant x-y_1-\dots-y_{n-1} \leqslant \frac{x}{n} + \frac{n-1}{n}\eta <b .
\end{align*}
Thus,
\begin{equation*}
    \mathds{1}_{[a,b]}^{\ast n}(x) \geqslant \int_{\left[\frac{x}{n}-\frac{\eta}{n}, \frac{x}{n} + \frac{\eta}{n}\right]^{n-1}} 1 \diff y_1 \dots  \diff y_{n-1} > 0,
\end{equation*}
which is what we wanted to prove.
\end{proof}

Let $n, \delta$ and $\beta$ be given by Lemma \ref{lem:sum_variables_smoothing}. Let $z \in \mathbb{R}$ and let $A \subset (0, \alpha)$ be a measurable set. Then,
\begin{equation*}
    Q^{(n)}(z, A) \geqslant \delta\int_{\beta}^{\beta+\alpha} \mathds{1}_A \left(z - y + \alpha\ceil*{-\frac{z}{\alpha} + \frac{y}{\alpha}}\right) \diff y
    = \delta \int_0^{\alpha} \mathds{1}_A(y) \diff y,
\end{equation*}
the last equality follows from a similar computation as in the proof of invariance of the uniform measure. Hence, a Markov chain $\Phi$ with $Q^{(n)}$ as transition kernel is $\phi$-irreducible and a T-chain, $\phi$ being the Lebesgue measure, see \cite[Parts 1 and 2]{meyn1992stability}. Endowing $[0, \alpha)$ with the topology of the torus (i.e. with a distance $D(x,y)=\min(|x-y|, \alpha - |x-y|)$) makes $[0, \alpha)$ compact. Furthermore, this topology has the same Borel sets as the standard one. In this topology, $[0,\alpha)$ is compact, hence $\Phi$ is bounded in probability \cite[Topological stability condition 2]{meyn1992stability}. Thus, by \cite[Corollary following Theorem 2.1]{meyn1992stability}, $\Phi$ is positive Harris recurrent, which is what we wanted to prove.

\subsection{Proof of Proposition \ref{gain tick}}\label{proofproptick}

We provide only a sketch of the proof, as the computations are essentially the same as in the proof of Proposition \ref{gain}. In particular, we omit the introduction of the limit order volume $\varepsilon$ and directly work in the asymptotic regime where $\varepsilon$ tends to zero. We analyze the gain of passive sell orders, noting that the gain for passive buy orders can be derived analogously.

First, we compute $G^d_{inf}(i)$, which denotes the gain of a new order placed at the $i^{th}$ limit when the trade is initiated by an informed trader, knowing that $Q^i > L^d(i)$.*,
$$G^d_{inf}(i) = d + (i-1)\alpha - \mathbb{E}[B|B > d + (i-1)\alpha ].$$
Second, we compute $G^d_{noise}(i)$, which denotes the gain of a new order placed at the $i^{th}$ limit when the trade is initiated by a noise trade, knowing that $Q^u > L^d(i)$,
$$G^d_{noise}(i) = d + (i-1)\alpha. $$
Therefore, $G^d(i)$ is given by   
\begin{align*}
G^d(i) & =  G^d_{inf}(i) \mathbb{P}[\nu = 1|Q > L^d(i)]+ G^d_{noise}(i) \mathbb{P}[\nu =0|Q > L^d(i)]\\
& = d + (i-1)\alpha - \frac{r\mathbb{E}[B\mathds{1}_{B > d + (i-1)\alpha}]}{\mathbb{P}[Q > L(d + (i-1)\alpha])}\\
& = d + (i-1)\alpha - \frac{r\mathbb{E}[B\mathds{1}_{B > d + (i-1)\alpha}]}{r\mathbb{P}[B > d + (i-1)\alpha] + (1-r)\mathbb{P}[Q^u > L(d + (i-1)\alpha)]}.\\
\end{align*}

\subsection{Proof of Theorem \ref{spread_lob_tick}}\label{slt}

We consider the ask side.  We first show that the spread is positive and finite.  Then we prove that beyond the spread, market makers insert limit orders on all possible limit prices.

We showed in the null-tick size case that there exists $\mu$ such that for all $x \le \mu, L(x)=0$ and  for all $x > \mu, L(x) >0$. The LOB being now discrete, the previous findings remain true for $k_r^d$ instead of $\mu$ where $k_r^d$ satisfies 

$$k_r^d=\text{min}\{k\in \mathbb{N}^+|d+(k-1)\alpha > \mu \}.$$

Equivalently,
$$k_r^d= 1 + \ceil*{\frac{\mu - d}{\alpha}} .$$
Similarly, for the first non-empty limit at the bid side, we get
$$k_l^d= \ceil*{\frac{\mu + d}{\alpha}}.$$

From Equation \eqref{eq_def_descrete}, the spread is equal to $(k_r^d+k_l^d) \alpha - \alpha$. Thus, the conditional constrained bid-ask spread $\phi_{\alpha}^d$, given the value of $d$, is given by
$$\phi_{\alpha}^d=\alpha \left(\ceil*{\frac{\mu - d}{\alpha}} + \ceil*{\frac{\mu + d}{\alpha}}\right).$$

Under Assumption \ref{zeroprofit tick}, we have for any $i \ge k_r^d$, $$L(d+(i-1)\alpha)=F_{Q^u}^{-1}\left(\frac{1}{1-r}-\frac{r}{1-r}\E[][\text{max}\left(\frac{B}{d+(i-1)\alpha},1\right)]\right). $$

We deduce that the cumulative LOB is unique and increasing beyond the spread. 

\subsection{Proof of Corollary \ref{cor_spread}}\label{pcs}

The parameter $d$ being approximately uniformly distributed between $[0,\alpha)$, we can compute the average value of the
constrained bid-ask spread by integrating $\phi_{\alpha}^d$

$$\phi_{\alpha}=\int_0^{\alpha} \lceil\frac{\mu-s}{\alpha}\rceil + \lceil \frac{\mu+s}{\alpha} \rceil \text{d}s.$$
Let $u:=\frac{\mu}{\alpha}$. We have

$$\phi_u=\alpha \int_0^{1} \lceil u-x \rceil + \lceil u+x \rceil \text{d}x. $$

We decompose $u$ as $u=u_i+u_f$, where $u_i$ represents the integer part of $u$. We obtain

\begin{align*}
    \phi_{\alpha}&=\alpha \int_0^{1} \lceil u_i+u_f-x \rceil + \lceil u_i+u_f+x \rceil \text{d}x\\
\phi_{\alpha}&=\alpha\left(\int_0^{u_f}(u_i+1)\text{d}x+\int_{u_f}^{1}u_i\text{d}x+\int_0^{1-u_f}(u_i+1)\text{d}x+\int_{(1-u_f)}^{1}(u_i+2)\text{d}x\right)\\
\phi_{\alpha}&=\alpha\left(u_f(u_i+1)+(1-u_f)u_i+(1-u_f)(u_i+1)+u_f(u_i+2)\right)\\
\phi_{\alpha}&=\alpha(2u_i+2u_f+1) =\alpha+2\mu =\alpha+\phi.
\end{align*}

\subsection{Proof of Proposition \ref{prop:likelihood}}
\label{sec:proof_formula_likelihood}

Throughout the proof, we use the following notations: $\Delta \tilde{P} \defeq \tilde{P}(t+\Delta t) - \tilde{P}(t)$ and $\Delta P \defeq P(t+\Delta t) - P(t)$. Observe that

\begin{align*}
    \Proba[][\Delta \tilde{P} = k\alpha]
    &= \Proba[][-d(t) + \Delta P \in ((k-1)\alpha, k\alpha)]\\
    &= \frac{1}{\alpha} \int_{0}^{\alpha} \Proba[][-u + \Delta P \in ((k-1)\alpha, k\alpha)] \diff u.
\end{align*}
By the Lévy-Khintchine formula, the characteristic function of $\Delta P$ is given by
\begin{equation*}
    \E[][\exp\left(iz \Delta P\right)] = \exp\left( \lambda^i \Delta t \left(\hat{f}_{B}(z) -1\right)\right),\quad z \in \mathbb{R}.
\end{equation*}
The characteristic function is not necessarily integrable, we only know it is bounded by $\exp(2\lambda^i \Delta t)$. Thus, we cannot invert it directly, instead, we will use test functions.

Let $\phi \in C^{\infty}(\mathbb{R})$ supported in $[0, 1]$, non-negative-valued and with integral equal to 1. For $n \in \mathbb{N}^*$, we define $\chi_n$ by
\begin{equation*}
    \chi_n (x) = \int_{-n(x-(k-1)\alpha)}^{\infty} \phi(y) \diff y - \int^{n(x-k\alpha)}_{-\infty} \phi(y) \diff y, \quad x \in \mathbb{R}.
\end{equation*}
Note that $\chi_n \in C^{\infty} (\mathbb{R})$, $\chi_n$'s support is included in $\left[(k-1)\alpha - \frac{1}{n}, k\alpha + \frac{1}{n}\right]$, and $\chi_n$ takes the value 1 in $\left[(k-1)\alpha, k\alpha\right]$, is non-decreasing on $\left[(k-1)\alpha - \frac{1}{n}, (k-1)\alpha\right]$ and non-increasing on $\left[k\alpha , k\alpha + \frac{1}{n}\right]$. For $x \in \left[(k-1)\alpha - \frac{1}{n}, (k-1)\alpha\right]$, $\chi_n'(x)=n\phi(-n(x-(k-1)\alpha))$ and $\chi_n''(x)=-n^2\phi'(-n(x-(k-1)\alpha))$. For $x \in \left[k\alpha , k\alpha + \frac{1}{n}\right]$, $\chi_n'(x)=-n\phi(n(x-k\alpha))$ and $\chi_n''(x)=n^2\phi'(n(x-k\alpha))$.

By dominated convergence,
\begin{equation*}
    \Proba[][\Delta\tilde{P} = k\alpha]
    = \lim_{n \to \infty}
    \frac{1}{\alpha} \int_{0}^{\alpha} \E[][\chi_n\left(-u + \Delta P\right)]\diff u.
\end{equation*}

Fix $n \in \mathbb{N}^*$. By the Fourier inversion formula applied to the test function $\chi_n(-u+\cdot)$,
\begin{equation*}
    \int_0^{\alpha}\E[][\chi_n\left(-u + \Delta P\right)]\diff u
    = \int_0^{\alpha} \int_{\mathbb{R}} e^{\lambda^i \Delta t \left(\hat{f}_{B}(z) -1\right)}\left(\frac{1}{2\pi}\int_{\mathbb{R}} \chi_n(-u + y)e^{-izy}\diff y\right) \diff z \diff u.
\end{equation*}

Since
\begin{align*}
    \int_0^{\alpha} \int_{\mathbb{R}} \Bigg|e^{\lambda^i \Delta t \left(\hat{f}_{B}(z) -1\right)}&\left(\frac{1}{2\pi}\int_{\mathbb{R}} \chi_n(-u + y)e^{-izy}\diff y\right)\Bigg| \diff z \diff u\\
    &\leqslant \frac{ e^{2\lambda^i \Delta t}}{2\pi} \int_0^{\alpha} \int_{\mathbb{R}} \left|\left(\int_{\mathbb{R}} \chi_n(y)e^{-izy}e^{-izu}\diff y\right)\right|\diff z \diff u\\
    &\leqslant \frac{ e^{2\lambda^i \Delta t}}{2\pi}\alpha \int_{\mathbb{R}} \left|\int_{\mathbb{R}} \chi_n(y)e^{-izy}\diff y\right|\diff z\\
    &< \infty,
\end{align*}
(the last inequality holds because $z \mapsto \int_{\mathbb{R}} \chi_n(z)e^{-izy}\diff y$ belongs to the Schwartz space, and it is therefore integrable), we can apply the Fubini theorem to deduce
\begin{align*}
    \int_0^{\alpha}\E[][\chi_n\left(-u + \Delta P\right)]\diff u
    &= \int_{\mathbb{R}}  e^{\lambda^i \Delta t \left(\hat{f}_{B}(z) -1\right)}\int_0^{\alpha}e^{-izu}\left(\frac{1}{2\pi}\int_{\mathbb{R}} \chi_n( y)e^{-izy}\diff y\right) \diff u\diff z\\
    &=\int_{\mathbb{R}}  e^{\lambda^i \Delta t \left(\hat{f}_{B}(z) -1\right)}\frac{e^{-\frac{i\alpha z}{2}}}{z}\sin\left(\frac{z \alpha}{2}\right)\left(\frac{1}{\pi}\int_{\mathbb{R}} \chi_n( y)e^{-izy}\diff y\right) \diff z.
\end{align*}

Let $z \in \mathbb{R}$. We have
\begin{align*}
    \int_{(k-1) \alpha}^{k \alpha} \chi_n( y)e^{-izy}\diff y
    = \int_{(k-1) \alpha}^{k \alpha} e^{-izy}\diff y
    = 2 \frac{e^{-ik\alpha z+\frac{i\alpha z}{2}}}{z}\sin\left(\frac{\alpha z}{2}\right).
\end{align*}

Since $z \mapsto \frac{2}{\pi}e^{\lambda^i \Delta t \left(\hat{f}_{B}(z) -1\right)}\frac{e^{-ik\alpha z}}{z^2}\sin\left(\frac{z \alpha}{2}\right)^2$ is integrable, the following function is also integrable
\begin{equation*}
    z \mapsto e^{\lambda^i \Delta t \left(\hat{f}_{B}(z) -1\right)}\frac{e^{-\frac{i\alpha z}{2}}}{z}\sin\left(\frac{z \alpha}{2}\right)\left(\frac{1}{\pi}\int_{\mathbb{R} \setminus [(k-1)\alpha, k\alpha]} \chi_n(y)e^{-izy}\diff y\right).
\end{equation*}

Defining $h_n:z \mapsto \sqrt{|z|}\int_{\mathbb{R} \setminus [(k-1)\alpha, k\alpha]} \chi_n(y)e^{-izy}\diff y$, we have
\begin{equation*}
    \begin{split}
        \int_0^{\alpha}\E[][\chi_n\left(-u + \Delta P\right)]\diff u
        = \frac{2}{\pi}\int_{\mathbb{R}} &e^{\lambda^i \Delta t \left(\hat{f}_{B}(z) -1\right)}\frac{e^{-ik\alpha z}}{z^2}\sin\left(\frac{z \alpha}{2}\right)^2 \diff z\\
        &+ \frac{1}{\pi}\int_{\mathbb{R}} e^{\lambda^i \Delta t \left(\hat{f}_{B}(z) -1\right)}\frac{e^{-\frac{i\alpha z}{2}}}{|z|^{\frac{3}{2}}}\sin\left(\frac{z \alpha}{2}\right) h_n(z) \diff z.
    \end{split}
\end{equation*}

The goal now is to show that the second term tends to 0 as $n$ tends to infinity.

The function $z \mapsto \frac{1}{z}\sin\left(\frac{z \alpha}{2}\right)$ is bounded near 0 and $z \mapsto \frac{1}{\sqrt{|z|}}e^{\lambda^i \Delta t \left(\hat{f}_{B}(z) -1\right)}$ is integrable on $[-1, 1]$. Furthermore, $z \mapsto e^{\lambda^i \Delta t \left(\hat{f}_{B}(z) -1\right)}\frac{e^{-\frac{i\alpha z}{2}}}{|z|^{\frac{3}{2}}}\sin\left(\frac{z \alpha}{2}\right)$ is integrable on $\mathbb{R}\setminus [-1, 1]$. Hence, to show the desired result, it is sufficient to show that $h_n(z)$ is bounded uniformly in $n$ and $z$, and that for each $z \in \mathbb{R}^*$, $h_n(z) \to 0$ as $n$ goes to infinity. Dominated convergence then allows us to conclude.

Let $z \in \mathbb{R}^*$. Using integration by parts, and the fact that $\chi_n\left((k-1)\alpha - \frac{1}{n}\right) = \chi_n\left(k\alpha + \frac{1}{n}\right) = 0$, $\chi'_n\left((k-1)\alpha - \frac{1}{n}\right) = \chi'_n\left(k\alpha + \frac{1}{n}\right) = 0$, $\chi_n\left((k-1)\alpha\right) = \chi_n\left(k\alpha\right) = 1$, $\chi'_n\left((k-1)\alpha \right) = \chi'_n\left(k\alpha\right) = 0$, we obtain
\begin{align}
    \label{eq:hn_integration_parts}
    h_n(z) &= \frac{i}{\sqrt{|z|}} \left(e^{-iz(k-1)\alpha} - e^{-izk\alpha} - \int_{(k-1)\alpha-\frac{1}{n}}^{(k-1) \alpha} \chi'_n(y)e^{-izy}\diff y- \int_{k\alpha}^{k \alpha + \frac{1}{n}} \chi'_n(y)e^{-izy}\diff y\right)\\
    &= \frac{1}{\sqrt{|z|}} \left(ie^{-iz(k-1)\alpha} - ie^{-izk\alpha} - \frac{1}{z}\int_{(k-1)\alpha-\frac{1}{n}}^{(k-1) \alpha} \chi''_n(y)e^{-izy}\diff y- \frac{1}{z}\int_{k\alpha}^{k \alpha + \frac{1}{n}} \chi''_n(y)e^{-izy}\diff y\right).\nonumber
\end{align}
Since $\int_{(k-1)\alpha-\frac{1}{n}}^{(k-1) \alpha} \chi'_n(y)\diff y = 1$ and  $\int_{k\alpha}^{k \alpha + \frac{1}{n}} \chi'_n(y)\diff y = -1$, \eqref{eq:hn_integration_parts} can be rewritten as
\begin{equation*}
    h_n(z) = \frac{i}{\sqrt{|z|}} \left(e^{-iz(k-1)\alpha}\int_0^{\frac{1}{n}}n\phi(1-ny)(1-e^{-izy + \frac{iz}{n}}) \diff y - e^{-izk\alpha} \int_{0}^{\frac{1}{n}} n\phi(ny)(1-e^{-izy})\diff y\right).
\end{equation*}
For all $y \in \mathbb{R}$, $|\chi''_n(y)| \leqslant n^2 \sup|\phi'|$, $\int_0^1 \phi(y)\diff y = 1$ and $\phi \geqslant 0$. The following two inequalities follow
\begin{align}
    \label{eq:hn_ineq_small_z}
    |h_n(z)| &\leqslant \frac{2}{\sqrt{|z|}}\sup_{-\frac{1}{n} \leqslant y \leqslant \frac{1}{n}}\left|1 - e^{-izy} \right| 
    \leqslant 2 \frac{\sqrt{|z|}}{n},\\
    \label{eq:hn_ineq_big_z}
    |h_n(z)| &\leqslant \frac{2}{\sqrt{|z|}} + \frac{2n}{|z|^{\frac{3}{2}}}\sup|\phi'|.
\end{align}
From \eqref{eq:hn_ineq_small_z}, we deduce that $\lim\limits_{n \to \infty} h_n(z) = 0$ and that for $z \leqslant n^{\frac{2}{3}}$, $|h_n(z)| \leqslant 2$. From \eqref{eq:hn_ineq_big_z}, we have that for $z \geqslant n^{\frac{2}{3}}$, $|h_n(z)| \leqslant 2 + 2 \sup|\phi'|$.

\subsection{Proof of Proposition \ref{prop:identifiability}}
\label{sec:proof_identifiability}

Rearranging the integrals, and noticing that $z \mapsto e^{\lambda_1 \left(e^{-\sigma_1^{a_1} |z|^{a_1}} -1\right)}\frac{1}{z^2}\sin\left(\frac{z \alpha}{2}\right)^2 \cos\left(k \alpha z\right)$ is an even function on $\mathbb{R}$, we have that for all $k \in \mathbb{N}$,
\begin{equation*}
    \begin{split}
        \int_{-\frac{\pi}{\alpha}}^{\frac{\pi}{\alpha}}
    \Bigg(\sum_{n=-\infty}^{\infty} &e^{\lambda_1 \left(e^{-\sigma_1^{a_1} \left|z + \frac{2n\pi}{\alpha}\right|^{a_1}} -1\right)}\frac{\sin\left(\frac{z \alpha}{2}\right)^2}{\left(z + \frac{2n\pi}{\alpha}\right)^2}\Bigg)\cos\left(k \alpha z\right) \diff z \\
    &=\int_{-\frac{\pi}{\alpha}}^{\frac{\pi}{\alpha}}
    \left(\sum_{n=-\infty}^{\infty} e^{\lambda_2 \left(e^{-\sigma_2^{a_2} \left|z + \frac{2n\pi}{\alpha}\right|^{a_1}} -1\right)}\frac{\sin\left(\frac{z \alpha}{2}\right)^2}{\left(z + \frac{2n\pi}{\alpha}\right)^2}\right)\cos\left(k \alpha z\right) \diff z .
    \end{split}
\end{equation*}
The integrands having identical Fourier series (they are even functions therefore the Fourier coefficients in sine are zero) and being continuous, they are equal, that is for all $z \in \left[-\frac{\pi}{\alpha},\frac{\pi}{\alpha}\right]$,
\begin{equation*}
    \sum_{n=-\infty}^{\infty} e^{\lambda_1 \left(e^{-\sigma_1^{a_1} \left|z + \frac{2n\pi}{\alpha}\right|^{a_1}} -1\right)}\frac{\sin\left(\frac{z \alpha}{2}\right)^2}{\left(z + \frac{2n\pi}{\alpha}\right)^2}
    =
    \sum_{n=-\infty}^{\infty} e^{\lambda_2 \left(e^{-\sigma_1^{a_2} \left|z + \frac{2n\pi}{\alpha}\right|^{a_2}} -1\right)}\frac{\sin\left(\frac{z \alpha}{2}\right)^2}{\left(z + \frac{2n\pi}{\alpha}\right)^2}.
\end{equation*}
This leads to the equality
\begin{equation}
    \label{eq:equality_periodized}
    \begin{split}
        e^{\lambda_1 \left(e^{-\sigma_1^{a_1} |z|^{a_1}} -1\right)}&+
    z^2\sum_{n \in \mathbb{Z}\setminus \{0\}} \frac{e^{\lambda_1 \left(e^{-\sigma_1^{a_1} \left|z + \frac{2n\pi}{\alpha}\right|^{a_1}} -1\right)}}{\left(z + \frac{2n\pi}{\alpha}\right)^2} \\
    &=
    e^{\lambda_2 \left(e^{-\sigma_2^{a_2} |z|^{a_2}} -1\right)}+
    z^2\sum_{n \in \mathbb{Z}\setminus \{0\}} \frac{e^{\lambda_2 \left(e^{-\sigma_2^{a_2} \left|z + \frac{2n\pi}{\alpha}\right|^{a_2}} -1\right)}}{\left(z + \frac{2n\pi}{\alpha}\right)^2}
    \end{split}
\end{equation}
for all $z \in \left[-\frac{\pi}{\alpha},\frac{\pi}{\alpha}\right]$. Doing an asymptotic expansion near $0^+$ of \eqref{eq:equality_periodized}, noting that the second term of both sides of the equality is $O_{z \to 0^{+}}\left(z^{2}\right)$, we have
\begin{equation*}
    1 - \lambda_1 \sigma_1^{a_1} z^{a_1} + o_{z \to 0^{+}}\left(z^{a_1}\right) = 1 - \lambda_2 \sigma_2^{a_2} z^{a_2} + o_{z \to 0^{+}}\left(z^{a_2}\right).
\end{equation*}
By identification, we get $a_1 = a_2$, which will now be denoted by $a$, and $b \defeq \lambda_1 \sigma_1^{a} = \lambda_2 \sigma_2^{a}$.

Suppose that $a < 1$. Taking the asymptotic expansion of \eqref{eq:equality_periodized} a step further, one has
\begin{equation*}
    1 - b z^{a} + \frac{z^{2a}}{2}\left(\lambda_1 \sigma_1^{2a} + \lambda_1^2 \sigma_1^{2a}\right) + o_{z \to 0^{+}}\left(z^{2a}\right) = 1 - b z^{a} + \frac{z^{2a}}{2}\left(\lambda_2 \sigma_2^{2a} + \lambda_2^2 \sigma_2^{2a}\right) + o_{z \to 0^{+}}\left(z^{2a}\right).
\end{equation*}
Identifying the terms in $z^{2a}$, we obtain $b \sigma_1^a + b^2 = b \sigma_2^a + b^2$, hence $\sigma_1 = \sigma_2$ and, from the formula of $b$, it follows that $\lambda_1 = \lambda_2$.

Suppose now that $a=1$. Doing the asymptotic expansion of \eqref{eq:equality_periodized} and cancelling the terms of order 1 and $z$ which have been shown to be equal on the left and right-hand side, one gets
\begin{equation*}
    \begin{split}
        \frac{z^{2}}{2}\left(\lambda_1 \sigma_1^{2} + \lambda_1^2 \sigma_1^{2}\right) + 
    &z^2\sum_{n \in \mathbb{Z}\setminus \{0\}} \frac{e^{\lambda_1 \left(e^{-\sigma_1^{a} \left(\frac{2n\pi}{\alpha}\right)^{a}} -1\right)}}{\left(\frac{2n\pi}{\alpha}\right)^2}
     + o_{z \to 0^{+}}\left(z^{2}\right) \\
    &= \frac{z^{2}}{2}\left(\lambda_2 \sigma_2^{2} + \lambda_2^2 \sigma_2^{2}\right) +
    z^2\sum_{n \in \mathbb{Z}\setminus \{0\}} \frac{e^{\lambda_2 \left(e^{-\sigma_2^{a} \left(\frac{2n\pi}{\alpha}\right)^{a}} -1\right)}}{\left(\frac{2n\pi}{\alpha}\right)^2} + o_{z \to 0^{+}}\left(z^{2}\right).
    \end{split}
\end{equation*}
Using the definition of $b$ and $g$ from Lemma \ref{lem:decreasing_sum}, the equality becomes
\begin{equation*}
    \frac{b^2}{2\lambda_1} + \sum_{n \in \mathbb{Z}\setminus \{0\}} \left(\frac{\alpha}{2 \pi n}\right)^2g_{\left(\frac{2n\pi}{\alpha}\right)^{a}}(\lambda_1)
    =
    \frac{b^2}{2\lambda_2} + \sum_{n \in \mathbb{Z}\setminus \{0\}} \left(\frac{\alpha}{2 \pi n}\right)^2g_{\left(\frac{2n\pi}{\alpha}\right)^{a}}(\lambda_2).
\end{equation*}
By Lemma \ref{lem:decreasing_sum}, $\lambda \mapsto \frac{b^2}{2\lambda} + \sum_{n \in \mathbb{Z}\setminus \{0\}} \left(\frac{\alpha}{2 \pi n}\right)^2g_{\left(\frac{2n\pi}{\alpha}\right)^{a}}(\lambda)$ is strictly decreasing hence injective. Thus, $\lambda_1 = \lambda_2$. Then, by the definition of $b$, $\sigma_1 = \sigma_2$.

In the case $a \in (1, 2)$, the asymptotic expansion of \eqref{eq:equality_periodized} leads to
\begin{equation*}
    \begin{split}
    z^2\sum_{n \in \mathbb{Z}\setminus \{0\}} \frac{e^{\lambda_1 \left(e^{-\sigma_1^{a} \left(\frac{2n\pi}{\alpha}\right)^{a}} -1\right)}}{\left(\frac{2n\pi}{\alpha}\right)^2}
     + o_{z \to 0^{+}}\left(z^{2}\right) 
    = 
    z^2\sum_{n \in \mathbb{Z}\setminus \{0\}} \frac{e^{\lambda_2 \left(e^{-\sigma_2^{a} \left(\frac{2n\pi}{\alpha}\right)^{a}} -1\right)}}{\left(\frac{2n\pi}{\alpha}\right)^2} + o_{z \to 0^{+}}\left(z^{2}\right),
    \end{split}
\end{equation*}
and we can conclude exactly as in the case $a=1$.

\begin{lemma}
    \label{lem:decreasing_sum}
    Let $c\in (0,\infty)$. Then, $g_{c}:\lambda \in (0, \infty) \mapsto \exp\left(\lambda\left(e^{-\frac{c}{\lambda}} - 1\right)\right)$ is strictly decreasing.
\end{lemma}
\begin{proof}
    Let $\lambda \in (0, \infty)$. The function $g_c$ is differentiable at $\lambda$. The inequality $e^x > 1 + x$, $x \neq 0$, yields
    \begin{equation*}
        g_c'(\lambda) = g_c(\lambda)\left(e^{-\frac{c}{\lambda} }- 1+ \frac{c}{\lambda}e^{-\frac{c}{\lambda}}\right)
        = g_c(\lambda)e^{-\frac{c}{\lambda}}
        \left(1 + \frac{c}{\lambda} - e^{-\frac{c}{\lambda}}\right)
        < 0.
    \end{equation*}
\end{proof}

\subsection{Proof of Proposition \ref{prop:next_trade_continuous_prices}}
\label{sec:proof_next_trade_continuous_prices}

In the continuous transactions prices framework, a transaction is triggered by one of the following events:
\begin{itemize}
    \item A trade by a noise trader.
    \item A jump of the efficient price with absolute value bigger than $\mu$.
\end{itemize}
When an event occurs (trade by a noise trader or jump of the efficient price of any size), the probability of it triggering a trade is thus $(1-r) + r \Proba[][|B| > \mu]$. By thinning, the counting process of trades is therefore a Poisson process with intensity $(\lambda^u + \lambda^i)\left((1-r) + r\Proba[][|B| > \mu]\right)$. Hence, the time of the first trade verifies
\begin{equation*}
    \E[][\tau] = \frac{1}{(\lambda^u + \lambda^i)\left((1-r) + r\Proba[][|B| > \mu]\right)}.
\end{equation*}
By the definition of $\mu$ in \eqref{equation},
    \begin{equation*}
        \frac{1+r}{2r} = \Proba[][B < \mu] + \frac{1}{\mu}\E[][B \mathds{1}_{\{B > \mu\}}] =
        \frac{1}{2} + \frac{1}{2}\Proba[][|B| < \mu] + \frac{1}{2\mu}\E[][|B| \mathds{1}_{\{|B| > \mu\}}],
    \end{equation*}
    which leads to
    \begin{equation*}
        r\Proba[][|B| > \mu] = r - 1 + \frac{r}{\mu}\E[][|B| \mathds{1}_{\{|B| > \mu\}}].
    \end{equation*}
Combined with the fact that $r (\lambda^u + \lambda^i) = \lambda^i$, we conclude that
\begin{equation*}
    \E[][\tau] = \frac{\mu}{\lambda^i \E[][|B| \mathds{1}_{\{|B| > \mu\}}]}.
\end{equation*}

\subsection{Proofs for expected time until next trade in the model with a nonzero tick size}

\subsubsection{Proof of Theorem \ref{thm:equation_trade_time}}
\label{sec:fredholm}

Let $d \in [0,\alpha)$. When $\tilde{P}(t) - P(t)$ is at $d$, a trade occurs if one of the following events happen:
\begin{itemize}
    \item A trade by a noise trader.
    \item A jump of the efficient price above $d + \alpha\ceil*{\frac{\mu-d}{\alpha}}$ (ask informed trade).
    \item A jump of the efficient price below $d - \alpha\ceil*{\frac{\mu+d}{\alpha}}$ (bid informed trade).
\end{itemize}
If the efficient price jumps by $z \in \left[d + \alpha\ceil*{\frac{\mu-d}{\alpha}}, d - \alpha\ceil*{\frac{\mu+d}{\alpha}}\right]$, which happens with probability $r f_B(z) \diff z$, then after the jump, $d(t)$ jumps to $d - z + \alpha\ceil*{\frac{z-d}{\alpha}}$. The first event arrival time is on average $\frac{1}{\lambda^u + \lambda^i}$. Thus, by the strong Markov property,
\begin{equation*}
    \E[][\tau^d_{\alpha}] = \underbrace{\frac{1}{\lambda^u + \lambda^i}}_{\text{Mean arrival time of the first event}}
    + \underbrace{r \int_{d - \alpha\ceil*{\frac{\mu + d}{\alpha}}}^{d + \alpha\ceil*{\frac{\mu - d}{\alpha}}} \E[][\tau_{\alpha}^{d - z + \alpha \ceil*{\frac{z - d}{\alpha}}}] f_B(z) \diff z}_{\text{If the first event was not a trade, mean waiting time for a trade at this point.}}
\end{equation*}

\subsubsection{Preliminary lemmas}
We begin by proving a preliminary lemma that is mainly computational but will simplify the formulas of interest.
\begin{lemma}
    \label{lem:fractional_intergal}
    Let $f:[0,\alpha) \to \mathbb{R}$ be a measurable function. Then,
    \begin{equation*}
        \int_0^{\alpha} \int_{y-\alpha\ceil*{\frac{\mu + y}{\alpha}}}^{y+\alpha\ceil*{\frac{\mu - y}{\alpha}}} f\left(-z + y + \alpha\ceil*{\frac{z-y}{\alpha}}\right)f_B(z)\diff z \diff y =
        \int_0^{\alpha} f(\alpha-y) \int_{-\mu-\alpha}^{\mu}f_B(z+y)\diff z \diff y.
    \end{equation*}
\end{lemma}

\begin{proof}
    There exist $m \in \mathbb{N}$ and $p \in [0, \alpha)$ such that $\mu = m \alpha + p$. Denoting by $I$ the left-hand side of the desired equality, we have
    \begin{align*}
        I &= \int_0^{\alpha} \int_{-\alpha\ceil*{\frac{\mu + y}{\alpha}}}^{\alpha\ceil*{\frac{\mu - y}{\alpha}}} f\left(-z + \alpha\ceil*{\frac{z}{\alpha}}\right)f_B(z+y)\diff z \diff y\\
        &=\begin{aligned}[t]{}\int_0^{\alpha}\sum_{i=-m}^m & \int_{\alpha (i-1)}^{\alpha i} f(-z+\alpha i)f_B(z+y)\diff z \diff y + \int_0^{p}\int_{\alpha m}^{\alpha m + \alpha} f(-z+\alpha m + \alpha) f_B(z+y)\diff z \diff y \\
            &+ \int_{\alpha - p}^{\alpha}\int_{-\alpha m - 2\alpha}^{-\alpha m - \alpha} f(-z-\alpha m - \alpha)f_B(z+y)\diff z \diff y. \end{aligned}
    \end{align*}
    Let $i \in \{-m,\dots,m\}$. Then, by the change of variables $z - \alpha(i-1) \to z$, $y + \alpha(i-1)z$,
    \begin{align*}
        \int_0^{\alpha} \int_{\alpha (i-1)}^{\alpha i} f(-z+\alpha i)f_B(z+y)\diff z \diff y
        &= \int_0^{\alpha} f(\alpha-z)\int_{\alpha (i-1)}^{\alpha i} f_B(z+y)\diff y \diff z.
    \end{align*}
    In the same fashion,
    \begin{equation*}
        \int_0^{p}\int_{\alpha m}^{\alpha m + \alpha} f(-z+\alpha m + \alpha) f_B(z+y)\diff z \diff y 
        = \int_0^{\alpha} f(\alpha-z)\int_{\alpha m}^{\alpha m + p} f_B(z+y)\diff y \diff z
    \end{equation*}
    and
    \begin{equation*}
        \int_{\alpha - p}^{\alpha}\int_{-\alpha m - 2\alpha}^{-\alpha m - \alpha} f(-z-\alpha m - \alpha)f_B(z+y)\diff z \diff y
        = \int_0^{\alpha} f(\alpha-z)\int_{-\alpha m-\alpha-p}^{-\alpha m -\alpha} f_B(z+y)\diff y \diff z.
    \end{equation*}
    Hence,
    \begin{equation*}
        I =
        \int_0^{\alpha} f(\alpha-z) \int_{-\alpha m-\alpha-p}^{\alpha m + p}f_B(z+y)\diff y \diff z.
    \end{equation*}
\end{proof}

From Lemma \ref{lem:fractional_intergal} and Theorem \ref{thm:equation_trade_time}, we deduce immediately that
\begin{equation*}
    u(\alpha) = \frac{1}{\lambda^u + \lambda^i}
    + \frac{r}{\alpha}\int_{0}^{\alpha}\E[][\tau_{\alpha}^{\alpha - y }]\int_{-\mu-\alpha+y}^{\mu+y} f_B(z)\diff z \diff y.
\end{equation*}
By the bid-ask symmetry of our model, $\E[][\tau_{\alpha}^{\alpha - y }] = \E[][\tau_{\alpha}^{y}]$, hence the previous equation becomes
\begin{equation}
    \label{eq:average_fredholm}
    u(\alpha) = \frac{1}{\lambda^u + \lambda^i}
    + \frac{2r}{\alpha}\int_{0}^{\alpha}\E[][\tau_{\alpha}^{y }]\int_{0}^{\mu+y} f_B(z)\diff z \diff y.
\end{equation}

\begin{lemma}
    \label{lem:bounds_time_trade}
    Let $y \in [0,\alpha)$. Then,
    \begin{equation*}
        \frac{1}{(\lambda^u + \lambda^i)\left(1 - r\Proba[][|B| < \mu ]\right)}
        \leqslant \E[][\tau_{\alpha}^y] \leqslant \frac{1}{(\lambda^u + \lambda^i)\left(1 - r\Proba[][|B| < \mu +\alpha]\right)}
    \end{equation*}
\end{lemma}

\begin{remark}
    The upper bound comes from the fact that an uninformed trade or a jump of $|B|$ larger than $\mu + \alpha$ will always trigger a trade. The lower bound comes from the fact that an informed trade must be associated to a jump of $|B|$ larger than $\mu$.
\end{remark}

\begin{proof}
    Let $y \in [0,\alpha)$. Define $M \defeq \sup_{d\in[0,\alpha)}\E[][\tau_{\alpha}^d]$ and $M'\defeq \inf_{d\in[0,\alpha)}\E[][\tau_{\alpha}^d]$. These two quantities are finite by Equation \eqref{eq:noisy_trade_bound}.
    
    Since $y - \alpha\ceil*{\frac{\mu + y}{\alpha}} \geqslant -\mu - \alpha$ and $y + \alpha\ceil*{\frac{\mu - y}{\alpha}} \leqslant \mu + \alpha$, Theorem \ref{thm:equation_trade_time} implies that
    \begin{equation*}
        \E[][\tau_{\alpha}^y] \leqslant \frac{1}{\lambda^u + \lambda^i} + r \int_{-\mu-\alpha}^{\mu+\alpha}M f_B(z)\diff z = \frac{1}{\lambda^u + \lambda^i} + rM\Proba[][|B|<\mu+\alpha].
    \end{equation*}
    Taking the $\sup$ on the left-hand side, the upper bound follows.

    Since $y - \alpha\ceil*{\frac{\mu + y}{\alpha}} \leqslant -\mu $ and $y + \alpha\ceil*{\frac{\mu - y}{\alpha}} \geqslant \mu$, Theorem \ref{thm:equation_trade_time} implies that
    \begin{equation*}
        \E[][\tau_{\alpha}^y] \geqslant \frac{1}{\lambda^u + \lambda^i} + r \int_{-\mu}^{\mu}M' f_B(z)\diff z = \frac{1}{\lambda^u + \lambda^i} + rM'\Proba[][|B|<\mu].
    \end{equation*}
    Taking the $\inf$ on the right-hand side, the lower bound follows.
\end{proof}

\begin{corollary}
    \label{corol:closeness_trade_times}
    \begin{equation*}
        \sup_{y,y'\in[0, \alpha)}\left|\E[][\tau_{\alpha}^y] - \E[][\tau_{\alpha}^{y'}]\right|
        \leqslant \frac{r}{\lambda^u + \lambda^i}\frac{\Proba[][|B| \in [\mu, \mu+\alpha]]}{\left(1 - r\Proba[][|B| < \mu +\alpha]\right)\left(1 - r\Proba[][|B| < \mu]\right)}
        \xrightarrow[\alpha \to 0]{} 0.
    \end{equation*}
\end{corollary}

\subsection{Proof of Proposition \ref{prop:small_tick_trade_time}}
\label{sec:proof_small_tick_trade_time}

From Equation \eqref{eq:average_fredholm} and the fact that
\begin{equation*}
    \frac{1}{\alpha} \int_{0}^{\alpha}\left(\E[][\tau_{\alpha}^y] - u(\alpha)\right)\int_{\mu}^{\mu+y}f_B(z)\diff z \diff y = o_{\alpha \to 0}\left(\frac{1}{\alpha}\int_{0}^{\alpha}\int_{\mu}^{\mu+y}f_B(z)\diff z \diff y\right),
\end{equation*}
which is a direct consequence of Corollary \ref{corol:closeness_trade_times}, we have
\begin{equation*}
    u(\alpha)\left(1 - r \mathbb{P}[|B| \leqslant \mu] -  r \frac{2}{\alpha}\int_{0}^{\alpha}\int_{\mu}^{\mu+y}f_B(z)\diff z \diff y\right) = \frac{1}{\lambda^u+\lambda^i} + o_{\alpha\to 0}\left(\frac{1}{\alpha}\int_{0}^{\alpha}\int_{\mu}^{\mu+y}f_B(z)\diff z \diff y\right).
\end{equation*}
By Equation \eqref{equation},
\begin{equation*}
    1 - r \mathbb{P}[|B| \leqslant \mu] = \frac{r}{\mu}\E[][|B| \mathds{1}_{\{|B| > \mu\}}].
\end{equation*}
Hence,
\begin{equation*}
    \begin{split}
        u(\alpha) = \frac{\mu}{(\lambda^u+\lambda^i) r \E[][|B| \mathds{1}_{\{|B| > \mu\}}]}
    &\frac{1}{1 - \frac{2\mu}{\alpha \E[][|B| \mathds{1}_{\{|B| > \mu\}}]}\int_{0}^{\alpha}\int_{\mu}^{\mu+y}f_B(z)\diff z \diff y}\\
    & + o_{\alpha\to 0}\left(\frac{1}{\alpha}\int_{0}^{\alpha}\int_{\mu}^{\mu+y}f_B(z)\diff z \diff y\right).
    \end{split}
\end{equation*}
Expanding the second fraction, and using the fact that
\begin{equation*}
    \int_{0}^{\alpha}\int_{\mu}^{\mu+y}f_B(z)\diff z \diff y
    =\int_{0}^{\alpha}(\alpha-z)f_B(\mu + z)\diff z,
\end{equation*}we get the desired result.

\subsection{Proof of Proposition \ref{prop:large_tick_trade_time}}
\label{sec:proof_large_tick_trade_time}

From Equation \eqref{eq:average_fredholm}, we deduce
\begin{equation*}
    u(\alpha) = \frac{1}{\lambda^u + \lambda^i} + r u(\alpha) - v(\alpha),
\end{equation*}
where
\begin{equation*}
    v(\alpha) = \frac{2r}{\alpha}\int_0^{\alpha}\E[][\tau_{\alpha}^{y }]\int_{\mu+y}^{\infty} f_B(z)\diff z \diff y.
\end{equation*}
Let $\epsilon > 0$. We can rewrite $v(\alpha)$ as
\begin{equation*}
    v(\alpha) =
    \frac{2r}{\alpha}\int_{\epsilon\alpha}^{(1-\epsilon)\alpha}\E[][\tau_{\alpha}^{y }]\int_{\mu+y}^{\infty}f_B(z)\diff z \diff y
    + \frac{2r}{\alpha}\int_{[0,\epsilon\alpha]\cup[(1-\epsilon)\alpha, \alpha]}\E[][\tau_{\alpha}^{y }]\int_{\mu+y}^{\infty}f_B(z)\diff z \diff y.
\end{equation*}
Using \eqref{eq:noisy_trade_bound}, we obtain
\begin{equation*}
    v(\alpha) \leqslant
    \frac{2r}{\lambda^u}\int_{\mu+\alpha\epsilon}^{\infty}f_B(z)\diff z \diff y
    + \frac{4\epsilon r}{\lambda^u}.
\end{equation*}
We deduce $\limsup\limits_{\alpha \to \infty} v(\alpha) \leqslant \frac{2\epsilon r}{\lambda^u}$. Since this holds for all $\epsilon$ and $v(\alpha)$ is non-negative, it follows that $\lim\limits_{\alpha \to \infty} v(\alpha) = 0$. Thus,
\begin{equation*}
    \lim_{\alpha\to\infty}u(\alpha) = \frac{1}{(1-r)(\lambda^u+\lambda^i)} = \frac{1}{\lambda^u}.
\end{equation*}

\section{Stocks used in the study}
\label{appendix:stocks}
    \begin{scriptsize}
    \begin{longtable}{|>{\centering}p{0.3\textwidth}|>{\centering}p{0.2\textwidth}|>{\centering}p{0.12\textwidth}|c|}
    \hline
    \textbf{Company} & \textbf{Stock Exchange} & \textbf{Currency} & \textbf{Sector} \\ \hline
    ADP & Euronext Paris & EUR & Industrials \\ \hline
    AIR FRANCE -KLM & Euronext Paris & EUR & Industrials \\ \hline
    AIR LIQUIDE & Euronext Paris & EUR & Materials \\ \hline
    AIRBUS & Euronext Paris & EUR & Industrials \\ \hline
    AMG & NYSE & USD & Financials \\ \hline
    ATOS & Euronext Paris & EUR & Information Technology \\ \hline
    AXA & Euronext Paris & EUR & Financials \\ \hline
    AbbVie Inc. & NYSE & USD & Health Care \\ \hline
    Accenture plc & NYSE & USD & Information Technology \\ \hline
    Alibaba Group Holding Limited & NYSE & USD & Communication Services \\ \hline
    American Express Company & NYSE & USD & Financials \\ \hline
    BNP PARIBAS ACT.A & Euronext Paris & EUR & Financials \\ \hline
    BORR DRILLING & NYSE & USD & Energy \\ \hline
    Bank of America Corporation & NYSE & USD & Financials \\ \hline
    Berkshire Hathaway Inc. & NYSE & USD & Financials \\ \hline
    Best Buy Co., Inc. & NYSE & USD & Consumer Discretionary \\ \hline
    Block, Inc. & NYSE & USD & Information Technology \\ \hline
    CREDIT AGRICOLE & Euronext Paris & EUR & Financials \\ \hline
    Carnival Corporation & NYSE & USD & Consumer Discretionary \\ \hline
    Carvana Co. & NYSE & USD & Consumer Discretionary \\ \hline
    Caterpillar Inc. & NYSE & USD & Industrials \\ \hline
    Chevron Corporation & NYSE & USD & Energy \\ \hline
    Citigroup Inc. & NYSE & USD & Financials \\ \hline
    ConocoPhillips & NYSE & USD & Energy \\ \hline
    D.R. Horton, Inc. & NYSE & USD & Consumer Discretionary \\ \hline
    DANONE & Euronext Paris & EUR & Consumer Staples \\ \hline
    DASSAULT SYSTEMES & Euronext Paris & EUR & Information Technology \\ \hline
    DERICHEBOURG & Euronext Paris & EUR & Industrials \\ \hline
    Delta Air Lines, Inc. & NYSE & USD & Industrials \\ \hline
    Devon Energy Corporation & NYSE & USD & Energy \\ \hline
    DoorDash Inc & NYSE & USD & Consumer Discretionary \\ \hline
    EDENRED & Euronext Paris & EUR & Industrials \\ \hline
    ENGIE & Euronext Paris & EUR & Utilities \\ \hline
    EOG Resources, Inc. & NYSE & USD & Energy \\ \hline
    ESSILORLUXOTTICA & Euronext Paris & EUR & Consumer Discretionary \\ \hline
    Exxon Mobil Corporation & NYSE & USD & Energy \\ \hline
    FDJ & Euronext Paris & EUR & Consumer Discretionary \\ \hline
    FLOW TRADERS & Euronext Amsterdam & EUR & Financials \\ \hline
    FORVIA & Euronext Paris & EUR & Consumer Discretionary \\ \hline
    Fidelity National Information Services, Inc. & NYSE & USD & Information Technology \\ \hline
    Ford Motor Company & NYSE & USD & Consumer Discretionary \\ \hline
    Freeport-McMoRan Inc. & NYSE & USD & Materials \\ \hline
    General Motors Company & NYSE & USD & Consumer Discretionary \\ \hline
    HERMES INTL & Euronext Paris & EUR & Consumer Discretionary \\ \hline
    Halliburton Company & NYSE & USD & Energy \\ \hline
    JPMorgan Chase \& Co. & NYSE & USD & Financials \\ \hline
    Johnson \& Johnson & NYSE &USD & Health Care \\ \hline
    KERING & Euronext Paris & EUR & Consumer Discretionary \\ \hline
    KLEPIERRE & Euronext Paris & EUR & Real Estate \\ \hline
    L'OREAL & Euronext Paris & EUR & Consumer Discretionary \\ \hline
    LHYFE & Euronext Paris & EUR & Utilities \\ \hline
    LVMH & Euronext Paris & EUR & Consumer Discretionary \\ \hline
    Lamb Weston Holdings, Inc. & NYSE & USD & Consumer Staples \\ \hline
    Lowe's Companies, Inc. & NYSE & USD & Consumer Discretionary \\ \hline
    MICHELIN & Euronext Paris & EUR & Consumer Discretionary \\ \hline
    Marathon Petroleum Corporation & NYSE & USD & Energy \\ \hline
    Marsh \& McLennan Companies, Inc. & USD & NYSE & Financials \\ \hline
    Mastercard Incorporated & NYSE & USD & Information Technology \\ \hline
    McDonald's Corporation & NYSE & USD & Consumer Discretionary \\ \hline
    NIKE, Inc. & NYSE & USD & Consumer Discretionary \\ \hline
    NIO Inc. & NYSE & USD & Consumer Discretionary \\ \hline
    Norwegian Cruise Line Holdings Ltd. & NYSE & USD & Consumer Discretionary \\ \hline
    OCI & Euronext Amsterdam & EUR & Materials \\ \hline
    OKEA & Oslo Stock Exchange & NOK & Energy \\ \hline
    ORANGE & Euronext Paris & EUR & Communication Services \\ \hline
    ORPEA & Euronext Paris & EUR & Health Care \\ \hline
    Occidental Petroleum Corporation & NYSE & USD & Energy \\ \hline
    Oracle Corporation & NYSE & USD & Information Technology \\ \hline
    Pfizer Inc & NYSE & USD & Health Care \\ \hline
    REMY COINTREAU & Euronext Paris & EUR & Consumer Discretionary \\ \hline
    RENAULT & Euronext Paris & EUR & Consumer Discretionary \\ \hline
    Roblox Corporation & NYSE & USD & Communication Services \\ \hline
    Royal Caribbean Group & NYSE & USD & Consumer Discretionary \\ \hline
    SAFRAN & Euronext Paris & EUR & Industrials \\ \hline
    SAINT GOBAIN & Euronext Paris & EUR & Materials \\ \hline
    SANOFI & Euronext Paris & EUR & Health Care \\ \hline
    SCHNEIDER ELECTRIC & Euronext Paris & EUR & Industrials \\ \hline
    SOCIETE GENERALE & Euronext Paris & EUR & Financials \\ \hline
    STELLANTIS NV & Euronext Amsterdam & EUR & Consumer Discretionary \\ \hline
    STMICROELECTRONICS & Euronext Paris & EUR & Information Technology \\ \hline
    Salesforce, Inc. & NYSE & USD & Information Technology \\ \hline
    Schlumberger Limited & NYSE & USD & Energy \\ \hline
    Sea Limited & NYSE & USD & Communication Services \\ \hline
    Shopify Inc. & NYSE & USD & Information Technology \\ \hline
    Snowflake Inc. & NYSE & USD & Information Technology \\ \hline
    Synchrony Financial & NYSE & USD & Financials \\ \hline
    TOTALENERGIES & Euronext Paris & EUR & Energy \\ \hline
    Taiwan Semiconductor Manufacturing Company Ltd. & NYSE & USD & Information Technology \\ \hline
    The Boeing Company & NYSE & USD & Industrials \\ \hline
    The Coca-Cola Company & NYSE & USD & Consumer Staples \\ \hline
    The Home Depot, Inc. & NYSE & USD & Consumer Discretionary \\ \hline
    The Procter \& Gamble Company & NYSE & USD & Consumer Staples \\ \hline
    The Progressive Corporation & NYSE & USD & Financials \\ \hline
    The Walt Disney Company & NYSE & USD & Communication Services \\ \hline
    Uber Technologies, Inc. & NYSE & USD & Consumer Discretionary \\ \hline
    Union Pacific Corporation & NYSE & USD & Industrials \\ \hline
    VALEO & Euronext Paris & EUR & Consumer Discretionary \\ \hline
    VALLOUREC & Euronext Paris & EUR & Materials \\ \hline
    VINCI & Euronext Paris & EUR & Industrials \\ \hline
    Valero Energy Corporation & NYSE & USD & Energy \\ \hline
    Visa Inc. & NYSE & USD & Information Technology \\ \hline
    Wells Fargo \& Company & NYSE & USD & Financials \\ \hline
    Welltower Inc. & NYSE & USD & Real Estate \\ \hline
    \end{longtable}
\end{scriptsize}

\section{Estimation results}
\label{appendix:estimation_results}

The tables present the estimation results for the efficient price jump distribution, and the standard deviation of the Gaussian distribution fitted for the sizes of the trades by a noise trader. Additionally, the number of trading days used in the estimation dataset is reported. Some Euronext stocks have fewer trading days included, as certain days are discarded due to a different tick size.

\subsection{From 2022-10-01 to 2023-03-31}

\begin{scriptsize}
\begin{longtable}{|>{\centering}p{0.22\textwidth}|c|c|c|c|c|c|c|c|}
    \hline
    \textbf{Company} & $\mathbf{\hat{\lambda}^i}$ & $\mathbf{\hat{a}}$ & $\mathbf{\hat{\sigma}}$ & $\mathbf{\hat{\mu}}$ & $\mathbf{\hat{r}}$ & $\mathbf{\hat{\sigma}_{noise}}$ & $\mathbf{\alpha}$ & \textbf{days}\endhead  \hline
    ADP & 0.0149 & 1.73 & 0.0442 & 0.04047 & 0.583 & 71.1 & 0.05 & 129 \\ \hline
AIR FRANCE -KLM & 0.0422 & 1.69 & 0.0004 & 0.00029 & 0.491 & 11562.8 & 0.0005 & 129 \\ \hline
AIR LIQUIDE & 0.0638 & 1.86 & 0.0185 & 0.00506 & 0.22 & 202.3 & 0.02 & 129 \\ \hline
AIRBUS & 0.0579 & 1.82 & 0.0179 & 0.00418 & 0.184 & 308.5 & 0.02 & 115 \\ \hline
AMG & 0.0309 & 1.76 & 0.0111 & 0.00969 & 0.577 & 554.7 & 0.02 & 129 \\ \hline
ATOS & 0.0351 & 1.72 & 0.0055 & 0.0045 & 0.534 & 1262.9 & 0.005 & 82 \\ \hline
AXA & 0.0604 & 1.78 & 0.0035 & 0.00091 & 0.197 & 1371.7 & 0.005 & 129 \\ \hline
AbbVie Inc. & 0.084 & 1.85 & 0.0217 & 0.01737 & 0.57 & 292.2 & 0.01 & 124 \\ \hline
Accenture plc & 0.0763 & 1.82 & 0.0474 & 0.07728 & 0.846 & 639.6 & 0.01 & 124 \\ \hline
Alibaba Group Holding Limited & 0.075 & 1.73 & 0.0216 & 0.00657 & 0.223 & 516.9 & 0.01 & 124 \\ \hline
American Express Company & 0.082 & 1.8 & 0.0276 & 0.03985 & 0.797 & 336.0 & 0.01 & 124 \\ \hline
BNP PARIBAS ACT.A & 0.0666 & 1.7 & 0.008 & 0.00196 & 0.178 & 627.4 & 0.01 & 101 \\ \hline
BORR DRILLING & 0.0264 & 1.41 & 0.0233 & 0.04119 & 0.65 & 4904.0 & 0.01 & 66 \\ \hline
Bank of America Corporation & 0.0548 & 1.79 & 0.0072 & 0.00014 & 0.015 & 2713.1 & 0.01 & 124 \\ \hline
Berkshire Hathaway Inc. & 0.0865 & 1.82 & 0.0377 & 0.04741 & 0.752 & 318.4 & 0.01 & 124 \\ \hline
Best Buy Co., Inc. & 0.0788 & 1.87 & 0.0179 & 0.01406 & 0.568 & 292.6 & 0.01 & 124 \\ \hline
Block, Inc. & 0.081 & 1.8 & 0.0263 & 0.01915 & 0.517 & 458.7 & 0.01 & 124 \\ \hline
CREDIT AGRICOLE & 0.0701 & 1.81 & 0.0013 & 0.00051 & 0.311 & 1987.3 & 0.001 & 65 \\ \hline
Carnival Corporation & 0.0411 & 1.88 & 0.0053 & 5e-05 & 0.007 & 4329.7 & 0.01 & 124 \\ \hline
Carvana Co. & 0.0476 & 1.58 & 0.01 & 0.00243 & 0.157 & 916.4 & 0.01 & 123 \\ \hline
Caterpillar Inc. & 0.0809 & 1.82 & 0.0396 & 0.05117 & 0.764 & 493.6 & 0.01 & 124 \\ \hline
Chevron Corporation & 0.0912 & 1.87 & 0.029 & 0.01398 & 0.378 & 291.1 & 0.01 & 123 \\ \hline
Citigroup Inc. & 0.0655 & 1.79 & 0.0093 & 0.00031 & 0.026 & 652.1 & 0.01 & 124 \\ \hline
ConocoPhillips & 0.0862 & 1.87 & 0.0266 & 0.01462 & 0.424 & 374.7 & 0.01 & 123 \\ \hline
D.R. Horton, Inc. & 0.0756 & 1.81 & 0.019 & 0.01507 & 0.552 & 315.6 & 0.01 & 123 \\ \hline
DANONE & 0.0484 & 1.85 & 0.0072 & 0.0023 & 0.254 & 618.3 & 0.01 & 61 \\ \hline
DASSAULT SYSTEMES & 0.0696 & 1.8 & 0.0056 & 0.00257 & 0.345 & 534.7 & 0.005 & 129 \\ \hline
DERICHEBOURG & 0.0107 & 1.75 & 0.0031 & 0.00347 & 0.674 & 1079.2 & 0.005 & 91 \\ \hline
Delta Air Lines, Inc. & 0.0636 & 1.85 & 0.0088 & 0.00038 & 0.035 & 449.9 & 0.01 & 124 \\ \hline
Devon Energy Corporation & 0.0817 & 1.85 & 0.017 & 0.00571 & 0.268 & 459.3 & 0.01 & 123 \\ \hline
DoorDash Inc & 0.0699 & 1.79 & 0.0231 & 0.02335 & 0.652 & 446.8 & 0.01 & 124 \\ \hline
EDENRED & 0.0258 & 1.8 & 0.0104 & 0.00546 & 0.39 & 370.0 & 0.02 & 89 \\ \hline
ENGIE & 0.0591 & 1.81 & 0.0019 & 0.00064 & 0.254 & 1625.7 & 0.002 & 129 \\ \hline
EOG Resources, Inc. & 0.0849 & 1.88 & 0.031 & 0.02711 & 0.619 & 371.8 & 0.01 & 123 \\ \hline
ESSILOR\-LUXOTTICA & 0.0472 & 1.86 & 0.0306 & 0.00863 & 0.226 & 249.7 & 0.05 & 129 \\ \hline
Exxon Mobil Corporation & 0.0818 & 1.87 & 0.0205 & 0.00464 & 0.185 & 475.0 & 0.01 & 124 \\ \hline
FDJ & 0.015 & 1.63 & 0.0083 & 0.0097 & 0.64 & 195.1 & 0.01 & 129 \\ \hline
FLOW TRADERS & 0.0069 & 1.68 & 0.0121 & 0.01319 & 0.636 & 639.6 & 0.02 & 55 \\ \hline
FORVIA & 0.053 & 1.8 & 0.0054 & 0.00353 & 0.47 & 1036.0 & 0.005 & 107 \\ \hline
Fidelity National Information Services, Inc. & 0.0728 & 1.78 & 0.0155 & 0.01192 & 0.532 & 349.9 & 0.01 & 124 \\ \hline
Ford Motor Company & 0.0341 & 1.87 & 0.0047 & 1e-05 & 0.002 & 6150.4 & 0.01 & 123 \\ \hline
Freeport-McMoRan Inc. & 0.0697 & 1.84 & 0.0109 & 0.00048 & 0.036 & 467.7 & 0.01 & 123 \\ \hline
General Motors Company & 0.0641 & 1.86 & 0.0097 & 0.00053 & 0.045 & 558.0 & 0.01 & 124 \\ \hline
HERMES INTL & 0.0577 & 1.85 & 0.2931 & 0.08708 & 0.238 & 30.3 & 0.5 & 129 \\ \hline
Halliburton Company & 0.0682 & 1.87 & 0.0116 & 0.00112 & 0.079 & 446.5 & 0.01 & 124 \\ \hline
JPMorgan Chase \& Co. & 0.0824 & 1.77 & 0.0198 & 0.00769 & 0.291 & 348.7 & 0.01 & 124 \\ \hline
Johnson \& Johnson & 0.0815 & 1.84 & 0.0182 & 0.00747 & 0.321 & 272.9 & 0.01 & 124 \\ \hline
KERING & 0.0621 & 1.79 & 0.0916 & 0.02402 & 0.202 & 57.4 & 0.1 & 91 \\ \hline
KLEPIERRE & 0.0361 & 1.84 & 0.006 & 0.00409 & 0.5 & 884.2 & 0.01 & 107 \\ \hline
L'OREAL & 0.0692 & 1.85 & 0.0551 & 0.01511 & 0.22 & 77.8 & 0.05 & 129 \\ \hline
LHYFE & 0.0036 & 1.1 & 0.0032 & 0.02287 & 0.589 & 7937.6 & 0.001 & 129 \\ \hline
LVMH & 0.0721 & 1.83 & 0.1074 & 0.01726 & 0.128 & 44.8 & 0.1 & 129 \\ \hline
Lamb Weston Holdings, Inc. & 0.0474 & 1.81 & 0.0186 & 0.02353 & 0.752 & 321.4 & 0.01 & 123 \\ \hline
Lowe's Companies, Inc. & 0.0773 & 1.8 & 0.0377 & 0.055 & 0.804 & 409.8 & 0.01 & 124 \\ \hline
MICHELIN & 0.0586 & 1.84 & 0.0049 & 0.00205 & 0.326 & 759.2 & 0.005 & 129 \\ \hline
Marathon Petroleum Corporation & 0.0817 & 1.86 & 0.0277 & 0.02507 & 0.629 & 496.0 & 0.01 & 124 \\ \hline
Marsh \& McLennan Companies, Inc. & 0.0613 & 1.8 & 0.0273 & 0.04465 & 0.842 & 342.7 & 0.01 & 124 \\ \hline
Mastercard Incorporated & 0.081 & 1.81 & 0.0542 & 0.09013 & 0.851 & 705.6 & 0.01 & 124 \\ \hline
McDonald's Corporation & 0.0796 & 1.84 & 0.0307 & 0.03765 & 0.75 & 405.7 & 0.01 & 124 \\ \hline
NIKE, Inc. & 0.0868 & 1.85 & 0.0198 & 0.00824 & 0.325 & 311.5 & 0.01 & 124 \\ \hline
NIO Inc. & 0.0386 & 1.78 & 0.0067 & 9e-05 & 0.011 & 3248.2 & 0.01 & 123 \\ \hline
Norwegian Cruise Line Holdings Ltd. & 0.0526 & 1.86 & 0.0072 & 0.00026 & 0.029 & 622.4 & 0.01 & 124 \\ \hline
OCI & 0.0176 & 1.74 & 0.0153 & 0.00928 & 0.425 & 454.2 & 0.02 & 129 \\ \hline
OKEA & 0.0068 & 1.73 & 0.0372 & 0.04029 & 0.658 & 3843.1 & 0.05 & 129 \\ \hline
ORANGE & 0.0651 & 1.88 & 0.001 & 0.0004 & 0.306 & 1914.4 & 0.001 & 96 \\ \hline
ORPEA & 0.0258 & 1.29 & 0.0033 & 0.00637 & 0.564 & 3906.5 & 0.002 & 70 \\ \hline
Occidental Petroleum Corporation & 0.0789 & 1.87 & 0.0169 & 0.00464 & 0.222 & 471.6 & 0.01 & 124 \\ \hline
Oracle Corporation & 0.0706 & 1.84 & 0.0138 & 0.00384 & 0.221 & 308.5 & 0.01 & 124 \\ \hline
Pfizer Inc & 0.0558 & 1.81 & 0.008 & 0.00019 & 0.018 & 689.8 & 0.01 & 124 \\ \hline
REMY COINTREAU & 0.0306 & 1.8 & 0.0423 & 0.02983 & 0.501 & 137.0 & 0.1 & 129 \\ \hline
RENAULT & 0.0676 & 1.79 & 0.0072 & 0.00322 & 0.334 & 547.6 & 0.005 & 129 \\ \hline
Roblox Corporation & 0.0698 & 1.79 & 0.0155 & 0.00485 & 0.24 & 459.9 & 0.01 & 124 \\ \hline
Royal Caribbean Group & 0.0794 & 1.82 & 0.0192 & 0.01342 & 0.503 & 342.6 & 0.01 & 124 \\ \hline
SAFRAN & 0.0546 & 1.83 & 0.0187 & 0.00543 & 0.229 & 238.4 & 0.02 & 120 \\ \hline
SAINT GOBAIN & 0.0719 & 1.83 & 0.0069 & 0.00311 & 0.345 & 408.8 & 0.005 & 67 \\ \hline
SANOFI & 0.0673 & 1.85 & 0.0113 & 0.00295 & 0.21 & 293.7 & 0.01 & 125 \\ \hline
SCHNEIDER ELECTRIC & 0.0752 & 1.85 & 0.0214 & 0.00648 & 0.241 & 194.6 & 0.02 & 129 \\ \hline
SOCIETE GENERALE & 0.0609 & 1.75 & 0.0042 & 0.00127 & 0.226 & 1154.0 & 0.005 & 123 \\ \hline
STELLANTIS NV & 0.0656 & 1.79 & 0.0023 & 0.00119 & 0.384 & 2326.8 & 0.002 & 129 \\ \hline
STMICRO\-ELECTRONICS & 0.0703 & 1.73 & 0.0062 & 0.0028 & 0.326 & 664.7 & 0.005 & 129 \\ \hline
Salesforce, Inc. & 0.0833 & 1.82 & 0.0347 & 0.02532 & 0.523 & 504.1 & 0.01 & 123 \\ \hline
Schlumberger Limited & 0.0772 & 1.87 & 0.0136 & 0.00249 & 0.149 & 480.6 & 0.01 & 124 \\ \hline
Sea Limited & 0.074 & 1.79 & 0.0241 & 0.02454 & 0.652 & 517.0 & 0.01 & 123 \\ \hline
Shopify Inc. & 0.0759 & 1.79 & 0.0151 & 0.00303 & 0.156 & 732.0 & 0.01 & 124 \\ \hline
Snowflake Inc. & 0.0825 & 1.78 & 0.0545 & 0.08507 & 0.82 & 800.6 & 0.01 & 124 \\ \hline
Synchrony Financial & 0.0616 & 1.83 & 0.0094 & 0.00133 & 0.114 & 442.7 & 0.01 & 124 \\ \hline
TOTALENERGIES & 0.0661 & 1.85 & 0.0091 & 0.00125 & 0.112 & 716.9 & 0.01 & 125 \\ \hline
Taiwan Semiconductor Manufacturing Company Ltd. & 0.077 & 1.8 & 0.0147 & 0.00287 & 0.152 & 472.3 & 0.01 & 124 \\ \hline
The Boeing Company & 0.0871 & 1.79 & 0.0377 & 0.04635 & 0.733 & 530.1 & 0.01 & 124 \\ \hline
The Coca-Cola Company & 0.0663 & 1.88 & 0.008 & 0.00028 & 0.03 & 482.9 & 0.01 & 124 \\ \hline
The Home Depot, Inc. & 0.0851 & 1.81 & 0.0506 & 0.07316 & 0.802 & 632.7 & 0.01 & 124 \\ \hline
The Procter \& Gamble Company & 0.0836 & 1.88 & 0.0182 & 0.00722 & 0.318 & 344.9 & 0.01 & 124 \\ \hline
The Progressive Corporation & 0.0692 & 1.8 & 0.0225 & 0.02539 & 0.702 & 347.2 & 0.01 & 124 \\ \hline
The Walt Disney Company & 0.0843 & 1.85 & 0.0193 & 0.00867 & 0.349 & 414.6 & 0.01 & 123 \\ \hline
Uber Technologies, Inc. & 0.0622 & 1.8 & 0.0099 & 0.00039 & 0.031 & 556.1 & 0.01 & 123 \\ \hline
Union Pacific Corporation & 0.0735 & 1.8 & 0.0348 & 0.0481 & 0.784 & 485.3 & 0.01 & 124 \\ \hline
VALEO & 0.053 & 1.79 & 0.0049 & 0.0037 & 0.528 & 920.8 & 0.005 & 99 \\ \hline
VALLOUREC & 0.0344 & 1.72 & 0.0045 & 0.0035 & 0.515 & 1245.7 & 0.005 & 125 \\ \hline
VINCI & 0.065 & 1.87 & 0.011 & 0.0035 & 0.256 & 240.8 & 0.01 & 71 \\ \hline
Valero Energy Corporation & 0.0853 & 1.87 & 0.0342 & 0.03606 & 0.697 & 436.1 & 0.01 & 124 \\ \hline
Visa Inc. & 0.0853 & 1.84 & 0.0313 & 0.0209 & 0.491 & 394.5 & 0.01 & 124 \\ \hline
Wells Fargo \& Company & 0.0585 & 1.79 & 0.0093 & 0.00032 & 0.027 & 668.7 & 0.01 & 124 \\ \hline
Welltower Inc. & 0.0586 & 1.86 & 0.0163 & 0.0176 & 0.703 & 277.7 & 0.01 & 124 \\ \hline
    \end{longtable}
\end{scriptsize}

\subsection{From 2023-04-01 to 2023-09-30}

\begin{scriptsize}
\begin{longtable}{|>{\centering}p{0.22\textwidth}|c|c|c|c|c|c|c|c|}
    \hline
    \textbf{Company} & $\mathbf{\hat{\lambda}^i}$ & $\mathbf{\hat{a}}$ & $\mathbf{\hat{\sigma}}$ & $\mathbf{\hat{\mu}}$ & $\mathbf{\hat{r}}$ & $\mathbf{\hat{\sigma}_{noise}}$ & $\mathbf{\alpha}$ & \textbf{days} \endhead  \hline
    ADP & 0.012 & 1.88 & 0.0424 & 0.02564 & 0.463 & 123.3 & 0.1 & 127 \\ \hline
    AIR FRANCE -KLM & 0.0364 & 1.73 & 0.0004 & 0.00027 & 0.476 & 10952.5 & 0.0005 & 105 \\ \hline
    AIR LIQUIDE & 0.055 & 1.92 & 0.0185 & 0.00582 & 0.261 & 187.8 & 0.02 & 127 \\ \hline
    AIRBUS & 0.0591 & 1.86 & 0.0171 & 0.00574 & 0.268 & 282.7 & 0.02 & 127 \\ \hline
    AMG & 0.0277 & 1.57 & 0.0092 & 0.01048 & 0.596 & 506.5 & 0.01 & 127 \\ \hline
    ATOS & 0.0316 & 1.58 & 0.0037 & 0.00403 & 0.591 & 1158.7 & 0.005 & 82 \\ \hline
    AXA & 0.0558 & 1.92 & 0.0038 & 0.0013 & 0.286 & 1363.8 & 0.005 & 127 \\ \hline
    AbbVie Inc. & 0.0736 & 1.87 & 0.0178 & 0.01516 & 0.605 & 236.7 & 0.01 & 124 \\ \hline
    Accenture plc & 0.0606 & 1.83 & 0.0437 & 0.07104 & 0.85 & 594.5 & 0.01 & 124 \\ \hline
    Alibaba Group Holding Limited & 0.0707 & 1.7 & 0.0148 & 0.0038 & 0.185 & 481.2 & 0.01 & 124 \\ \hline
    American Express Company & 0.0722 & 1.86 & 0.0232 & 0.03087 & 0.79 & 266.9 & 0.01 & 124 \\ \hline
    BNP PARIBAS ACT.A & 0.0682 & 1.88 & 0.0092 & 0.00229 & 0.205 & 750.2 & 0.01 & 127 \\ \hline
    BORR DRILLING & 0.0095 & 1.66 & 0.0429 & 0.04888 & 0.646 & 1660.0 & 0.05 & 122 \\ \hline
    Bank of America Corporation & 0.0413 & 1.88 & 0.0061 & 6e-05 & 0.009 & 2795.7 & 0.01 & 124 \\ \hline
    Berkshire Hathaway Inc. & 0.0735 & 1.89 & 0.0329 & 0.04612 & 0.82 & 253.3 & 0.01 & 124 \\ \hline
    Best Buy Co., Inc. & 0.066 & 1.89 & 0.0137 & 0.00955 & 0.522 & 235.3 & 0.01 & 124 \\ \hline
    Block, Inc. & 0.0734 & 1.82 & 0.0172 & 0.0094 & 0.41 & 317.1 & 0.01 & 124 \\ \hline
    CREDIT AGRICOLE & 0.0584 & 1.87 & 0.0015 & 0.0006 & 0.326 & 2327.3 & 0.002 & 127 \\ \hline
    Carnival Corporation & 0.0421 & 1.82 & 0.0055 & 7e-05 & 0.011 & 2215.6 & 0.01 & 124 \\ \hline
    Carvana Co. & 0.0565 & 1.43 & 0.0185 & 0.01759 & 0.446 & 1054.6 & 0.01 & 124 \\ \hline
    Caterpillar Inc. & 0.0732 & 1.81 & 0.0386 & 0.05421 & 0.793 & 501.2 & 0.01 & 124 \\ \hline
    Chevron Corporation & 0.0772 & 1.9 & 0.0228 & 0.01092 & 0.382 & 334.5 & 0.01 & 124 \\ \hline
    Citigroup Inc. & 0.0576 & 1.84 & 0.008 & 0.00022 & 0.022 & 495.9 & 0.01 & 124 \\ \hline
    ConocoPhillips & 0.0827 & 1.9 & 0.0191 & 0.01082 & 0.443 & 284.1 & 0.01 & 124 \\ \hline
    D.R. Horton, Inc. & 0.0713 & 1.87 & 0.0201 & 0.01988 & 0.668 & 310.1 & 0.01 & 124 \\ \hline
    DANONE & 0.0462 & 1.87 & 0.0064 & 0.00276 & 0.338 & 503.4 & 0.01 & 127 \\ \hline
    DASSAULT SYSTEMES & 0.0571 & 1.83 & 0.0052 & 0.00288 & 0.417 & 475.1 & 0.005 & 127 \\ \hline
    DERICHEBOURG & 0.0045 & 1.74 & 0.0031 & 0.00344 & 0.675 & 906.7 & 0.005 & 82 \\ \hline
    Delta Air Lines, Inc. & 0.0598 & 1.87 & 0.0085 & 0.00025 & 0.024 & 367.3 & 0.01 & 124 \\ \hline
    Devon Energy Corporation & 0.0712 & 1.9 & 0.0108 & 0.00148 & 0.114 & 408.6 & 0.01 & 124 \\ \hline
    DoorDash Inc & 0.0611 & 1.84 & 0.0213 & 0.02302 & 0.697 & 379.9 & 0.01 & 124 \\ \hline
    EDENRED & 0.0233 & 1.79 & 0.01 & 0.00521 & 0.384 & 378.4 & 0.02 & 127 \\ \hline
    ENGIE & 0.0492 & 1.84 & 0.0019 & 0.00089 & 0.356 & 1567.9 & 0.002 & 127 \\ \hline
    EOG Resources, Inc. & 0.0821 & 1.88 & 0.0215 & 0.02273 & 0.702 & 313.3 & 0.01 & 124 \\ \hline
    ESSILOR\-LUXOTTICA & 0.058 & 1.86 & 0.0226 & 0.01081 & 0.372 & 133.0 & 0.02 & 127 \\ \hline
    Exxon Mobil Corporation & 0.0736 & 1.9 & 0.0177 & 0.00324 & 0.153 & 421.3 & 0.01 & 124 \\ \hline
    FDJ & 0.0074 & 1.7 & 0.0097 & 0.01034 & 0.637 & 258.6 & 0.02 & 127 \\ \hline
    FLOW TRADERS & 0.0051 & 1.66 & 0.009 & 0.00839 & 0.563 & 544.3 & 0.02 & 62 \\ \hline
    FORVIA & 0.033 & 1.81 & 0.0073 & 0.00483 & 0.482 & 791.8 & 0.01 & 69 \\ \hline
    Fidelity National Information Services, Inc. & 0.0579 & 1.88 & 0.0121 & 0.00678 & 0.434 & 281.9 & 0.01 & 124 \\ \hline
    Ford Motor Company & 0.0203 & 1.93 & 0.0048 & 1e-05 & 0.001 & 5284.3 & 0.01 & 124 \\ \hline
    Freeport-McMoRan Inc. & 0.0561 & 1.89 & 0.009 & 0.00028 & 0.026 & 396.0 & 0.01 & 124 \\ \hline
    General Motors Company & 0.0507 & 1.87 & 0.0081 & 0.00017 & 0.018 & 568.7 & 0.01 & 124 \\ \hline
    HERMES INTL & 0.0686 & 1.89 & 0.289 & 0.13054 & 0.36 & 16.7 & 0.2 & 117 \\ \hline
    Halliburton Company & 0.0605 & 1.9 & 0.0093 & 0.00047 & 0.042 & 426.7 & 0.01 & 124 \\ \hline
    JPMorgan Chase \& Co. & 0.079 & 1.87 & 0.0175 & 0.00625 & 0.286 & 303.3 & 0.01 & 124 \\ \hline
    Johnson \& Johnson & 0.0741 & 1.82 & 0.0171 & 0.00697 & 0.314 & 298.8 & 0.01 & 124 \\ \hline
    KERING & 0.0524 & 1.87 & 0.0839 & 0.02964 & 0.283 & 63.1 & 0.1 & 82 \\ \hline
    KLEPIERRE & 0.0271 & 1.87 & 0.0057 & 0.00338 & 0.452 & 846.0 & 0.01 & 127 \\ \hline
    L'OREAL & 0.0638 & 1.92 & 0.0554 & 0.01741 & 0.26 & 66.3 & 0.05 & 127 \\ \hline
    LHYFE & 0.0009 & 1.56 & 0.0072 & 0.01651 & 0.831 & 1196.7 & 0.01 & 127 \\ \hline
    LVMH & 0.0658 & 1.89 & 0.1109 & 0.01459 & 0.109 & 41.7 & 0.1 & 127 \\ \hline
    Lamb Weston Holdings, Inc. & 0.0396 & 1.73 & 0.0178 & 0.02537 & 0.764 & 346.8 & 0.01 & 124 \\ \hline
    Lowe's Companies, Inc. & 0.0674 & 1.85 & 0.03 & 0.0404 & 0.792 & 322.2 & 0.01 & 124 \\ \hline
    MICHELIN & 0.0369 & 1.91 & 0.0051 & 0.00217 & 0.342 & 1015.4 & 0.01 & 127 \\ \hline
    Marathon Petroleum Corporation & 0.0755 & 1.88 & 0.0262 & 0.02556 & 0.665 & 411.9 & 0.01 & 124 \\ \hline
    Marsh \& McLennan Companies, Inc. & 0.0567 & 1.85 & 0.0217 & 0.03569 & 0.86 & 249.6 & 0.01 & 124 \\ \hline
    Mastercard Incorporated & 0.0679 & 1.84 & 0.0445 & 0.07186 & 0.851 & 607.2 & 0.01 & 124 \\ \hline
    McDonald's Corporation & 0.0743 & 1.9 & 0.0263 & 0.03206 & 0.771 & 311.8 & 0.01 & 124 \\ \hline
    NIKE, Inc. & 0.0769 & 1.88 & 0.0161 & 0.00596 & 0.299 & 271.4 & 0.01 & 124 \\ \hline
    NIO Inc. & 0.0264 & 1.74 & 0.0059 & 5e-05 & 0.006 & 3777.9 & 0.01 & 124 \\ \hline
    Norwegian Cruise Line Holdings Ltd. & 0.0433 & 1.87 & 0.0067 & 0.00011 & 0.013 & 552.3 & 0.01 & 124 \\ \hline
    OCI & 0.0138 & 1.58 & 0.009 & 0.00908 & 0.556 & 429.4 & 0.01 & 127 \\ \hline
    OKEA & 0.0075 & 1.65 & 0.018 & 0.02639 & 0.735 & 3557.3 & 0.02 & 122 \\ \hline
    ORANGE & 0.0447 & 1.83 & 0.0013 & 0.00058 & 0.348 & 2550.3 & 0.002 & 126 \\ \hline
    ORPEA & 0.0084 & 1.12 & 0.0007 & 0.00271 & 0.495 & 19206.2 & 0.0005 & 83 \\ \hline
    Occidental Petroleum Corporation & 0.0681 & 1.89 & 0.0111 & 0.00169 & 0.126 & 427.1 & 0.01 & 124 \\ \hline
    Oracle Corporation & 0.0742 & 1.78 & 0.0166 & 0.00662 & 0.3 & 332.4 & 0.01 & 124 \\ \hline
    Pfizer Inc & 0.0381 & 1.85 & 0.0067 & 0.00004 & 0.005 & 952.5 & 0.01 & 124 \\ \hline
    REMY COINTREAU & 0.0285 & 1.72 & 0.0304 & 0.03045 & 0.618 & 113.4 & 0.05 & 127 \\ \hline
    RENAULT & 0.0597 & 1.79 & 0.0066 & 0.00388 & 0.425 & 486.6 & 0.005 & 127 \\ \hline
    Roblox Corporation & 0.0626 & 1.81 & 0.0127 & 0.00425 & 0.258 & 323.4 & 0.01 & 124 \\ \hline
    Royal Caribbean Group & 0.0718 & 1.84 & 0.0212 & 0.02019 & 0.644 & 306.9 & 0.01 & 124 \\ \hline
    SAFRAN & 0.0575 & 1.86 & 0.0176 & 0.00717 & 0.321 & 200.2 & 0.02 & 127 \\ \hline
    SAINT GOBAIN & 0.0556 & 1.87 & 0.0084 & 0.00362 & 0.343 & 423.4 & 0.01 & 124 \\ \hline
    SANOFI & 0.0676 & 1.9 & 0.0114 & 0.00405 & 0.288 & 226.6 & 0.01 & 79 \\ \hline
    SCHNEIDER ELECTRIC & 0.0709 & 1.91 & 0.0207 & 0.0067 & 0.267 & 172.4 & 0.02 & 127 \\ \hline
    SOCIETE GENERALE & 0.0604 & 1.83 & 0.004 & 0.00148 & 0.291 & 1191.2 & 0.005 & 127 \\ \hline
    STELLANTIS NV & 0.0661 & 1.89 & 0.0023 & 0.00115 & 0.395 & 1866.9 & 0.002 & 127 \\ \hline
    STMICRO\-ELECTRONICS & 0.07 & 1.79 & 0.0062 & 0.00286 & 0.348 & 579.1 & 0.005 & 127 \\ \hline
    Salesforce, Inc. & 0.0869 & 1.86 & 0.0319 & 0.03468 & 0.708 & 383.1 & 0.01 & 124 \\ \hline
    Schlumberger Limited & 0.071 & 1.91 & 0.0118 & 0.00118 & 0.085 & 374.5 & 0.01 & 124 \\ \hline
    Sea Limited & 0.0627 & 1.76 & 0.017 & 0.01416 & 0.556 & 449.7 & 0.01 & 124 \\ \hline
    Shopify Inc. & 0.0725 & 1.86 & 0.0176 & 0.00352 & 0.163 & 521.0 & 0.01 & 124 \\ \hline
    Snowflake Inc. & 0.0833 & 1.79 & 0.0432 & 0.06019 & 0.782 & 568.3 & 0.01 & 124 \\ \hline
    Synchrony Financial & 0.0498 & 1.88 & 0.0075 & 0.00033 & 0.037 & 401.7 & 0.01 & 124 \\ \hline
    TOTALENERGIES & 0.0586 & 1.91 & 0.0082 & 0.00149 & 0.152 & 1017.5 & 0.01 & 127 \\ \hline
    Taiwan Semiconductor Manufacturing Company Ltd. & 0.0743 & 1.8 & 0.0137 & 0.00304 & 0.173 & 330.3 & 0.01 & 124 \\ \hline
    The Boeing Company & 0.0778 & 1.78 & 0.0326 & 0.04245 & 0.752 & 450.0 & 0.01 & 124 \\ \hline
    The Coca-Cola Company & 0.0462 & 1.92 & 0.0069 & 0.00012 & 0.015 & 522.7 & 0.01 & 124 \\ \hline
    The Home Depot, Inc. & 0.078 & 1.87 & 0.0378 & 0.04916 & 0.784 & 439.3 & 0.01 & 124 \\ \hline
    The Procter \& Gamble Company & 0.0758 & 1.93 & 0.0159 & 0.00697 & 0.359 & 260.9 & 0.01 & 124 \\ \hline
    The Progressive Corporation & 0.0578 & 1.79 & 0.0206 & 0.02414 & 0.713 & 332.7 & 0.01 & 124 \\ \hline
    The Walt Disney Company & 0.0725 & 1.84 & 0.0131 & 0.00331 & 0.201 & 354.6 & 0.01 & 124 \\ \hline
    Uber Technologies, Inc. & 0.06 & 1.84 & 0.0108 & 0.00057 & 0.043 & 435.1 & 0.01 & 124 \\ \hline
    Union Pacific Corporation & 0.0665 & 1.84 & 0.028 & 0.03948 & 0.804 & 371.3 & 0.01 & 124 \\ \hline
    VALEO & 0.0387 & 1.8 & 0.0044 & 0.00389 & 0.593 & 955.9 & 0.005 & 102 \\ \hline
    VALLOUREC & 0.0282 & 1.77 & 0.0036 & 0.00332 & 0.601 & 1160.2 & 0.005 & 118 \\ \hline
    VINCI & 0.0501 & 1.91 & 0.0133 & 0.00471 & 0.291 & 320.9 & 0.02 & 126 \\ \hline
    Valero Energy Corporation & 0.0789 & 1.85 & 0.0267 & 0.02791 & 0.687 & 351.7 & 0.01 & 124 \\ \hline
    Visa Inc. & 0.0831 & 1.87 & 0.0248 & 0.02045 & 0.588 & 337.5 & 0.01 & 124 \\ \hline
    Wells Fargo \& Company & 0.0553 & 1.86 & 0.0081 & 0.00013 & 0.014 & 662.5 & 0.01 & 124 \\ \hline
    Welltower Inc. & 0.0556 & 1.92 & 0.0157 & 0.0172 & 0.729 & 238.3 & 0.01 & 124 \\ \hline
\end{longtable}
\end{scriptsize}

\subsection{Failed calibration of a piecewise constant $f_{Q^u}$}
\label{appendix:failed_histogram}
\begin{table}[h!]
    \centering
    \begin{scriptsize}
    \begin{tabular}{|c|c|c|c|}
        \hline
        \textbf{Company} & \textbf{tick size} & \textbf{spread (ticks)} & \textbf{days} \\ \hline
        Accenture plc & 0.01 & 16.46 & 124 \\ \hline
        American Express Company & 0.01 & 8.97 & 124 \\ \hline
        BORR DRILLING & 0.01 & 9.24 & 66 \\ \hline
        Berkshire Hathaway Inc. & 0.01 & 10.48 & 124 \\ \hline
        Caterpillar Inc. & 0.01 & 11.23 & 124 \\ \hline
        DoorDash Inc & 0.01 & 5.67 & 124 \\ \hline
        EOG Resources, Inc. & 0.01 & 6.42 & 123 \\ \hline
        LHYFE & 0.001 & 46.73 & 129 \\ \hline
        Lamb Weston Holdings, Inc. & 0.01 & 5.71 & 123 \\ \hline
        Lowe's Companies, Inc. & 0.01 & 12.0 & 124 \\ \hline
        Marathon Petroleum Corporation & 0.01 & 6.01 & 124 \\ \hline
        Marsh \& McLennan Companies, Inc. & 0.01 & 9.93 & 124 \\ \hline
        Mastercard Incorporated & 0.01 & 19.03 & 124 \\ \hline
        McDonald's Corporation & 0.01 & 8.53 & 124 \\ \hline
        ORPEA & 0.002 & 7.37 & 70 \\ \hline
        Salesforce, Inc. & 0.01 & 6.06 & 123 \\ \hline
        Sea Limited & 0.01 & 5.91 & 123 \\ \hline
        Snowflake Inc. & 0.01 & 18.01 & 124 \\ \hline
        The Boeing Company & 0.01 & 10.27 & 124 \\ \hline
        The Home Depot, Inc. & 0.01 & 15.63 & 124 \\ \hline
        The Progressive Corporation & 0.01 & 6.08 & 124 \\ \hline
        Union Pacific Corporation & 0.01 & 10.62 & 124 \\ \hline
        Valero Energy Corporation & 0.01 & 8.21 & 124 \\ \hline
        \end{tabular}
    \end{scriptsize}
    \caption{Stocks where the piecewise constant calibration of $f_{Q^u}$ failed, with their spread. Period: 2022-10-01 to 2023-03-31.}
    \label{table:failed_histogram_period_1}
\end{table}

\begin{table}[h!]
    \centering
    \begin{scriptsize}
    \begin{tabular}{|c|c|c|c|}
        \hline
        \textbf{Company} & \textbf{tick size} & \textbf{spread (ticks)} & \textbf{days} \\ \hline
        Accenture plc & 0.01 & 15.21 & 124 \\ \hline
American Express Company & 0.01 & 7.17 & 124 \\ \hline
Berkshire Hathaway Inc. & 0.01 & 10.22 & 124 \\ \hline
Caterpillar Inc. & 0.01 & 11.84 & 124 \\ \hline
DoorDash Inc & 0.01 & 5.6 & 124 \\ \hline
LHYFE & 0.01 & 4.3 & 127 \\ \hline
Lamb Weston Holdings, Inc. & 0.01 & 6.07 & 124 \\ \hline
Lowe's Companies, Inc. & 0.01 & 9.08 & 124 \\ \hline
Marathon Petroleum Corporation & 0.01 & 6.11 & 124 \\ \hline
Marsh \& McLennan Companies, Inc. & 0.01 & 8.14 & 124 \\ \hline
Mastercard Incorporated & 0.01 & 15.37 & 124 \\ \hline
McDonald's Corporation & 0.01 & 7.41 & 124 \\ \hline
ORPEA & 0.0005 & 11.84 & 83 \\ \hline
Salesforce, Inc. & 0.01 & 7.94 & 124 \\ \hline
Snowflake Inc. & 0.01 & 13.04 & 124 \\ \hline
The Boeing Company & 0.01 & 9.49 & 124 \\ \hline
The Home Depot, Inc. & 0.01 & 10.83 & 124 \\ \hline
The Progressive Corporation & 0.01 & 5.83 & 124 \\ \hline
Union Pacific Corporation & 0.01 & 8.9 & 124 \\ \hline
Valero Energy Corporation & 0.01 & 6.58 & 124 \\ \hline
Welltower Inc. & 0.01 & 4.44 & 124 \\ \hline
        \end{tabular}
    \end{scriptsize}
    \caption{Stocks where the piecewise constant calibration of $f_{Q^u}$ failed, with their spread. Period: 2023-04-01 to 2023-09-31.}
    \label{table:failed_histogram_period_2}
\end{table}

\begin{figure}[h!]
    \centering
    \begin{subfigure}[b]{0.45\textwidth}
        \centering
        \includegraphics[width=\textwidth,page=2]{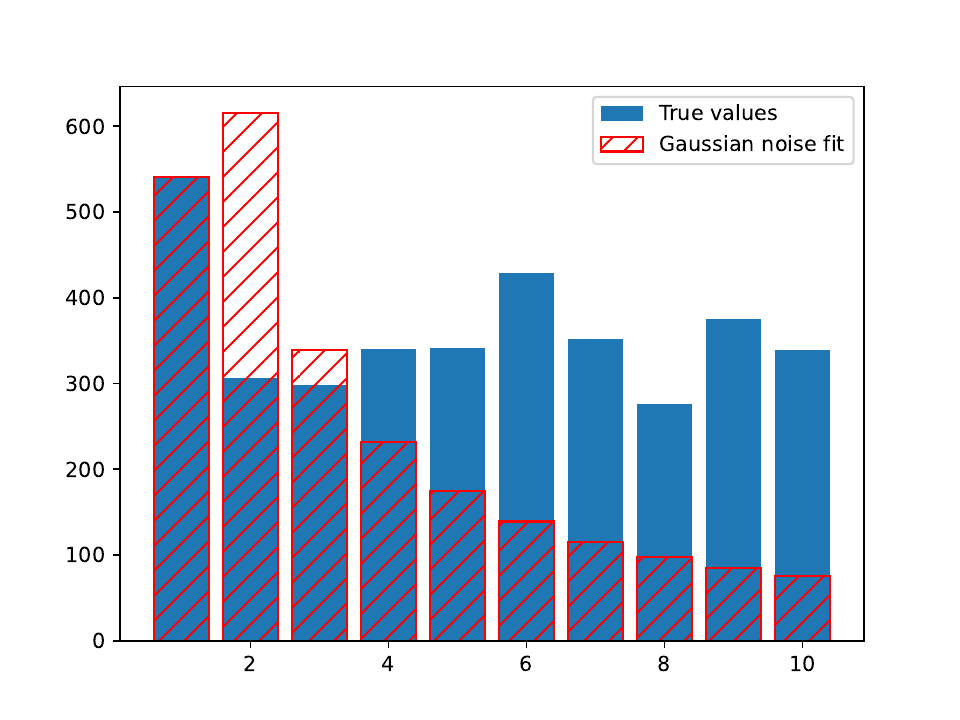}
        \caption{LHYFE}
    \end{subfigure}
    \hfill
    \begin{subfigure}[b]{0.45\textwidth}
        \centering
        \includegraphics[width=\textwidth,page=2]{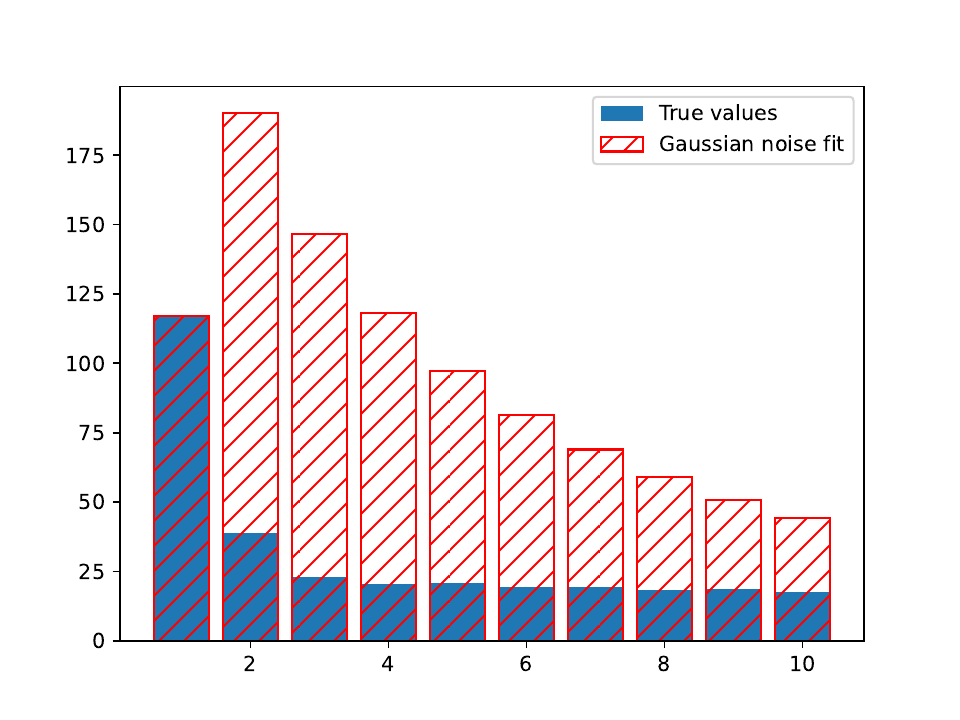}
        \caption{Mastercard Incorporated}
    \end{subfigure}

    \caption{Mean volume pending on each pile up to 10 ticks from the best (in blue) and the mean volumes given by the model when $Q^u$ follows a normal distribution calibrated on the volume pending at the best price. Period: 2023-04-01 to 2023-09-30.}
    \label{fig:failed_histogram_LOB_shape}
\end{figure}

\FloatBarrier

\section{Queue position valuation estimation}
\label{appendix:qpv_estimation}

Here we report the estimated mean queue position values (QPV) of the best piles.

\subsection{From 2022-10-01 to 2023-03-31}

\begin{scriptsize}
\begin{longtable}{|>{\centering}p{0.3\textwidth}|c|c|c|c|}
    \hline
    \textbf{Company} & \textbf{spread} & \textbf{QPV} & \textbf{tick size} & \textbf{days} \endhead  \hline
    ADP & 0.1309 & 0.0433 & 0.05 & 129 \\ \hline
AIR FRANCE -KLM & 0.0011 & 0.0004 & 0.0005 & 129 \\ \hline
AIR LIQUIDE & 0.0301 & 0.0126 & 0.02 & 129 \\ \hline
AIRBUS & 0.0284 & 0.0121 & 0.02 & 115 \\ \hline
AMG & 0.0394 & 0.0151 & 0.02 & 129 \\ \hline
ATOS & 0.014 & 0.0045 & 0.005 & 82 \\ \hline
AXA & 0.0068 & 0.003 & 0.005 & 129 \\ \hline
AbbVie Inc. & 0.0447 & 0.0118 & 0.01 & 124 \\ \hline
Accenture plc & 0.1646 & 0.0287 & 0.01 & 124 \\ \hline
Alibaba Group Holding Limited & 0.0231 & 0.008 & 0.01 & 124 \\ \hline
American Express Company & 0.0897 & 0.0184 & 0.01 & 124 \\ \hline
BNP PARIBAS ACT.A & 0.0139 & 0.006 & 0.01 & 101 \\ \hline
BORR DRILLING & 0.0924 & 0.0221 & 0.01 & 66 \\ \hline
Bank of America Corporation & 0.0103 & 0.0051 & 0.01 & 124 \\ \hline
Berkshire Hathaway Inc. & 0.1048 & 0.0209 & 0.01 & 124 \\ \hline
Best Buy Co., Inc. & 0.0381 & 0.0106 & 0.01 & 124 \\ \hline
Block, Inc. & 0.0483 & 0.0126 & 0.01 & 124 \\ \hline
CREDIT AGRICOLE & 0.002 & 0.0007 & 0.001 & 65 \\ \hline
Carnival Corporation & 0.0101 & 0.005 & 0.01 & 124 \\ \hline
Carvana Co. & 0.0149 & 0.0062 & 0.01 & 123 \\ \hline
Caterpillar Inc. & 0.1123 & 0.0219 & 0.01 & 124 \\ \hline
Chevron Corporation & 0.038 & 0.0109 & 0.01 & 123 \\ \hline
Citigroup Inc. & 0.0106 & 0.0052 & 0.01 & 124 \\ \hline
ConocoPhillips & 0.0392 & 0.0111 & 0.01 & 123 \\ \hline
D.R. Horton, Inc. & 0.0401 & 0.0111 & 0.01 & 123 \\ \hline
DANONE & 0.0146 & 0.0062 & 0.01 & 61 \\ \hline
DASSAULT SYSTEMES & 0.0101 & 0.0037 & 0.005 & 129 \\ \hline
DERICHEBOURG & 0.0119 & 0.0042 & 0.005 & 91 \\ \hline
Delta Air Lines, Inc. & 0.0108 & 0.0052 & 0.01 & 124 \\ \hline
Devon Energy Corporation & 0.0214 & 0.0076 & 0.01 & 123 \\ \hline
DoorDash Inc & 0.0567 & 0.0137 & 0.01 & 124 \\ \hline
EDENRED & 0.0309 & 0.0131 & 0.02 & 89 \\ \hline
ENGIE & 0.0033 & 0.0013 & 0.002 & 129 \\ \hline
EOG Resources, Inc. & 0.0642 & 0.015 & 0.01 & 123 \\ \hline
ESSILORLUXOTTICA & 0.0673 & 0.0298 & 0.05 & 129 \\ \hline
Exxon Mobil Corporation & 0.0193 & 0.0072 & 0.01 & 124 \\ \hline
FDJ & 0.0294 & 0.0093 & 0.01 & 129 \\ \hline
FLOW TRADERS & 0.0464 & 0.0164 & 0.02 & 55 \\ \hline
FORVIA & 0.0121 & 0.0041 & 0.005 & 107 \\ \hline
Fidelity National Information Services, Inc. & 0.0338 & 0.01 & 0.01 & 124 \\ \hline
Ford Motor Company & 0.01 & 0.005 & 0.01 & 123 \\ \hline
Freeport-McMoRan Inc. & 0.011 & 0.0053 & 0.01 & 123 \\ \hline
General Motors Company & 0.0111 & 0.0053 & 0.01 & 124 \\ \hline
HERMES INTL & 0.6742 & 0.299 & 0.5 & 129 \\ \hline
Halliburton Company & 0.0122 & 0.0056 & 0.01 & 124 \\ \hline
JPMorgan Chase \& Co. & 0.0254 & 0.0085 & 0.01 & 124 \\ \hline
Johnson \& Johnson & 0.0249 & 0.0083 & 0.01 & 124 \\ \hline
KERING & 0.148 & 0.0622 & 0.1 & 91 \\ \hline
KLEPIERRE & 0.0182 & 0.0072 & 0.01 & 107 \\ \hline
L'OREAL & 0.0802 & 0.0324 & 0.05 & 129 \\ \hline
LHYFE & 0.0467 & 0.0114 & 0.001 & 129 \\ \hline
LVMH & 0.1345 & 0.0588 & 0.1 & 129 \\ \hline
Lamb Weston Holdings, Inc. & 0.0571 & 0.0135 & 0.01 & 123 \\ \hline
Lowe's Companies, Inc. & 0.12 & 0.0228 & 0.01 & 124 \\ \hline
MICHELIN & 0.0091 & 0.0035 & 0.005 & 129 \\ \hline
Marathon Petroleum Corporation & 0.0601 & 0.0143 & 0.01 & 124 \\ \hline
Marsh \& McLennan Companies, Inc. & 0.0993 & 0.0195 & 0.01 & 124 \\ \hline
Mastercard Incorporated & 0.1903 & 0.0324 & 0.01 & 124 \\ \hline
McDonald's Corporation & 0.0853 & 0.0177 & 0.01 & 124 \\ \hline
NIKE, Inc. & 0.0265 & 0.0087 & 0.01 & 124 \\ \hline
NIO Inc. & 0.0102 & 0.0051 & 0.01 & 123 \\ \hline
Norwegian Cruise Line Holdings Ltd. & 0.0105 & 0.0051 & 0.01 & 124 \\ \hline
OCI & 0.0386 & 0.0146 & 0.02 & 129 \\ \hline
OKEA & 0.1306 & 0.0436 & 0.05 & 129 \\ \hline
ORANGE & 0.0018 & 0.0007 & 0.001 & 96 \\ \hline
ORPEA & 0.0147 & 0.0038 & 0.002 & 70 \\ \hline
Occidental Petroleum Corporation & 0.0193 & 0.0072 & 0.01 & 124 \\ \hline
Oracle Corporation & 0.0177 & 0.0068 & 0.01 & 124 \\ \hline
Pfizer Inc & 0.0104 & 0.0051 & 0.01 & 124 \\ \hline
REMY COINTREAU & 0.1597 & 0.0676 & 0.1 & 129 \\ \hline
RENAULT & 0.0114 & 0.004 & 0.005 & 129 \\ \hline
Roblox Corporation & 0.0197 & 0.0073 & 0.01 & 124 \\ \hline
Royal Caribbean Group & 0.0368 & 0.0105 & 0.01 & 124 \\ \hline
SAFRAN & 0.0309 & 0.0127 & 0.02 & 120 \\ \hline
SAINT GOBAIN & 0.0112 & 0.0039 & 0.005 & 67 \\ \hline
SANOFI & 0.0159 & 0.0065 & 0.01 & 125 \\ \hline
SCHNEIDER ELECTRIC & 0.033 & 0.0132 & 0.02 & 129 \\ \hline
SOCIETE GENERALE & 0.0075 & 0.0032 & 0.005 & 123 \\ \hline
STELLANTIS NV & 0.0044 & 0.0016 & 0.002 & 129 \\ \hline
STMICROELECTRONICS & 0.0106 & 0.0038 & 0.005 & 129 \\ \hline
Salesforce, Inc. & 0.0606 & 0.0149 & 0.01 & 123 \\ \hline
Schlumberger Limited & 0.015 & 0.0062 & 0.01 & 124 \\ \hline
Sea Limited & 0.0591 & 0.0141 & 0.01 & 123 \\ \hline
Shopify Inc. & 0.0161 & 0.0065 & 0.01 & 124 \\ \hline
Snowflake Inc. & 0.1801 & 0.032 & 0.01 & 124 \\ \hline
Synchrony Financial & 0.0127 & 0.0057 & 0.01 & 124 \\ \hline
TOTALENERGIES & 0.0125 & 0.0057 & 0.01 & 125 \\ \hline
Taiwan Semiconductor Manufacturing Company Ltd. & 0.0157 & 0.0064 & 0.01 & 124 \\ \hline
The Boeing Company & 0.1027 & 0.0209 & 0.01 & 124 \\ \hline
The Coca-Cola Company & 0.0106 & 0.0052 & 0.01 & 124 \\ \hline
The Home Depot, Inc. & 0.1563 & 0.0283 & 0.01 & 124 \\ \hline
The Procter \& Gamble Company & 0.0244 & 0.0082 & 0.01 & 124 \\ \hline
The Progressive Corporation & 0.0608 & 0.0142 & 0.01 & 124 \\ \hline
The Walt Disney Company & 0.0273 & 0.0088 & 0.01 & 123 \\ \hline
Uber Technologies, Inc. & 0.0108 & 0.0052 & 0.01 & 123 \\ \hline
Union Pacific Corporation & 0.1062 & 0.0209 & 0.01 & 124 \\ \hline
VALEO & 0.0124 & 0.0042 & 0.005 & 99 \\ \hline
VALLOUREC & 0.012 & 0.0041 & 0.005 & 125 \\ \hline
VINCI & 0.017 & 0.0067 & 0.01 & 71 \\ \hline
Valero Energy Corporation & 0.0821 & 0.0175 & 0.01 & 124 \\ \hline
Visa Inc. & 0.0518 & 0.0133 & 0.01 & 124 \\ \hline
Wells Fargo \& Company & 0.0106 & 0.0052 & 0.01 & 124 \\ \hline
Welltower Inc. & 0.0452 & 0.0116 & 0.01 & 124 \\ \hline
    \end{longtable}
\end{scriptsize}

\subsection{From 2023-04-01 to 2023-09-30}

\begin{scriptsize}
\begin{longtable}{|>{\centering}p{0.3\textwidth}|c|c|c|c|}
    \hline
    \textbf{Company} & \textbf{spread} & \textbf{QPV} & \textbf{tick size} & \textbf{days} \endhead  \hline
    ADP & 0.1513 & 0.0657 & 0.1 & 127 \\ \hline
    AIR FRANCE -KLM & 0.001 & 0.0004 & 0.0005 & 105 \\ \hline
    AIR LIQUIDE & 0.0316 & 0.0129 & 0.02 & 127 \\ \hline
    AIRBUS & 0.0315 & 0.0129 & 0.02 & 127 \\ \hline
    AMG & 0.031 & 0.0096 & 0.01 & 127 \\ \hline
    ATOS & 0.0131 & 0.0044 & 0.005 & 82 \\ \hline
    AXA & 0.0076 & 0.0032 & 0.005 & 127 \\ \hline
    AbbVie Inc. & 0.0403 & 0.0109 & 0.01 & 124 \\ \hline
    Accenture plc & 0.1521 & 0.0266 & 0.01 & 124 \\ \hline
    Alibaba Group Holding Limited & 0.0176 & 0.0068 & 0.01 & 124 \\ \hline
    American Express Company & 0.0717 & 0.0154 & 0.01 & 124 \\ \hline
    BNP PARIBAS ACT.A & 0.0146 & 0.0062 & 0.01 & 127 \\ \hline
    BORR DRILLING & 0.1478 & 0.0466 & 0.05 & 122 \\ \hline
    Bank of America Corporation & 0.0101 & 0.005 & 0.01 & 124 \\ \hline
    Berkshire Hathaway Inc. & 0.1022 & 0.0193 & 0.01 & 124 \\ \hline
    Best Buy Co., Inc. & 0.0291 & 0.0091 & 0.01 & 124 \\ \hline
    Block, Inc. & 0.0288 & 0.0091 & 0.01 & 124 \\ \hline
    CREDIT AGRICOLE & 0.0032 & 0.0013 & 0.002 & 127 \\ \hline
    Carnival Corporation & 0.0101 & 0.005 & 0.01 & 124 \\ \hline
    Carvana Co. & 0.0452 & 0.0127 & 0.01 & 124 \\ \hline
    Caterpillar Inc. & 0.1184 & 0.0226 & 0.01 & 124 \\ \hline
    Chevron Corporation & 0.0318 & 0.0097 & 0.01 & 124 \\ \hline
    Citigroup Inc. & 0.0104 & 0.0051 & 0.01 & 124 \\ \hline
    ConocoPhillips & 0.0316 & 0.0096 & 0.01 & 124 \\ \hline
    D.R. Horton, Inc. & 0.0498 & 0.0124 & 0.01 & 124 \\ \hline
    DANONE & 0.0155 & 0.0065 & 0.01 & 127 \\ \hline
    DASSAULT SYSTEMES & 0.0108 & 0.0038 & 0.005 & 127 \\ \hline
    DERICHEBOURG & 0.0119 & 0.0042 & 0.005 & 82 \\ \hline
    Delta Air Lines, Inc. & 0.0105 & 0.0051 & 0.01 & 124 \\ \hline
    Devon Energy Corporation & 0.013 & 0.0058 & 0.01 & 124 \\ \hline
    DoorDash Inc & 0.056 & 0.0134 & 0.01 & 124 \\ \hline
    EDENRED & 0.0304 & 0.013 & 0.02 & 127 \\ \hline
    ENGIE & 0.0038 & 0.0014 & 0.002 & 127 \\ \hline
    EOG Resources, Inc. & 0.0555 & 0.0132 & 0.01 & 124 \\ \hline
    ESSILORLUXOTTICA & 0.0416 & 0.015 & 0.02 & 127 \\ \hline
    Exxon Mobil Corporation & 0.0165 & 0.0066 & 0.01 & 124 \\ \hline
    FDJ & 0.0407 & 0.0154 & 0.02 & 127 \\ \hline
    FLOW TRADERS & 0.0368 & 0.0145 & 0.02 & 62 \\ \hline
    FORVIA & 0.0197 & 0.0074 & 0.01 & 69 \\ \hline
    Fidelity National Information Services, Inc. & 0.0236 & 0.008 & 0.01 & 124 \\ \hline
    Ford Motor Company & 0.01 & 0.005 & 0.01 & 124 \\ \hline
    Freeport-McMoRan Inc. & 0.0106 & 0.0051 & 0.01 & 124 \\ \hline
    General Motors Company & 0.0103 & 0.0051 & 0.01 & 124 \\ \hline
    HERMES INTL & 0.4611 & 0.1589 & 0.2 & 117 \\ \hline
    Halliburton Company & 0.0109 & 0.0053 & 0.01 & 124 \\ \hline
    JPMorgan Chase \& Co. & 0.0225 & 0.0079 & 0.01 & 124 \\ \hline
    Johnson \& Johnson & 0.0239 & 0.0081 & 0.01 & 124 \\ \hline
    KERING & 0.1593 & 0.0651 & 0.1 & 82 \\ \hline
    KLEPIERRE & 0.0168 & 0.0068 & 0.01 & 127 \\ \hline
    L'OREAL & 0.0848 & 0.0335 & 0.05 & 127 \\ \hline
    LHYFE & 0.043 & 0.012 & 0.01 & 127 \\ \hline
    LVMH & 0.1292 & 0.0575 & 0.1 & 127 \\ \hline
    Lamb Weston Holdings, Inc. & 0.0607 & 0.0143 & 0.01 & 124 \\ \hline
    Lowe's Companies, Inc. & 0.0908 & 0.0182 & 0.01 & 124 \\ \hline
    MICHELIN & 0.0143 & 0.0063 & 0.01 & 127 \\ \hline
    Marathon Petroleum Corporation & 0.0611 & 0.0142 & 0.01 & 124 \\ \hline
    Marsh \& McLennan Companies, Inc. & 0.0814 & 0.0165 & 0.01 & 124 \\ \hline
    Mastercard Incorporated & 0.1537 & 0.0268 & 0.01 & 124 \\ \hline
    McDonald's Corporation & 0.0741 & 0.0156 & 0.01 & 124 \\ \hline
    NIKE, Inc. & 0.0219 & 0.0077 & 0.01 & 124 \\ \hline
    NIO Inc. & 0.0101 & 0.005 & 0.01 & 124 \\ \hline
    Norwegian Cruise Line Holdings Ltd. & 0.0102 & 0.0051 & 0.01 & 124 \\ \hline
    OCI & 0.0282 & 0.0091 & 0.01 & 127 \\ \hline
    OKEA & 0.0728 & 0.0212 & 0.02 & 122 \\ \hline
    ORANGE & 0.0032 & 0.0013 & 0.002 & 126 \\ \hline
    ORPEA & 0.0059 & 0.0015 & 0.0005 & 83 \\ \hline
    Occidental Petroleum Corporation & 0.0134 & 0.0059 & 0.01 & 124 \\ \hline
    Oracle Corporation & 0.0232 & 0.008 & 0.01 & 124 \\ \hline
    Pfizer Inc & 0.0101 & 0.005 & 0.01 & 124 \\ \hline
    REMY COINTREAU & 0.1109 & 0.0401 & 0.05 & 127 \\ \hline
    RENAULT & 0.0128 & 0.0042 & 0.005 & 127 \\ \hline
    Roblox Corporation & 0.0185 & 0.007 & 0.01 & 124 \\ \hline
    Royal Caribbean Group & 0.0504 & 0.0126 & 0.01 & 124 \\ \hline
    SAFRAN & 0.0343 & 0.0136 & 0.02 & 127 \\ \hline
    SAINT GOBAIN & 0.0172 & 0.0068 & 0.01 & 124 \\ \hline
    SANOFI & 0.0181 & 0.0069 & 0.01 & 79 \\ \hline
    SCHNEIDER ELECTRIC & 0.0334 & 0.0133 & 0.02 & 127 \\ \hline
    SOCIETE GENERALE & 0.008 & 0.0033 & 0.005 & 127 \\ \hline
    STELLANTIS NV & 0.0043 & 0.0015 & 0.002 & 127 \\ \hline
    STMICROELECTRONICS & 0.0107 & 0.0038 & 0.005 & 127 \\ \hline
    Salesforce, Inc. & 0.0794 & 0.017 & 0.01 & 124 \\ \hline
    Schlumberger Limited & 0.0124 & 0.0056 & 0.01 & 124 \\ \hline
    Sea Limited & 0.0383 & 0.0108 & 0.01 & 124 \\ \hline
    Shopify Inc. & 0.017 & 0.0067 & 0.01 & 124 \\ \hline
    Snowflake Inc. & 0.1304 & 0.0248 & 0.01 & 124 \\ \hline
    Synchrony Financial & 0.0107 & 0.0052 & 0.01 & 124 \\ \hline
    TOTALENERGIES & 0.013 & 0.0058 & 0.01 & 127 \\ \hline
    Taiwan Semiconductor Manufacturing Company Ltd. & 0.0161 & 0.0065 & 0.01 & 124 \\ \hline
    The Boeing Company & 0.0949 & 0.0196 & 0.01 & 124 \\ \hline
    The Coca-Cola Company & 0.0102 & 0.0051 & 0.01 & 124 \\ \hline
    The Home Depot, Inc. & 0.1083 & 0.0208 & 0.01 & 124 \\ \hline
    The Procter \& Gamble Company & 0.0239 & 0.0081 & 0.01 & 124 \\ \hline
    The Progressive Corporation & 0.0583 & 0.0138 & 0.01 & 124 \\ \hline
    The Walt Disney Company & 0.0166 & 0.0066 & 0.01 & 124 \\ \hline
    Uber Technologies, Inc. & 0.0111 & 0.0053 & 0.01 & 124 \\ \hline
    Union Pacific Corporation & 0.089 & 0.0179 & 0.01 & 124 \\ \hline
    VALEO & 0.0128 & 0.0043 & 0.005 & 102 \\ \hline
    VALLOUREC & 0.0116 & 0.0041 & 0.005 & 118 \\ \hline
    VINCI & 0.0294 & 0.0125 & 0.02 & 126 \\ \hline
    Valero Energy Corporation & 0.0658 & 0.015 & 0.01 & 124 \\ \hline
    Visa Inc. & 0.0509 & 0.0128 & 0.01 & 124 \\ \hline
    Wells Fargo \& Company & 0.0103 & 0.0051 & 0.01 & 124 \\ \hline
    Welltower Inc. & 0.0444 & 0.0114 & 0.01 & 124 \\ \hline
    \end{longtable}
\end{scriptsize}

\end{document}